\documentclass{iopart}

\usepackage{graphicx}
\usepackage{bm}
\usepackage{amssymb}

\newcommand{\bra}[1]{\langle #1|}
\newcommand{\ket}[1]{|#1\rangle}

\begin{document}

\topical{Spin effects in single electron tunneling}

\author{J. Barna\'s$^{1,2}$ and I. Weymann$^1$}
\address{$^1$ Department of Physics, Adam
Mickiewicz University, 61-614 Pozna\'n, Poland}
\address{$^2$ Institute of Molecular Physics, Polish
Academy of Sciences, 60-179 Pozna\'n, Poland}
\ead{barnas@amu.edu.pl}\ead{weymann@amu.edu.pl}

\date{\today}

\begin{abstract}
An important consequence of the discovery of giant
magnetoresistance in metallic magnetic multilayers is a broad
interest in spin dependent effects in electronic transport through
magnetic nanostructures. An example of such systems are tunnel
junctions -- single-barrier planar junctions or more complex ones.
In this review we present and discuss recent theoretical results
on electron and spin transport through ferromagnetic mesoscopic
junctions including two or more barriers. Such systems are also
called ferromagnetic single-electron transistors. We start from
the situation when the central part of a device has the form of a
magnetic (or nonmagnetic) metallic nanoparticle. Transport
characteristics reveal then single-electron charging effects,
including the Coulomb staircase, Coulomb blockade, and Coulomb
oscillations. Single-electron ferromagnetic transistors based on
semiconductor quantum dots and large molecules (especially carbon
nanotubes) are also considered. The main emphasis is placed on the
spin effects due to spin-dependent tunnelling through the
barriers, which gives rise to spin accumulation and tunnel
magnetoresistance. Spin effects also occur in the current-voltage
characteristics, (differential) conductance, shot noise, and
others. Transport characteristics in the two limiting situations
of weak and strong coupling are of particular interest. In the
former case we distinguish between the sequential tunnelling and
cotunnelling regimes. In the strong coupling regime we concentrate
on the Kondo phenomenon, which in the case of transport through
quantum dots or molecules leads to an enhanced conductance and to
a pronounced zero-bias Kondo peak in the differential conductance.
\end{abstract}


\maketitle

\tableofcontents


\section{Introduction}


Since last few decades one can observe a common pursuit towards
miniaturization of electronic systems. According to the empirical
Moore's law, the number of transistors per microchip is doubled
every two years. It is however obvious that the possibility of
further miniaturization will be stopped in the near future due to
the loss of chips' stability when the device components achieve
critical dimensions in the nanometer range. Thus, the most
challenging task facing contemporary science and technology is to
implement structures, alternative to the silicon-based devices,
whose size could be reduced further.

The same tendency can also be observed in the development of hard
discs, whose memory cells become smaller and smaller every year.
This generates new challenges related to information
reading/writing. An important step towards further miniaturization
was the discovery of the giant magnetoresistance (GMR) effect. The
GMR was discovered in 1988 in artificially layered metallic
structures consisting of ferromagnetic 3d films separated by
nonmagnetic metallic layers \cite{fertGMR,binasch89}. It turned
out that electrical resistance of magnetic metallic multilayers
depends on their magnetic state, and usually drops when magnetic
configuration varies from antiparallel alignment to the parallel
one \cite{fertGMR,binasch89,barnasGMR90,barnasGMR90a}. The GMR
effect occurs for current flowing in the film plane as well as
perpendicularly to it. This phenomenon turned out to be very
useful for applications in highly sensitive read heads, and
allowed reading information from smaller memory cells using their
weak magnetic field.

The discovery of GMR initiated broad interest in spin polarized
electronic transport in nanoscopic systems. It turned out that
electron spin provides an additional degree of freedom, which
considerably broadens the range of applications of mesoscopic
systems in novel electronic devices. The spin-based
nanoelectronics -- called now spin electronics or shortly
spintronics --  is a relatively new area of mesoscopic physics
dealing with the interplay of charge and spin degrees of freedom
\cite{review:spintronics:wolf01,book:ziese,gregg02,book:maekawa02,book:loss02,
review:spintronics:zutic04,review:spintronics:dyakonov04,book:maekawa06}.
Although a lot of theoretical and experimental works on the
spintronic properties of mesoscopic systems have been carried out,
this field is still in an early stage of development.

An effect similar to the aforementioned GMR also occurs when
nonmagnetic metallic layer in a trilayer structure is replaced
with a nonmagnetic insulating barrier, and the current flows owing
to the phenomenon of quantum-mechanical tunnelling through the
barrier. This effect was discovered in 1975 in ferromagnetic
planar junctions by Julli\'ere \cite{julliere}, and is of current
interest due to applications in magnetic storage technology
(Magnetic Random Access Memories) and in spin-electronics devices
\cite{review:spintronics:wolf01,book:ziese,gregg02,book:maekawa02,book:loss02,
review:spintronics:zutic04,review:spintronics:dyakonov04,book:maekawa06}.
As in the case of GMR, the tunnel magnetoresistance (TMR) consists
in a decrease (increase is also possible) in the junction
resistance when magnetic configuration of the junction changes
from the antiparallel to parallel one. Tunneling in complex
junctions, particularly in mesoscopic ones, where charging effects
become important, is still not fully explored. A specific kind of
such systems are double-barrier junctions with a small central
electrode (called an island). Such systems are known as
single-electron transistors (SETs), mainly because electrons in a
biased device flow one by one through the system and the transfer
of single electrons can be controlled by a gate voltage.
Electronic transport in such devices was extensively studied in
the past decade, but mainly in the nonmagnetic limit
\cite{book:averin91,book:SingleChargeTunneling,
book:datta95,book:ferry97, book:kouwenhoven97,book:schon98,
book:likharev99,delft01,book:heinzel03}.

Recent experiments on magnetic nano-structured materials revealed
new phenomena associated with the interplay of ferromagnetism and
discrete charging effects. A typical example is a ferromagnetic
single-electron transistor, i.e., a small metallic nanoparticle
(semiconducting quantum dots or molecules are also possible)
coupled by tunnel barriers to ferromagnetic electrodes
\cite{seneor_review,takanashi_review}. First ferromagnetic
single-electron transistors were fabricated by Ono et al.
\cite{ono96,ono97} and later by Br\"uckl et al. \cite{bruckl98}.
Transport in ferromagnetic single-electron transistors with
nonmagnetic metallic islands -- both normal and superconducting --
was also measured
\cite{chenPRL02,chenJAP02,shyuJAP03,johanssonPRL03}. One should
bear in mind, that the interplay of spin and charge effects was
already studied long time ago in granular systems, in which
magnetic nanoparticles were randomly dispersed in a nonmagnetic
matrix \cite{abeles}. Recently granular films were investigated
again by several groups \cite{schelpPRB97,fettarPRB02,mitaniPRL98,
imamuraPRB00,yakushijiAPL01,yakushijiJAP02,ernult04,yakushijiN05},
but both the size of the grains and also their location were
strictly controlled.

In the case of sufficiently large metallic islands (but still in
the nanometer range), discrete structure of the electronic states
in the grain is not resolved and is irrelevant. To observe the
discrete electronic states in transport characteristics one should
either diminish size of the metallic nanoparticles
\cite{deshmukh02}, or use semiconducting quantum dots based on
two-dimensional electron gas \cite{hamayaAPL07a,hamayaAPL07b,
hamayaAPL07c,hamayaPRB08}. An alternative strategy is to use
ferromagnetic semiconducting materials instead of metallic ones as
the electrodes \cite{chyePRB02}. Magnetic impurities in the middle
of the tunnel barrier of a ferromagnetic tunnel junction
\cite{jansenAPL99,tanoueJAP00,fertAPL06,parkinNL08} also can be
considered as quantum dots with a very strong Coulomb interaction.
Another group of single-electron devices are molecular
ferromagnetic transistors \cite{pasupathy04} and, especially,
ferromagnetically contacted carbon nanotubes
\cite{tsukagoshi99,zhaoAPL02,zhaoJAP02,jensen03,
sahooNP05,jensenPRB05,manPRB06,liuPRB06,
nagabhiravaAPL06,cottetSST06,schonenbergerSST06}.

In the following we will review basic transport characteristics of
ferromagnetic single-electron transistors in the sequential
tunnelling, cotunneling, and strong coupling (Kondo) regimes. In
particular, we will consider such properties of the device like
charge and spin currents, tunnel magnetoresistance, spin
accumulation, shot noise, Kondo effect, and others. In section 2
we review basic principles of single-electron transport in the
case of a large metallic island attached to ferromagnetic leads.
The size of the island is however small enough so that the
charging energy is the dominant energy scale in the system. The
limits of fast and slow spin relaxation in the island are also
discussed. Then we consider shot noise and the role of discrete
electronic structure in such devices. In section 3 we consider
devices based on double metallic islands. In turn, in section 4 we
deal with electronic transport through single-level quantum dots
(QDs) in the sequential tunnelling and cotunneling regimes.
Transport through multi-level QDs, including transport through
two-level quantum dots and carbon nanotubes attached to
ferromagnetic leads is discussed in section 5. Transport through
quantum dots in the Kondo regime is briefly addressed in section
6. Final conclusions are in section 7.


\subsection{Basic concepts}


The systems considered in this review consist of a mesoscopic
central part (island) coupled by tunnel barriers to external
ferromagnetic leads. The central part is characterized by an
addition energy, which corresponds to the energy needed for adding
a single electron, and includes contributions from the
electrostatic charging energy and the discrete single-particle
level separation. If the charging energy is the most relevant
energy scale, the systems exhibit the single-electron charging
effects \cite{book:SingleChargeTunneling,fulton87,mullen88,
likharev88,geerlings90,meir91,meir92,devoret92}. An electron can
tunnel to the central part only when the energy provided by the
transport voltage surpasses the corresponding charging energy;
otherwise, the current is exponentially suppressed and the system
is in the Coulomb blockade regime
\cite{book:averin91,averin86,beenakker91}. Once the bias voltage
is larger than the threshold voltage, the electrons can tunnel one
by one through the system leading to the step-like current-voltage
characteristics -- so-called Coulomb staircase. The blockade can
also be overcome by applying a gate voltage $V_g$ that leads to
sawtooth-like variation of electric current with $V_g$ --
so-called Coulomb oscillations
\cite{book:SingleChargeTunneling,fulton87,mullen88,meir91,kouvenhoven91}.

When the electrodes are made of a ferromagnetic material, the
system exhibits further interesting phenomena resulting from the
interplay of charge and spin degrees of freedom
\cite{barnas98,takahashi98}. In particular, the tunnelling current
flowing through the system depends on the relative alignment of
the magnetic moments of ferromagnetic electrodes, giving rise to
the TMR effect \cite{ono96,ono97}, which is described
quantitatively by the ratio
\begin{equation}\label{Eq:TMR}
  {\rm TMR} = \frac{R_{\rm AP}-R_{\rm P}}{R_{\rm P}} =
    \frac{I_{\rm P}-I_{\rm AP}}{I_{\rm AP}}\;,
\end{equation}
where $R_{\rm P}$ and $R_{\rm AP}$ denote the total system
resistance in the parallel and antiparallel magnetic
configurations, respectively, and $I_{\rm P}$, $I_{\rm AP}$ are
the corresponding currents. A simple theoretical model of TMR was
introduced by Julli\`{e}re \cite{julliere}, who considered a
single planar ferromagnetic tunnel junction and showed that TMR in
such a device is given by ${\rm TMR^{Jull}} = 2p_{\rm L}p_{\rm R}/
(1-p_{\rm L}p_{\rm R})$, where the spin polarization $p_{r}$ of
the lead $r$ ($r=L,R$ for the left/right lead) is defined as
\begin{equation} \label{Eq:polarization}
   p_{r}=\frac{\rho_{r+} - \rho_{r-}}{\rho_{r+} + \rho_{r-}}\,,
\end{equation}
with $\rho_{r\pm}$ being the spin-dependent density of states of
lead $r$ for the spin-majority $(+)$ and spin-minority $(-)$
electrons.

\begin{figure}[t]
  \begin{center}
  \includegraphics[width=0.45\columnwidth]{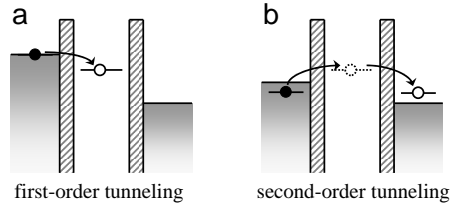}
  \caption{\label{Sec1:Fig1}
  A sketch illustrating (a) a single first-order (sequential tunnelling) process,
  and (b) a second-order (cotunneling)  transport process.}
  \end{center}
\end{figure}

\subsection{Transport regimes}

In the following considerations we will distinguish between the
three different transport regimes:

{\it Sequential tunnelling} -- In the regime of weak coupling
between the island and leads, and out of the Coulomb blockade,
electron transport is dominated by processes of the first order in
the coupling parameter. Electrons flow then consecutively one by
one due to the tunnelling events, and the transition rate from an
initial state $|i\rangle$ to a final state $|f\rangle$ can be
determined from the Fermi golden-rule,
\begin{equation}\label{Sec1:Eq:seqrate}
 \alpha_{ i \rightarrow f}=\frac{2 \pi}{ \hbar} \left|
 \langle i|H_{ T}|f \rangle\right|^2 \delta(\varepsilon_{ i} - \varepsilon_{ f}
 ),
\end{equation}
where $H_{T}$ is the relevant tunnelling Hamiltonian, and
$\varepsilon_{i}$ ($\varepsilon_{f}$) is the energy of the initial
(final) state. This transport regime is known as the sequential
tunnelling regime \cite{book:averin91,book:SingleChargeTunneling}.
An example of a single first-order process is sketched in
Fig.~\ref{Sec1:Fig1}a.

{\it Cotunneling} -- Although the sequential tunnelling in the
Coulomb blockade regime is exponentially suppressed, the current
still flows due to higher-order tunnelling processes involving
tunnelling of a single, two or more electrons {\it via}
intermediate virtual states
\cite{odintsov89,averin90,geerligs90cot}. These processes are
known as cotunneling. An exemplary cotunneling process is
illustrated in Fig.~\ref{Sec1:Fig1}b.

The second order perturbation theory gives the cotunneling rate
from an initial state $|i\rangle$ to a final state $|f\rangle$
\cite{averin90}
\begin{eqnarray}\label{Sec1:Eq:cotrate}
 \alpha_{ i \rightarrow f}=\frac{2 \pi}{ \hbar} \left|
 \sum_q \frac{ \langle i|H_{ T}|q \rangle \langle q|H_{T}|f \rangle }
 {\varepsilon_{ i} - \varepsilon_{ q}}
 \right|^2 \delta(\varepsilon_{ i} - \varepsilon_{ f} )
\end{eqnarray}
where the summation is over all virtual states $|q\rangle$, and
$\varepsilon_{\alpha}$ is the energy od the state $\alpha$
($\alpha =i,f,q$). In the Coulomb blockade regime this tunnelling
rate is only algebraically suppressed, contrary to the sequential
transport processes which are then suppressed exponentially.
Because of that even at low temperatures and in the strong Coulomb
blockade regime the rates of cotunneling processes do not vanish.
The second-order corrections become also important on resonance
for intermediate coupling strengths.

{\it Strong coupling} -- For strong coupling of the metallic
island to electrodes, the tunnelling processes lead to logarithmic
corrections to conductance, and the perturbation theory fails at
the degeneracy points of two consecutive charge states. In the
case of quantum dots additionally the Kondo effect appears at low
temperatures (below the Kondo temperature $T_{K}$, $T \lesssim
T_{K} $) leading to an enhanced conductance in the linear response
regime \cite{book:hewson}.


\section{Ferromagnetic single-electron transistors based on
metallic nanoparticles}


In this section we review spin-polarized transport in a metallic
ferromagnetic single-electron transistor (FM SET). The device
consists of a metallic nanoparticle as the central electrode
(island), which is coupled through tunnel barriers to external
reservoirs of spin polarized electrons. A gate voltage is attached
capacitively to the island, which allows to control position of
the corresponding Fermi level.

Electronic transport in nonmagnetic SETs was already extensively
studied in the past two decades
\cite{book:averin91,book:SingleChargeTunneling,kastner92,
devoret98,korotkov90,averinPRB91, amman91,hanna91,
whan96,hirvi96,melsen97,park00,weymannJJAP03}. Recently, the
attention was also drawn to electron tunnelling in magnetic
systems
\cite{ono96,ono97,barnas98,takahashi98,moodera95,moodera96,imamura00},
which was stimulated by recent progress in nanotechnology. It has
been shown theoretically that some qualitatively new effects may
arise from the interplay of charging effects and spin degrees of
freedom. These include, for example, oscillations of TMR with
increasing bias voltage, spin accumulation, enhancement of TMR in
the Coulomb blockade regime, etc.
\cite{barnas98,takahashi98,barnasEL98,majumdar98,
brataas99,korotkov99,martinekPRB02,weymannPSSb03}. The enhancement
of TMR in the cotunneling regime and the oscillations of TMR as a
function of the transport voltage have also been observed
experimentally
\cite{yakushijiJAP02,ernult04,takanashi00,mitani98}.

The capacitance $C$ of few-nanometer-size particles is of the
order of $10^{-18}$ F \cite{ernult04,deshmukh02,pashkin00}.
Consequently, the corresponding charging energy, $E_{C}=e^2/2C$,
establishes a new relevant energy scale. If the charging energy is
larger than the thermal energy, $E_{ C}\gg k_{\rm B}T$, where $T$
denotes temperature and $k_{\rm B}$ is the Boltzmann constant, the
effects due to discreteness of charge become observable in
transport characteristics
\cite{book:averin91,book:SingleChargeTunneling}.

\begin{figure}[t]
  \begin{center}
  \includegraphics[width=0.8\columnwidth]{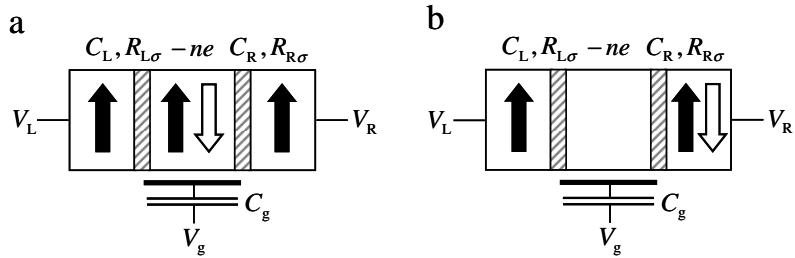}
  \caption{\label{Sec2:Fig1}
    Schematics of ferromagnetic single-electron transistors.
    The parallel and antiparallel magnetic configurations
    of the system are also specified.
    The island is separated from external electrodes by tunnel barriers.
    Each barrier is characterized by its spin-dependent resistance $R_{r\sigma}$ and
    capacitance $C_r$ ($r={L,R}$). The system is symmetrically
    biased, $V_{ L}=V/2$, $V_{ R}=-V/2$, and there is also
    a gate voltage $V_{g}$ applied to the island.}
  \end{center}
\end{figure}

The single-electron transistors considered in this section are
illustrated schematically in Fig.~\ref{Sec2:Fig1}, where part (a)
shows a device whose all three electrodes are ferromagnetic,
whereas part (b) shows a system with ferromagnetic source and
drain electrodes and a nonmagnetic island. Generally, magnetic
moments of the leads may form an arbitrary magnetic configuration.
However, we focuss on systems whose moments form either parallel
(P) or antiparallel (AP) magnetic configuration, as shown in
Fig.~\ref{Sec2:Fig1}. In the system shown in part (a) the magnetic
moments of external electrodes point in the same direction, while
magnetic moment of the island is either parallel or antiparallel
to them. In the system shown in part (b), magnetic moments of
external electrodes in the parallel configuration are aligned,
while in the antiparallel configuration they are anti-aligned. The
two magnetic alignments can be easily achieved by sweeping
magnetic field through the hysteresis loop, provided the
respective ferromagnetic components have different coercive
fields. One may also make use of exchange anisotropy to fix
magnetic moment of a particular layer and rotate magnetic moment
of the second layer with a weak magnetic field. Generally, there
is a chance that electron tunnelling through a barrier will change
its spin orientation. However, we consider only spin-conserving
tunnelling processes through the two barriers. First, we assume
that the islands are relatively large, so the effects due to
quantization of the corresponding energy levels can be neglected.
For smaller islands, however, the discrete energy spectrum may
modify transport characteristics
\cite{yakushijiN05,martinek99,barnasPRB00}, and this will be
considered later (see section \ref{discreteSET}).

To describe charge and spin transport we need to specify a model
Hamiltonian of the system. First, the electrostatic energy
required to add $n$ excess electrons to the island, while keeping
constant voltages $V_{ L}$ and $V_{ R}$ in the left and right
electrodes and the gate voltage $V_{ g}$, is given by \cite
{schoeller94}
\begin{eqnarray} \label{Eq:charging}
  H_{\rm ch}=E_{ C} (n-\frac{Q_g}{e})^2 \; ,
\end{eqnarray}
where a constant term (independent of $n$) is irrelevant and has
been dropped. Here, the total island's capacitance $C$ is the sum
of the capacitances of the left and right junctions and of the
gate, $C=C_{L}+C_{ R}+C_{ g}$. The external charge $Q_g \equiv
C_{L} V_{L}+C_{R} V_{R}+C_{g} V_{g}$ accounts for the effect of
applied voltages, and can be continuously tuned. The total
Hamiltonian of the device takes then the general form
\begin{equation} \label{Eq:hamiltonian1}
   H = \sum_{ r=L,R} H_{r} + H_{I}+H_{
   ch}+H_{ T} \equiv H_0+H_{ T} \; .
\end{equation}
Here $H_0$ describes the decoupled leads and island, while $H_{
T}$ takes into account the lead-island coupling. The ferromagnetic
leads and the island are described by
\begin{equation} \label{Eq:hamiltonian2}
   H_{ r} = \sum_{{\mathbf k}
   \sigma}\varepsilon_{r {\mathbf k} \sigma} c^\dagger_{ r {\mathbf k} \sigma} c_{ r {\mathbf k}
   \sigma} \; ,
\end{equation}
for $r=L,R,I$, where $c_{r {\mathbf k} \sigma}$ are the Fermi
operators for electrons with a wavevector ${\mathbf k}$ and spin
$\sigma$ in the electrodes and island ($r=L,R,I$), and
$\varepsilon_{r {\mathbf k} \sigma}$ is the corresponding
single-particle energy.
The last part of the Hamiltonian,
\begin{equation} \label{Eq:tunneling_metal}
   H_{T}=\sum_{r = {L,R}}\sum_{{\mathbf k} {\mathbf q} \sigma}
   t_{r {\mathbf k} {\mathbf q} \sigma }
   c^\dagger_{{r}{\mathbf k} \sigma}c_{I{\mathbf q}\sigma}+
   {\rm h.c.} \,
\end{equation}
describes tunnelling processes between the leads and island, with
$t_{r {\mathbf k} {\mathbf q} \sigma }$ being the relevant matrix
elements.


\subsection{Transport in the sequential tunnelling regime}


When the resistances of both tunnel barriers are much larger than
the quantum resistance, $R_{r}\gg R_{Q}\equiv h/e^2$, and the
system is not in the Coulomb blockade regime, transport is
dominated by sequential tunnelling processes. As a consequence,
the charge is well localized in the island and the {\it orthodox}
tunnelling theory is applicable
\cite{book:averin91,book:SingleChargeTunneling}.

In order to calculate electric current in the sequential transport
regime, one may use the method based on the master equation which
is a detailed balance of electrons tunnelling to and off the
island. In the stationary case the master equation reads

\begin{eqnarray}
  0&=&- \sum\limits_{\sigma} \left[ \Gamma _{L\sigma}^+(n)
  + \Gamma _{L\sigma}^-(n)+\Gamma_{R\sigma}^+(n)
  + \Gamma _{R\sigma}^-(n) \right] P(n,V)\nonumber \\
  &&+\sum\limits_{\sigma}
  \left[\Gamma _{L\sigma}^+(n-1)
  +\Gamma _{R\sigma}^+(n-1)\right]P(n-1,V)
  \nonumber\\
  &&+\sum\limits_{\sigma}
  \left[\Gamma _{L\sigma}^-(n+1)
  +\Gamma _{R\sigma}^-(n+1)\right]P(n+1,V),
\end{eqnarray}
where $P(n,V)$ is the probability to find the island in a state
with $n$ additional electrons when a bias voltage $V$
($V=V_L-V_R$) is applied, and $\Gamma _{r\sigma }^\pm (n)$ is the
spin-dependent rate for tunnelling of electrons with spin $\sigma$
from the lead $r$ to island (upper sign) and from the island to
lead $r$ (lower sign), when the island is occupied by $n$ excess
electrons. These tunnelling rates depend on the bias voltage (not
indicated explicitly), and can be expressed by means of the Fermi
golden rule as
\begin{equation}\label{Eq:SETrate}
  \Gamma _{r\sigma }^\pm (n)=
  \frac{1}{e^{2}R_{r\sigma} } \frac{\Delta
  E_{r\sigma}^\pm (n)}{\exp \left[
  \Delta E_{r\sigma}^\pm (n) / k_{\rm B}T
  \right] -1}\,,
\end{equation}
where $\Delta E_{r\sigma}^\pm (n)$ describes a change in the
electrostatic energy of the system caused by the corresponding
tunnelling event, when in the initial state there were $n$
additional electrons on the island. In the above equation
$R_{r\sigma}$ denotes the spin-dependent tunnel resistance of the
$r$-th barrier, given by
\begin{equation}\label{Eq:SETbarrierR}
  R_{r\sigma} = \frac{\hbar} {2\pi e^2 \rho_{r\sigma} \rho_{ I\sigma}
  |t_{r\sigma}|^2},
\end{equation}
where $t_{r {\mathbf k} {\mathbf q} \sigma}= t_{r\sigma}$ has been
assumed for simplicity. The spin dependence of the resistance is a
consequence of the spin-dependent density of electron states at
the Fermi level in the respective electrodes  and the
corresponding tunnelling matrix elements. In particular, in
Eq.~(\ref{Eq:SETbarrierR}) it is due to the spin-dependent density
of states of the $r$-th lead, $\rho_{r\sigma}$, spin-dependent
density of states of the island, $\rho_{I\sigma}$, as well as due
to the spin-dependent tunnelling matrix elements $t_{r\sigma}$.

In a stationary state, the net transition rate between the charge
states with $n$ and $n+1$ excess electrons on the island vanishes.
The probability $P(n,V)$ can be then determined from the following
recursion relation \cite{amman91,hanna91}
\begin{equation}\label{Eq:SETmaster}
  P(n+1,V) \sum_{ \sigma}  y^{ \sigma}(n+1) =P(n,V)
  \sum_{\sigma} x^{ \sigma}(n) \,,
\end{equation}
where $x^{\sigma }(n)=\sum_{r={L,R}}\Gamma _{r\sigma }^+(n)$ and
$y^{\sigma }(n)=\sum_{r={L,R}}\Gamma _{r\sigma }^-(n)$,
corresponding to transition rates for tunnelling to and off the
island, respectively.

Generally, energy of an electron after tunnelling event is relaxed
to the relevant Fermi level in a short time scale. One can assume
that the energy relaxation time is the shortest time scale,
shorter than the time between two successive tunnelling events.
However, such a restriction cannot be imposed on the spin
relaxation time which can be relatively long. In a general case, a
nonequilibrium magnetic moment may accumulate on the island due to
the spin dependence of tunnelling processes, which leads to spin
splitting of the corresponding Fermi level. For arbitrary spin
relaxation times, the splitting of the Fermi level can be
determined from the spin balance \cite{korotkov99,barnasJMMM99}
\begin{equation}\label{Eq:SETspinconser}
  \frac{1}{e}(I_{ R}^{\sigma }-I_{L}^{\sigma })
  -\frac{\rho_{I\sigma}\Omega _{I}}{\tau_{ sf}}
  \Delta E_{F}^{\sigma }=0 \,,
\end{equation}
where $\Omega _{I}$ is the island's volume, $-e$ is the electron
charge ($e>0$), $\tau_{ sf}$ denotes the spin relaxation time on
the island, while $I_{L}^{\sigma }$ and $I_{R}^{\sigma }$ are the
currents flowing through the left and right junctions in the spin
channel $\sigma$. The last term in Eq.~(\ref{Eq:SETspinconser})
takes into account intrinsic spin-flip processes on the island.
>From this condition it is possible to calculate self-consistently
the shifts of the Fermi levels due to spin accumulation for both
spin orientations.

The electric current flowing through the left junction can be
calculated from the following formula:
\begin{equation} \label{Eq:SETcurrent}
  I_{L} = \sum\limits_{\sigma }I_{L}^\sigma =-e\sum\limits_{\sigma }\sum_{n=-\infty }^{\infty
  }\left[ \Gamma _{L\sigma }^+(n)-\Gamma _{L\sigma }^-(n) \right] P(n,V)
  \,.
\end{equation}
Similar formula also holds for $I_R$. In the stationary state the
currents flowing through both junctions are equal,
$I_{L}=I_{R}\equiv I$.

In the following we discuss two limiting cases: the limit of fast
spin relaxation and the limit of slow spin relaxation on the
island. In the former case the spin of an electron tunnelling to
the island relaxes before a next tunnelling event takes place. In
the latter case, on the other hand, the electron spin is conserved
for a time much longer than the time between successive tunnelling
events.

\subsubsection{Fast spin relaxation}

In the limit of fast spin relaxation, SETs with nonmagnetic
islands behave like nonmagnetic junctions, exhibiting no TMR
effect. Therefore, the following discussion in this subsection is
limited to FM SETs with ferromagnetic islands, see
Fig.~\ref{Sec2:Fig1}a. Generally, one may assume that the
spin-dependent resistances for the parallel ($R_{r\sigma}^{\rm
P}$) and antiparallel ($R_{r\sigma}^{\rm AP}$) configurations
fulfill the condition $ R_{r\sigma}^{\rm P}R_{r\bar{\sigma}}^{\rm
P}= R_{r\sigma }^{\rm AP}R_{r \bar{\sigma}}^{\rm AP}$, where
$\bar{\sigma} \equiv -\sigma$. This formula follows from
Eq.~(\ref{Eq:SETbarrierR}) when assuming that all spin effects are
included into the spin dependent density of states (transfer
matrix element are independent of spin and magnetic
configuration). Moreover, since there is no spin accumulation on
the island in the limit of fast spin relaxation, there is no
associated spin splitting of the Fermi level, $\Delta
E_{F}^{\sigma }=0$.

\begin{figure}[t]
  \begin{center}
  \includegraphics[width=0.4\columnwidth]{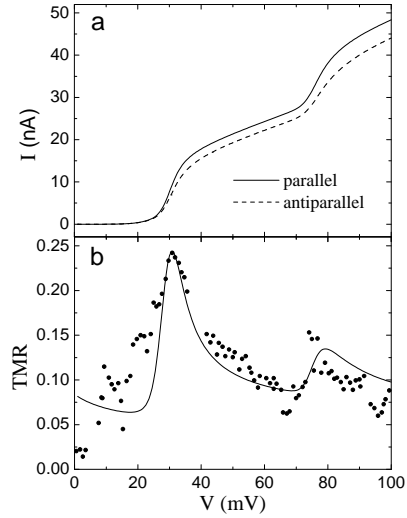}
  \caption{\label{Sec2:Fig2}
  Current in the parallel and antiparallel magnetic
  configurations (a) and the resulting TMR (b) as a function of
  the bias voltage for a FM SET with a ferromagnetic island
  in the limit of fast spin relaxation.
  The parameters are: $T=9$ K, $R_{L \uparrow}^{\rm P}=0.65$
  M$\Omega$, $R_{L \downarrow}^{\rm P}=0.065$ M$\Omega$,
  $R_{R \uparrow}^{\rm P}=5$  M$\Omega$,
  $R_{R \downarrow}^{\rm P}=2.5$ M$\Omega$,
  $R_{r \downarrow}^{\rm AP} = R_{r \uparrow}^{\rm AP}
  =(R_{r \uparrow}^{\rm P}R_{r \downarrow}^{\rm
  P})^{1/2}$ (for $r={L,R}$),
  $C_{L}=0.1$ aF,
  $C_{R}=1$ aF, $C_{g}=5.1$ aF, $V_{g}=0$, and
  the offset charge $Q_0=0.025e$.
  The dots in part (b) present the experimental data
  taken from Ref.~\cite{ernult04}.}
  \end{center}
\end{figure}

The current flowing through the system in the parallel and
antiparallel configurations is displayed in Fig.~\ref{Sec2:Fig2}a.
For both magnetic configurations, the $I-V$ curves reveal the
well-known Coulomb steps. Moreover, these two curves are
different; the current flowing in the parallel configuration is
generally larger than the current flowing in the antiparallel
configuration, see Fig.~\ref{Sec2:Fig2}a. This difference leads in
turn to nonzero TMR effect, as shown in Fig.~\ref{Sec2:Fig2}b. The
TMR effect has a component that oscillates as a function of the
bias voltage. The amplitude of these oscillations, however,
decreases as the transport voltage increases. For the parameters
assumed here, TMR reaches local maxima at the voltages
corresponding to the positions of Coulomb steps. However, this is
not a general rule, and for other parameters TMR can have local
minima at the Coulomb steps. The global maximum value of TMR in
Fig.~\ref{Sec2:Fig2}b appears at the first step, i.e., at the
threshold voltage. When the temperature increases, the effects due
to discrete charging, i.e. the Coulomb steps and enhancement of
TMR at the Coulomb steps become diminished and disappear at
$k_{\rm B}T\approx E_{C}$ \cite{weymannPSSb03}. The oscillatory
behavior of the TMR effect with increasing transport voltage was
observed experimentally, for example by Ernult {\it et
al.}~\cite{ernult04}, see the dots in Fig.~\ref{Sec2:Fig2}b. The
curves presented in Fig.~\ref{Sec2:Fig2} are calculated for the
parameters corresponding to those in Ref.~\cite{ernult04}. To get
a good agreement with experimental observations, a nonzero offset
charge $Q_0$ on the island (due to external charges) has been
assumed. As can be seen in Fig.~\ref{Sec2:Fig2}b, there is a
satisfactory agreement between the theoretical curve and
experimental data for voltages above the threshold, $V \gtrsim
30$mV, while in the Coulomb blockade regime transport calculated
using the sequential tunnelling approximation is not properly
described.

\subsubsection{Slow spin relaxation}

In the case of a FM SET with a nonmagnetic island, see
Fig.~\ref{Sec2:Fig1}b, a nonzero TMR in the sequential tunnelling
regime can exist only when the spin relaxation time is
sufficiently long, i.e. significantly longer than the time between
successive tunnelling events. We note that the longest spin
relaxation times were measured for aluminium and copper
\cite{lubzens76,johnson85} (for example, the relaxation time for
copper was estimated to be of the order of $10^{-7}$ s). If this
is the case, a nonequilibrium magnetic moment builds up on the
island due to spin accumulation. This moment leads a nonvanishing
TMR. In other words, the island becomes magnetized in a
nonequilibrium situation, and the created moment depends on the
bias and gate voltages.

\begin{figure}[t]
  \begin{center}
    \includegraphics[width=0.4\columnwidth]{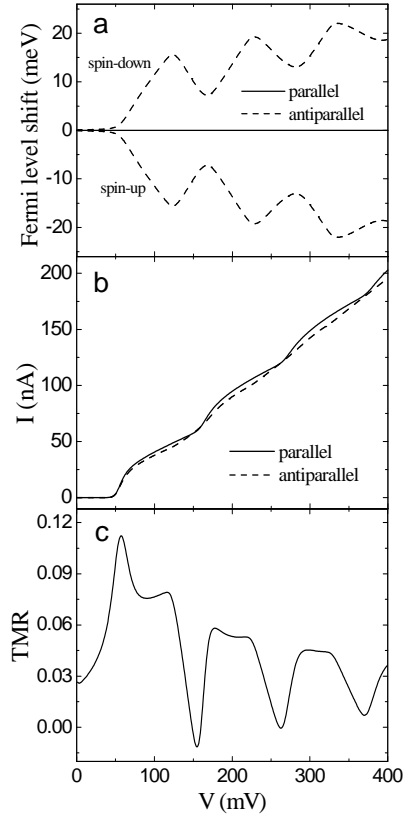}
    \caption{\label{Sec2:Fig3}
    Basic characteristics of a FM SET with a nonmagnetic island
    as a function of the
    bias voltage in the limit of slow spin relaxation on the island:
    (a) spin splitting of the Fermi level, (b) $I-V$ characteristics,
    and (c) tunnel magnetoresistance.
    The parameters are: $k_{\rm B}T=0.05\,E_{C}$,
    $R_{L \uparrow}^{\rm P}=5$ M$\Omega $,
    $R_{L \downarrow}^{\rm P}=2.5$ M$\Omega$,
    $R_{R \uparrow}^{\rm P}=0.3$ M$\Omega $,
    $R_{R \downarrow}^{\rm P}=0.15$ M$\Omega$,
    $C_{L}=C_{R}=C_{g}=1$ aF, and $V_{g}=0$.}
  \end{center}
\end{figure}

Since the density of states at the Fermi level in a nonmagnetic
island is independent of the spin orientation, one finds  $\Delta
E_{ F}^{\sigma}=-\Delta E_{F}^{\bar{\sigma}}$. Apart from this,
the resistances in the antiparallel configuration are $R_{L \sigma
}^{\rm AP}=R_{L \sigma}^{\rm P}$ and $R_{R \sigma}^{\rm AP} =R_{R
\bar{\sigma}} ^{\rm P}$, see also Fig.~\ref{Sec2:Fig1}b.

Typical transport characteristics of a FM SET with nonmagnetic
island in the limit of slow spin relaxation are shown in Fig.
\ref{Sec2:Fig3} as a function of the bias voltage. The splitting
of the Fermi level is displayed in Fig.~\ref{Sec2:Fig3}a for both
magnetic configurations. As one could expect, the splitting takes
place only in the antiparallel alignment, while in the parallel
configuration there is no spin accumulation. In the former case
the ratio of tunnelling rates for electrons with opposite spin
orientations becomes the same for electrons tunnelling to and off
the island only when a nonequilibrium magnetic moment is built on
the island. In the parallel configuration, however, this condition
is already fulfilled without any spin accumulation. However, this
is true only in the case when both junctions are characterized by
equal spin asymmetries. When the spin asymmetries of both
electrodes are different, spin accumulation also occurs in the
parallel configuration.

As illustrated in Fig.~\ref{Sec2:Fig3}a,  behavior of the Fermi
level splitting with increasing transport voltage can be
decomposed into two components. One component monotonously
increases, while the second one oscillates with increasing bias
voltage. This oscillatory behavior can be accounted for in the
following way. Let us assume that the voltage is slightly above
that corresponding to a certain Coulomb step and begins to
increase. Then, the spin accumulation also increases, until a
local maximum value is reached. The local maximum occurs at a
voltage, at which the chemical potential of the depleted spin
channel approaches the value which allows the next charge state in
the island. This, in turn, enhances tunnelling rate (onto the
island) of electrons corresponding to the depleted spin channel,
and consequently reduces the spin splitting of the Fermi level.
When the voltage increases further, a local minimum in the spin
accumulation is then reached at a voltage, where the chemical
potential of the second (accumulated) spin channel approaches the
value which allows the next charge state on the island. The same
scenario repeats at each Coulomb step leading to the oscillatory
component in the spin accumulation. In the sequential tunnelling
regime, where only the first-order tunnelling processes are taken
into account,  spin accumulation is exponentially small in the
Coulomb blockade regime, as can be seen in Fig.~\ref{Sec2:Fig3}a.

The current as a function of the bias voltage for the parallel and
antiparallel configurations is shown in Fig.~\ref{Sec2:Fig3}b. As
before, characteristic Coulomb steps are clearly visible.
Moreover, owing to different stationary spin accumulations in the
parallel and antiparallel configurations, the current in the
parallel configuration is also different from that in the
antiparallel configuration. This gives rise to the TMR effect
which is presented in Fig.~\ref{Sec2:Fig3}c. Now, the bias
dependence is more complex than it was in the case of FM SETs with
magnetic islands in the absence of spin accumulation. It is
interesting to note, that TMR can change sign in the vicinity of
the Coulomb steps in current-voltage curves, which is  a
consequence of different spin accumulations in the two magnetic
configurations.

Generally, it is more difficult to obey the slow spin relaxation
limit in ferromagnetic islands than in nonmagnetic ones. Anyway,
some experimental data show that this is achievable. In such a
case, the spin accumulation builds up in the magnetic island and
has significant influence on transport characteristics. In Refs.
\cite{barnasEL98,barnasJMMM99} it has been shown that spin
accumulation modifies the 'staircase'-like variation of the
electric current with the bias voltage. Since the spin
accumulation depends on the magnetic configuration of the
junction, this can also lead to an enhanced TMR effect. Moreover,
in some voltage regions TMR can change sign. In addition, the
Coulomb steps for the two magnetic configurations become slightly
shifted, as was also observed experimentally \cite{fertAPL06}. The
difference between current-voltage curves in the fast and slow
spin relaxation limits varies continuously with spin relaxation
rates. This difference was used very recently to evaluate spin
relaxation time in Co nanoparticles \cite{yakushijiN05}.


\subsection{Contribution to the conductance due to cotunneling processes}


The electric current and spin accumulation in the limit of
sequential tunnelling (see the preceding subsection) are
exponentially small in the Coulomb blockade regime. However,
charge transport may occur not only due to spin-dependent
sequential tunnelling, but also due to spin-dependent cotunneling
processes \cite{takahashi98,martinekPRB02,imamuraPRB99}, which
give dominant contribution in the Coulomb blockade regime and also
lead to spin accumulation \cite{imamuraPRB99,aronov76}. Close to
resonance (in the vicinity of the threshold voltage) both
sequential and cotunneling currents may be comparable. The current
$I$ is then equal to the sum of first, $I^{(1)}$, and second,
$I^{(2)}$, order contributions, $I=I^{(1)}+I^{(2)}$.

To calculate cotunneling current far from the resonance one could
use Eq.~(\ref{Sec1:Eq:cotrate}). This formula, however, cannot be
used when voltage approaches the threshold voltage (at resonance).
To calculate current and associated spin accumulation in the whole
voltage range, the real-time diagrammatic formalism
\cite{martinekPRB02,schoeller94,koenigPRL97,koenigPRB98} has been
used. The spin accumulation on the island (or equivalently spin
splitting of the electrochemical potential) is then determined
from the spin balance equation, similarly as in the sequential
tunnelling regime, see Eq.~(\ref{Eq:SETspinconser}).
\begin{figure}[t]
  \begin{center}
    \includegraphics[width=0.5\columnwidth]{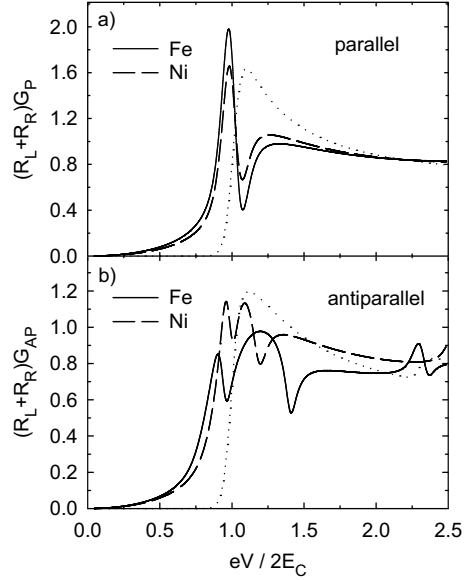}
   \caption{\label{Sec2:Fig4}
    Differential conductance in FM SETs with nonmagnetic islands as a function of
    the bias voltage $V$ in the (a) parallel and
    (b) antiparallel configurations,
    calculated in the whole (sequential and cotunneling) transport regime in the
    limit of slow spin relaxation in the island. The curves have been calculated
    for $k_{\rm B}T/E_C = 0.02$ and
    symmetric junctions with $R_{R\uparrow} = 5h/e^2$.
    The spin polarization is $p=0.23$
    and $p=0.40$ for Ni and Fe electrodes, respectively.
    The dotted line corresponds to the sequential
   tunnelling limit in the SET with Fe electrodes. (After Ref.~\cite{martinekPRB02})}
  \end{center}
\end{figure}

The differential conductance in the whole (sequential and
cotunneling) transport regime is shown in Fig.~\ref{Sec2:Fig4} for
SETs with Ni and Fe electrodes and for nonmagnetic islands (in the
limit of slow spin relaxation). Upper part corresponds to the
parallel configuration while the lower one to the antiparallel
one. First, we find well resolved splitting of the conductance
peaks in the antiparallel alignment, while no splitting can be
seen in the parallel configuration. This splitting in conductance
peaks is a direct consequence of the spin splitting of the
corresponding electrochemical potential of the island, and
therefore can be used to detect and measure spin accumulation. The
absence of conductance splitting in the parallel configuration is
simply a consequence of the absence of spin accumulation in this
configuration. Such a splitting of the conductance peaks in the
cotunneling regime was recently observed experimentally
\cite{fertAPL06}. Generally, there are several experimental
techniques by which the spin accumulation can be detected
indirectly \cite{johnson85,jedema01}. The peculiarities of
transport characteristics of FM SETs offer new possibilities.


\subsection{Discrete energy structure of the island \label{discreteSET}}


\begin{figure}[t]
  \begin{center}
   \includegraphics[width=0.53\columnwidth]{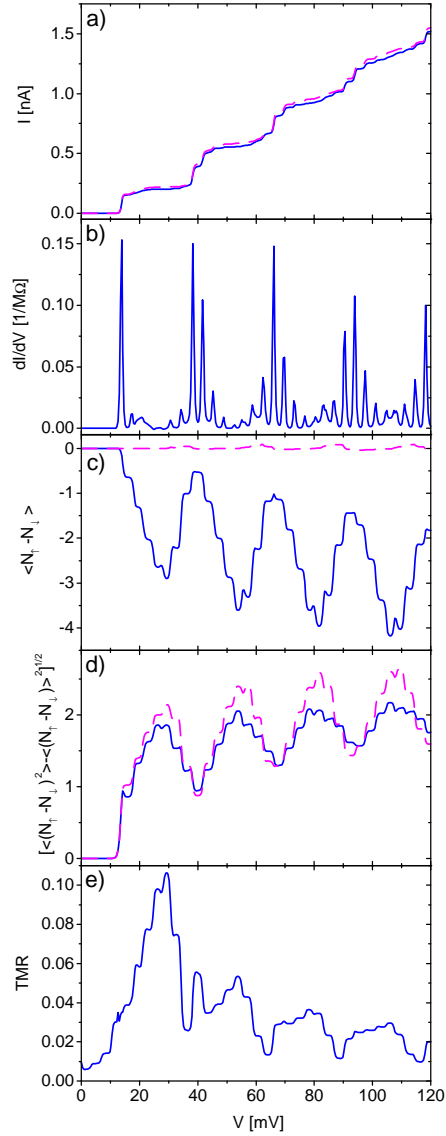}
   \caption{\label{Sec2:Fig6}
   (Color online) The influence of discrete electronic structure on transport
   characteristics in a FM SET with nonmagnetic island in the limit
   of slow spin relaxation.
   Voltage dependence of the tunnel current $I$ (a), the
   differential conductance $G = dI/dV$ (b), spin accumulation
   $\langle n_\uparrow -n_\downarrow \rangle$ (c), standard deviation
   $ [\langle (n_\uparrow -n_\downarrow )^2 \rangle - \langle
   n_\uparrow -n_\downarrow \rangle^2]^{1/2} $ (d), and tunnel
   magnetoresistance ${\rm TMR}$ (e), calculated at $T = 2.3 {\rm K}
   $. The solid and dashed curves in (a), (c) and (d) correspond to
   the antiparallel and parallel configurations, respectively. The
   other parameters are: $\Delta E = 3 {\rm meV }$, $C_{L}/
   C_{ L} = 5$, $E_C = 10{\rm meV }$, $R_{L \uparrow} = 2
   R_{L \downarrow} = 200 {\rm M } \Omega $, $R_{ R \uparrow}
   = 2 R_{R \downarrow} = 4 {\rm M} \Omega$ and $R_{R
   \uparrow} = 2 {\rm M} \Omega$ for the parallel alignment ($2
   R_{R \uparrow} = R_{R \downarrow}=4 {\rm M} \Omega$ for
   the antiparallel alignment). (After Ref.~\cite{martinek99})}
  \end{center}
\end{figure}
Discussion up to now was limited to the case when addition of a
single electron costed only an electrostatic energy. When,
however, the size of metallic island becomes reduced further,
electron spectrum of the island cannot be considered as a
continuous one, and discrete structure of the energy levels plays
an important role. Addition of an electron costs then not only the
electrostatic charging energy, but also the energy equal to the
level spacing $\Delta E$. In this subsection we describe briefly
the effect of discreteness on the transport characteristics. The
relevant approach was developed in Refs.
\cite{martinek99,barnasPRB00}. The electron transport in the
stationary state is then governed by the solution of the
generalized master equation \cite{martinek99,kuoPRB02}
\begin{eqnarray}\label{Seq2Eq:1}
   && 0  =  -  \left\{ \Gamma ( n_{\uparrow},  n_{\downarrow} )  +
   \Omega_{\uparrow ,   \downarrow} (n_{\uparrow} , n_{\downarrow})
   +   \Omega_{\downarrow, \uparrow} (n_{\uparrow} ,
   n_{\downarrow}) \right\}
   P(n_{\uparrow},   n_{\downarrow},V ) \nonumber \\
   &&+  \Gamma_{\uparrow}^{+}   (n_{\uparrow}   -    1,
   n_{\downarrow}) P( n_{\uparrow}   -    1,   n_{\downarrow},V)
   +   \Gamma_{\downarrow}^{+}   ( n_{\uparrow},
   n_{\downarrow}   -    1) P( n_{\uparrow},   n_{\downarrow}   -
   1,V) \nonumber
   \\
   &&+ \Gamma_{\uparrow}^{-}   (n_{\uparrow}   +    1,
   n_{\downarrow}) P( n_{\uparrow}   +    1,   n_{\downarrow},V)
   +   \Gamma_{\downarrow}^{-}   ( n_{\uparrow},
   n_{\downarrow}   +    1) P( n_{\uparrow},   n_{\downarrow}   +
   1,V) \nonumber
   \\
   &&+ \Omega_{\uparrow,\downarrow}(n_{\uparrow} - 1, n_{\downarrow}
   + 1) P( n_{\uparrow} - 1, n_{\downarrow} + 1,V ) \nonumber \\
   &&+ \Omega_{\downarrow,\uparrow}( n_{\uparrow} + 1, n_{\downarrow}
   - 1) P( n_{\uparrow} + 1, n_{\downarrow} - 1,V) ,
\end{eqnarray}
where $P(n_{\uparrow},n_{\downarrow},V)$ denotes the probability
to find $n_{\uparrow}$ and $n_{\downarrow}$ excess electrons on
the island ($ n = n_{\uparrow} + n_{\downarrow}$ is the total
number of excess electrons). The first term in
Eq.~(\ref{Seq2Eq:1}) describes how the probability of a given
charge and spin state decays due to electron tunnelling to and
from the island, whereas other terms describe the rate at which
this probability increases. The $\Omega$-terms account for
spin-flip relaxation processes. The coefficients entering
Eq.~(\ref{Seq2Eq:1}) are defined as $\Gamma_{\sigma}^{\pm}
(n_{\uparrow},n_{\downarrow}) = \sum_{r= L,R} \Gamma_{r
\sigma}^{\pm} (n_{\uparrow},n_{\downarrow}) $ and $ \Gamma (
n_{\uparrow}, n_{\downarrow} ) =  \sum_{\sigma} [\Gamma_{\sigma}^+
(n_{\uparrow},n_{\downarrow}) + \Gamma_{\sigma}^-
(n_{\uparrow},n_{\downarrow})] $, where $\Gamma_{r\sigma}^{\pm}
(n_{\uparrow},n_{\downarrow})$ are the tunnelling rates for
electrons with spin $\sigma$, tunnelling to ($+$) the island from
the lead ${r = L,R}$ or backward ($-$). These coefficients are
given by \cite{martinek99,barnasPRB00}
\begin{eqnarray}
   \label{Seq2Eq:2}
   && \Gamma^{\pm}_{ r \sigma}   (n_{\uparrow} ,
   n_{\downarrow}) =  \sum_{i}   \gamma^{ r}_{i \sigma
   }F^{\mp}_{\sigma}   (E_{i\sigma}   | n_{\uparrow} , n_{\downarrow}
   ) f^{\pm}   (E_{i\sigma}    +    E^{\pm}_r(n)    -    E_F) ,
   \nonumber
   \\
   && \Omega_{\sigma \bar{\sigma}}(n_{\uparrow},   n_{\downarrow} ) =
   \sum_{i}   \sum_{j}   \omega_{i{\sigma},j{ \bar{\sigma}}}
   F^{+}_{\sigma}   (E_{i\sigma} |
   n_{\uparrow} ,   n_{\downarrow} )
   F^{-}_{\bar{\sigma}}
   (E_{j\bar{\sigma}} | n_{\uparrow} ,   n_{\downarrow} ) .
\end{eqnarray}
Here, $f^{+}(E)$ is the Fermi function ($f^{-}=1-f^{+}$), whereas
$ F^{+}_{\sigma}(E_{i\sigma} | n_{\uparrow},n_{\downarrow}) $
($F^{-}_{\sigma}=1-F^{+}_{\sigma}$) describes the probability that
the energy level $E_{i\sigma}$ is occupied by an electron with
spin $\sigma$ for a particular configuration  $ (n_{\uparrow}
,n_{\downarrow})$.
The parameter $\gamma^{ r}_{i\sigma}$ is the bare tunnelling rate
of electrons between the lead ${ r}$ and the energy level
$E_{i\sigma}$ of the island, and
$\omega_{i{\sigma},j{\overline{\sigma}}}$ is the transition
probability from the state $ i{\sigma} $ to $j{\overline{\sigma}}$
of the island  due to the spin-flip processes.
The energies $E^{\pm}_{L} (n)$ and $E^{\pm}_{R} (n)$ are given by
$ E^{\pm}_{L} (n)=C_{R}/C \; e V +U^{\pm}(n)$ and $ E^{\pm}_{ R}
(n)=-C_{L}/C \; e V +U^{\pm}(n)$ where $U^{\pm}(n)= E_{ C} [
2(n-n_x) \pm 1 ] $ and $n_x = C_{g} V_{g}/e $.

>From the solution $P(n_{\uparrow},n_{\downarrow},V)$ of the master
equation (Eq.~\ref{Seq2Eq:1}), one can determine current flowing
through the island,
\begin{equation}\label{Seq2Eq:3}
   I_L = -e  \sum_{\sigma} \sum_{\; n_{\uparrow}
   ,n_{\downarrow}} P(n_{\uparrow} ,n_{\downarrow},V) \left\{
   \Gamma^{+}_{L \sigma}(n_{\uparrow} ,n_{\downarrow})  -
   \Gamma^{-}_{L \sigma}(n_{\uparrow} ,n_{\downarrow})
   \right\} .
\end{equation}
For further discussion we  assume that the discrete energy levels
$E_{i\sigma}$ are spin degenerate (nonmagnetic situation) and
equally separated with the level spacing $\Delta E$.

To emphasize the role of spin accumulation let us  assume that the
intrinsic spin relaxation time on the island is long enough to
neglect all intrinsic spin-flip processes. The corresponding $I-V$
characteristics for the parallel and antiparallel alignments are
shown in Fig.~\ref{Sec2:Fig6}a. In both cases the electric current
is blocked below a threshold voltage, and a typical 'Coulomb
staircase' appears above it, with additional small steps due to
the discrete levels of the island. The effects due to discrete
charging and discrete electronic structure are more clearly seen
in the differential conductance shown in Fig.~\ref{Sec2:Fig6}b,
where the small peaks correspond to new discrete levels taking
part in transport. The difference between the $I$-$V$
characteristics for the parallel and antiparallel configurations
is due to a different spin accumulation in both geometries. In
Fig.~\ref{Sec2:Fig6}c we see the average value of the difference
between the numbers of spin-up and spin-down excess electrons on
the island, $\langle M \rangle \equiv \langle n_\uparrow
-n_\downarrow \rangle$, i.e., the  spin accumulation. There is no
significant spin accumulation in the parallel configuration. The
number $ M \equiv n_\uparrow -n_\downarrow $ of spins accumulated
on the island fluctuates in time around its average value $\langle
M \rangle$, as shown in Fig.~\ref{Sec2:Fig6}d, where the standard
deviation $ ( \langle M^2 \rangle - \langle M \rangle^2)^{1/2}
\equiv [\langle (n_\uparrow -n_\downarrow )^2 \rangle - \langle
n_\uparrow -n_\downarrow \rangle^2]^{1/2}$ is plotted against the
voltage $V$. It is worth to note that although there is almost no
spin accumulation in the parallel configuration, the corresponding
fluctuations are relatively large.

The difference between the $I-V$ curves in the parallel and
antiparallel configurations leads to the tunnel magnetoresistance,
shown in Fig.~\ref{Sec2:Fig6}e. The broad peaks correspond to the
Coulomb steps, while the fine structure originates from the
discrete structure of the density of states of the island.

One can extend the presented formalism also to the case of a FM
SET with a ferromagnetic island \cite{brataasJJAP01} and calculate
nonequilibrium spin accumulation on the island (when spin
relaxation time is sufficiently long). Using similar approach
Inoue and Brataas \cite{inouePRB04} analyzed current-induced
magnetization reversal induced by spin accumulation rather than by
spin torque. They found that the magnetization reversal is
possible when a free energy change due to nonequilibrium spin
accumulation is comparable to the anisotropy energy.


\subsection{Shot noise in ferromagnetic SETs}


>From the application point of view, an important transport
characteristics of the system is the corresponding current noise.
Functionality of a system depends on the relevant noise to signal
ratio, which should be as small as possible. However, the noise is
also a  source of additional information on the system properties
like quantum and Coulomb correlations~\cite{blanterPR00}. The shot
noise in ferromagnetic single-electron transistors was studied
theoretically in Ref.~\cite{bulkaPRB99}, where the method
developed for spinless electrons in single-electron transistors
~\cite{korotkovPRB94,hershfieldPRB93,hankePRB93,hankePRB94,imamogluPRB93}
was extended to magnetic systems.

The time correlation function of any two quantities $X$ and $Y$
can be expressed as~\cite{book:vliet}
\begin{eqnarray}\label{Sec2Eq:4}\fl
   \langle X(t)Y(0) \rangle = \!\!\!\!\!\!
   \sum_{n'_{\uparrow},n'_{\downarrow};n_{\uparrow},n_{\downarrow}}
   \!\!\!\!\!\!
   X_{n'_{\uparrow},n'_{\downarrow}}P(n'_{\uparrow},n'_{\downarrow};
   t|n_{\uparrow},n_{\downarrow};0)
   Y_{n_{\uparrow},n_{\downarrow}}P^0(n_{\uparrow},n_{\downarrow})\;,
\end{eqnarray}
where, $P(n'_{\uparrow},n'_{\downarrow}; t|n_{\uparrow},
n_{\downarrow};0)$ is the conditional probability to find the
system in the final state with $n'_{\uparrow}$ and
$n'_{\downarrow}$ excess electrons at time $t$, if there was
$n_{\uparrow}$ and  $n_{\downarrow}$ excess electrons in the
initial time t=0. The probability $P^0$ can be determined from
Eq.~(\ref{Seq2Eq:1}).

\begin{figure}[t]
  \begin{center}
   \includegraphics[width=0.7\linewidth,clip]{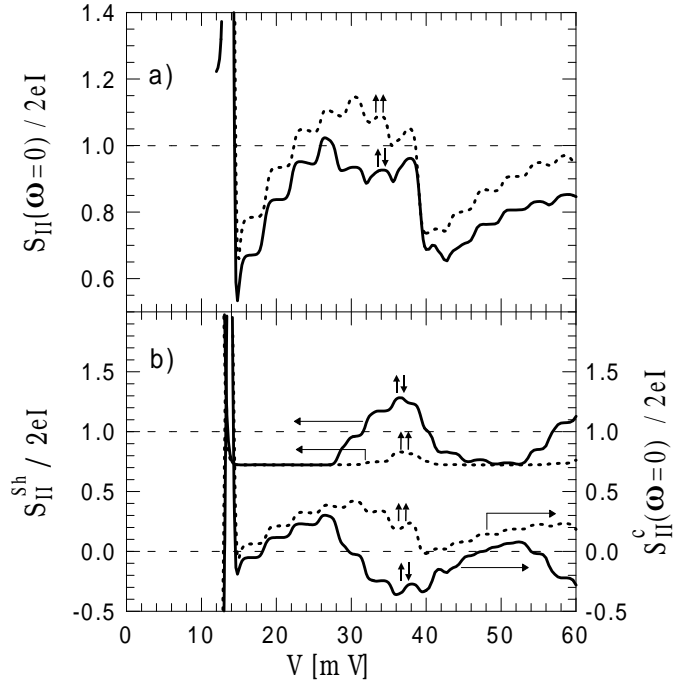}
   \caption{\label{Sec2:Fig7}
   Voltage dependence of the current shot noise at $\omega=0$ (a)
   in a FM SET with nonmagnetic island and
   in the slow spin relaxation limit for
   $4 R_{L \uparrow} = R_{L \downarrow} = 8 {\rm M } \Omega
   $, $R_{ R \uparrow} = 4 R_{R \downarrow} = 240 {\rm M}
   \Omega$ for the antiparallel alignment ($ R_{L \uparrow} = 4
   R_{L \downarrow} = 8 {\rm M} \Omega$, for the parallel
   configuration) (other parameters as in Fig.~\ref{Sec2:Fig6}). In
   part (b) $S_{II}(\omega=0)$ is split into two components:
   $S_{II}^{Sh}$ (upper curves) and $S_{II}^c(\omega=0)$ (lower
   curves). (After Ref.~\cite{bulkaPRB99})}
  \end{center}
\end{figure}

Following this procedure, the current-current correlation function
in a FM SET considered in the preceding subsection has been
calculated in Ref.~\cite{bulkaPRB99}. The corresponding Fourier
transform can be presented as
\begin{equation}\label{Sec2Eq:5}
   S_{II}(\omega)=S^{Sh}_{II}+S^c_{II}(\omega)\;,
\end{equation}
where $S^{Sh}_{II}$ is the Schottky value (the frequency
independent part), while the second term in Eq.~(\ref{Sec2Eq:5})
is the frequency dependent component.

Figure~\ref{Sec2:Fig7}a shows the bias dependence of the
zero-frequency current noise $S_{II}(\omega=0)$. The current noise
is smaller in the antiparallel configuration than in the parallel
one. This is because in the presence of spin accumulation (which
is significant only in the antiparallel alignment) the amplitude
of fluctuations is smaller. In Fig.~\ref{Sec2:Fig7}b
$S_{II}(\omega=0)$ is split into two parts; the frequency
independent component $S_{II}^{Sh}$ and the contribution
$S_{II}^c(\omega=0)$ arising from the frequency dependent part of
the current noise, see Eq.~(\ref{Sec2Eq:5}). The component
$S_{II}^{Sh}$ is almost constant $\approx 2eI(C_1^2+C_2^2)/C^2$ at
the plateaux of the $I-V$ curve and increases with opening of new
channels. Dynamical correlations between the currents are
described by $S_{II}^c(\omega)$. Its value in the limit $\omega
\to 0$ can  be positive between the $I-V$ steps and negative when
new channels become open. This is evident for the antiparallel
alignment at $V \approx 26{\rm mV}$, when opening a tunnelling
channel for electrons with $\sigma=\downarrow$ leads to negative
dynamical correlations. This effect is almost compensated by an
increase in $S_{II}^{Sh}$, and therefore one gets only a small
reduction of the current noise $S_{II}(\omega=0)$.

In the power spectrum of the current $S_{II}^c(\omega)$ one can
distinguish two distinct relaxation times, one in the high and
another one in the low frequency regions \cite{bulkaPRB99}. In a
wide voltage range the corresponding relaxation times are very
close to the effective relaxation times for the charge and spin
noise.

The asymmetry between the tunnelling channels for electrons with
the opposite spins leads to activation of the spin component in
the current noise. In Ref.~\cite{bulkaPRB99} the components
$S^c_{II\;charge}$ and $S^c_{II\;spin}$, corresponding to the
charge and spin noise, respectively, have been extracted from
$S^c_{II}$. It has been shown that the charge component is almost
constant whereas the spin component increases with spin
polarization $p$ of the leads, and for $p \to 1$ can be much
larger than the charge component. The analysis showed that both
charge and spin fluctuations are relevant for the shot noise in FM
SETs. It has been also pointed out that a super-Poissonian shot
noise,  the Fano factor $S_{II}(0)/2eI > 1$, can occur due to
lifting of the spin degeneracy.


\section{Transport in double-island devices}

\begin{figure}[t]
  \begin{center}
  \includegraphics[width=0.5\columnwidth]{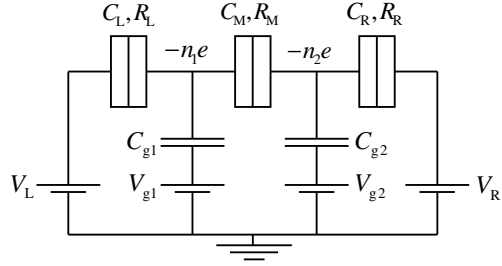}
  \caption{\label{Sec3:Fig1} Schematic of a double-island device.
  The first (1) and second (2) islands are capacitively
  coupled to two gate voltages, $V_{g1}$ and $V_{g2}$,
  and separated from each other and from the
  left and right electrodes by tunnel barriers.
  }
  \end{center}
\end{figure}

Some new features of transport characteristics appear in
spin-polarized electronic transport through double-island
structures \cite{yakushijiJAP02,yakushijiN05,imamura00,
takanashi00,weymannPRB06_DSET}. A typical double-island device,
shown schematically in Fig.~\ref{Sec3:Fig1}, consists of two
metallic islands separated from each other and from external
reservoirs by tunnel barriers. A gate voltage is additionally
attached capacitively to each island. Generally, both islands and
external electrodes can be either magnetic or nonmagnetic
\cite{ono96,yakushijiJAP02,shimadaPRB01}. In the following, we
consider the situations when at least two electrodes (external or
central) are ferromagnetic, and their magnetic moments form only
collinear (parallel and antiparallel) configurations. Apart from
this, we limit discussion to continuous density of states in the
islands.

Hamiltonian of the double-island systems is similar to that used
for single-island devices, but now the electrostatic energy is
given by  \cite{weymannPRB06_DSET,wiel03}
\begin{eqnarray}\label{Sec3Eq:Ech}
  H_{ch}&=&E_{C_1}\left( n_{1} - \frac{Q_{g1}}{e}
  \right)^2+ E_{C_2} \left( n_{2} - \frac{Q_{g2}}{e}
  \right)^2\nonumber\\
  &&+2E_{C_M}\left( n_{1} - \frac{Q_{g1}}{e} \right) \left(
  n_{2} -\frac{Q_{g2}}{e} \right)\,,
\end{eqnarray}
where $n_{1(2)}$ is the number of excess electrons on the first
(second) island and $Q_{g1(g2)}$ describes the respective charge
induced by applied voltages, $Q_{g1(g2)}= C_{L(R)}V_{L(R)}+
C_{g1(g2)} V_{g1(g2)}$. Apart from this, $E_{C_1}$ and $E_{C_2}$
denote the charging energies of the corresponding islands, whereas
$E_{C_M}$ describes electrostatic coupling of the islands,
\begin{eqnarray}
  E_{C_1(C_2)}&=&\frac{e^2}{2C_{1(2)}}\left( 1 -
  \frac{C_{M}^2} {C_1 C_2} \right)^{-1}\,, \\
  E_{C_M}&=&\frac{e^2}{2C_{M}}\left( \frac{C_1 C_2} {C_{M}^2}-1 \right)^{-1} \,,
\end{eqnarray}
with $C_{1(2)}$ being the total capacitance of the first (second)
island, $C_{1(2)}=C_{L(R)}+C_{g1(g2)}+C_{M}$, and $C_{M}$ denoting
the capacitance of the middle junction.

To find the stationary current flowing through the system one
needs then to know the probabilities $P(n_{1},n_{2},V)$ of finding
the system in the charge state with $n_{1}$ and $n_{2}$ additional
electrons on the first and second islands, respectively, when a
bias voltage $V$ is applied. These probabilities can be calculated
in a recursive way from the following steady-state master equation
\begin{eqnarray}\label{Sec3Eq:master}
  0&=&- \sum\limits_{\sigma} \big[ \Gamma _{L 1}^\sigma(n_1,n_2)
  + \Gamma _{1L}^\sigma(n_1,n_2)+\Gamma_{1 2}^\sigma(n_1,n_2)
  \nonumber\\
  && + \Gamma _{2 1}^\sigma(n_1,n_2)+\Gamma _{2
  R}^\sigma(n_1,n_2) + \Gamma _{R
  2}^\sigma(n_1,n_2) \big] P(n_1,n_2,V)\nonumber \\
  &&+\sum\limits_{\sigma} \Gamma _{L
  1}^\sigma(n_{1}-1,n_{2})P(n_{1}-1,n_{2},V)\nonumber\\
  &&+\sum\limits_{\sigma} \Gamma_{1 L}
  ^\sigma(n_{1}+1,n_{2})P(n_{1}+1,n_{2},V) \nonumber\\
  &&+\sum\limits_{\sigma}\Gamma _{1
  2}^\sigma(n_{1}+1,n_{2}-1)P(n_{1}+1,n_{2}-1,V)\nonumber\\
  &&+\sum\limits_{\sigma}\Gamma
  _{2 1}^\sigma(n_{1}-1,n_{2}+1)P(n_{1}-1,n_{2}+1,V) \nonumber\\
  &&+\sum\limits_{\sigma}\Gamma _{R
  2}^\sigma(n_{1},n_{2}-1)P(n_{1},n_{2}-1,V)\nonumber\\
  &&+\sum\limits_{\sigma}\Gamma _{2
  R}^\sigma(n_{1},n_{2}+1)P(n_{1},n_{2}+1,V) \,,
\end{eqnarray}
with the normalization condition $\sum_{n_1,n_2} P(n_1,n_2,V)=1$.
The corresponding transition rates are given by
Eq.~(\ref{Eq:SETrate}), with the respective changes in the total
system electrostatic energy. For example for tunnelling from the
first island to the second one, the change in the electrostatic
energy can be written as
\begin{equation}
  \Delta E^\sigma_{12}(n_{1},n_{2}) =
  E(n_{1}-1,n_{2}+1)-E(n_{1},n_{2})-\Delta E_{F1}^{\sigma}+\Delta E_{F2}^{\sigma}\,.
\end{equation}
Here, $\Delta E_{F1}^{\sigma}$ and $\Delta E_{F2}^{\sigma}$ denote
the corresponding shifts of the chemical potentials for spin
$\sigma$ in the first and second islands, respectively. The spin
asymmetry of tunnelling through the barrier $r$ will be
characterized in this section by the asymmetry factor $\alpha_r$,
defined as $\alpha_r=R_{r \uparrow}/R_{r \downarrow}$. This
asymmetry factor corresponds to the ratio of the respective
spin-dependent densities of states. In particular, for FM/NM
junctions it is given by $\alpha = \rho_\downarrow/\rho_\uparrow$,
whereas for FM/FM junctions $\alpha= \rho_\downarrow^2/
\rho_\uparrow^2$, provided that the two ferromagnetic electrodes
are built of the same material. We note that the relation between
spin dependent barrier resistances in the parallel and
antiparallel configurations of magnetic moments on the opposite
sides of the barrier is the same as that described in Sec.2. When
a bias voltage is applied to the system, a nonequilibrium spin
accumulation may appear on the islands. Generally, the shifts of
the Fermi level for both spin orientations are different. However,
one can assume that the ratio of the Fermi level shifts for the
spin-up and spin-down electrons, defined as $\beta_j = -\Delta
E^\uparrow_{Fj}/ \Delta E^\downarrow_{Fj}$, fulfills the relation
$\beta_j = \rho_{Ij \uparrow}/ \rho_{Ij \downarrow}$, for the
first ($j=1$) and second ($j=2$) island, respectively, with
$\rho_{Ij\sigma}$ being the spin-dependent density of states of
the $j$-th island. As a consequence, for nonmagnetic islands one
directly gets, $\beta_1=\beta_2=1$.

Electric current flowing through the system can be determined from
the following formula:
\begin{eqnarray}\label{Sec3Eq:current}\fl
  I_{\rm L}&=&
  -e\sum\limits_{\sigma}\sum\limits_{n_{1},n_{2}=-\infty}^\infty
  \left[\Gamma_{L 1}^\sigma(n_{1},n_{2})
  -\Gamma _{1 L}^\sigma(n_{1},n_{2})\right]
  P(n_{1},n_{2},V)\,,
\end{eqnarray}
which corresponds to the current flowing through the left
junction, but $I_L=I_R\equiv I$ in the stationary state.


\subsection{Fast spin relaxation: no spin accumulation}


\begin{figure}[t]
  \begin{center}
  \includegraphics[width=0.4\columnwidth]{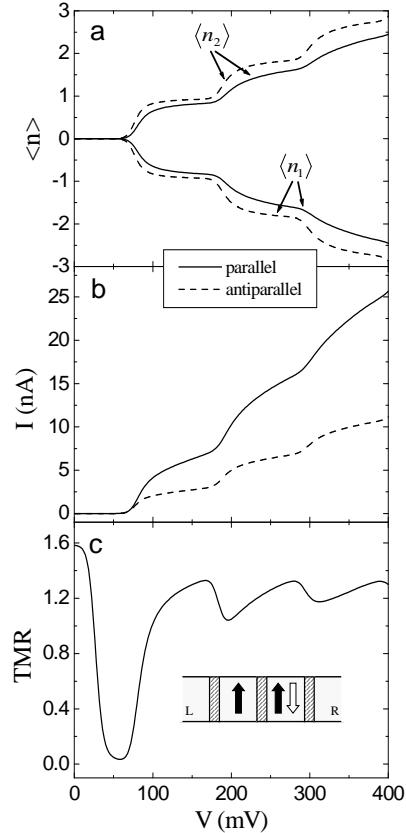}
  \caption{\label{Sec3:Fig2} The average electron number on the islands
    (a), currents (b) in the parallel and
    antiparallel configurations and the resulting TMR (c) as a
    function of the bias voltage. The parameters are:
    $C_{1}=C_{2}\equiv C =3 C_{M}=3$ aF, $k_{\rm B}T/ (e^2/2C)
    =0.05$. The spin asymmetries of resistances in the parallel
    configuration are $\alpha_{L}=\alpha_{R}=5$, $\alpha_{M}=25$,
    whereas the total junction resistances are $R_{L}^{\rm P}
    =R_{R}^{\rm P} = R_{M}^{\rm P}/10=1$
    M$\Omega$.
    In the antiparallel configuration, $R_{L \sigma}
    ^{\rm AP} = R_{L \sigma} ^{\rm P}$,
    $R_{M \uparrow}^{\rm AP}= R_{
    M \downarrow}^ {\rm AP} = (R_{M \uparrow}^{\rm P} R_{
    M \downarrow}^{\rm P})^{1/2}$, and $R_{R \sigma}
    ^{\rm AP} = R_{R \bar{\sigma}}^{\rm P}$.}
  \end{center}
\end{figure}

Consider first transport characteristics of a system built of two
ferromagnetic islands and nonmagnetic external electrodes, shown
in Fig.~\ref{Sec3:Fig2}. Different magnetic configurations of the
system are specified in the inset of Fig.~\ref{Sec3:Fig2}c. When
the spin relaxation time in the islands is much shorter than the
time between two successive tunnelling events, no spin
accumulation builds up on the islands. Since there is  an
asymmetry between the two barrier resistances, some charge
accumulates on the islands, as displayed in Fig.~\ref{Sec3:Fig2}a.
First of all, the magnitude of excess charge on the islands
increases with increasing voltage. For the barrier asymmetry
assumed in Fig.~\ref{Sec3:Fig2}, the electrons easier tunnel to
the second island from the right lead than out of the second
island to the first one. As a consequence, the electrons
accumulate on the second island. On the other hand, the electrons
easier tunnel out of the first island to the left lead than from
the second island to the first one. Thus, the number of excess
electrons on the first island decreases with increasing the bias
voltage -- there are holes accumulated on the first island. This
occurs in both magnetic configurations. The electric current
flowing through the system in the parallel and antiparallel
configurations is shown in Fig.~\ref{Sec3:Fig2}b. The Coulomb
steps due to discrete charging are clearly evident, and the
difference in currents flowing in the parallel and antiparallel
configurations gives rise to the tunnel magnetoresistance, shown
in Fig.~\ref{Sec3:Fig2}c. The TMR effect oscillates as a function
of the bias voltage and the amplitude of these oscillations
decreases as the voltage is increased in a similar way as in the
case of single-island FM SETs, see Fig.~\ref{Sec2:Fig2}b.
Moreover, some dips occur now in TMR at the voltages corresponding
to the steps in the current-voltage characteristics.

\begin{figure}[t]
 \begin{center}
  \includegraphics[width=0.4\columnwidth]{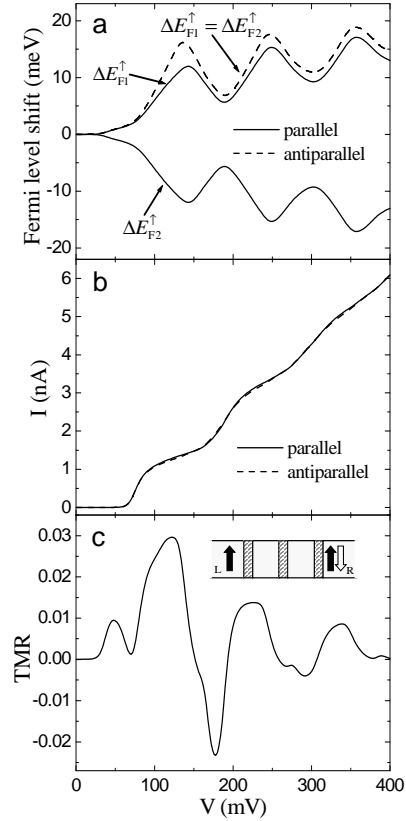}
  \caption{\label{Sec3:Fig3} The shifts of the Fermi levels
   for spin-up electrons (a), currents in the parallel
   and antiparallel configurations (b)
   and TMR (c) as a function of the bias voltage.
   The parameters are:
   $C_{1}=C_{2}\equiv C =3 C_{M}=3$ aF, $k_{\rm B}T/ (e^2/2C)
   =0.05$, $\beta_1=\beta_2=1$, $\tau_{sf 1}\rightarrow\infty$,
   $\tau_{sf 2}=\rightarrow\infty$,
   $\alpha_{L}=\alpha_{R}=5$, $\alpha_{ M}=1$,
   whereas $R_{L}^{\rm P}=R_{ R}^{\rm P} = R_{M}^{\rm P}/50=1$
   M$\Omega$.
   In the antiparallel configuration, $R_{L \sigma}
   ^{\rm AP} = R_{L \sigma} ^{\rm P}$, $R_{R \sigma}
   ^{\rm AP} = R_{R \bar{\sigma}} ^{\rm P}$.
   }
 \end{center}
\end{figure}


\subsection{Slow spin relaxation: spin accumulation}


When the spin relaxation time is longer than the time between two
successive tunnelling events, a nonequilibrium magnetic moment
appears on each island. The corresponding shifts of the Fermi
level due to spin accumulation can be calculated from spin current
conservation, see Eq. (\ref{Eq:SETspinconser}), written for each
island \cite{weymannPRB06_DSET}.

Figure~\ref{Sec3:Fig3} shows the shifts of the Fermi levels and
currents in the parallel and antiparallel configurations, as well
as the resulting TMR calculated in the limit of long spin
relaxation time for the system built of ferromagnetic electrodes
and nonmagnetic islands. The magnetic moments of external
electrodes can form either parallel or antiparallel
configurations, as illustrated in the inset of
Fig.~\ref{Sec3:Fig3}c. First of all, the nonequilibrium spin
accumulation, shown in Fig.~\ref{Sec3:Fig3}a, exists not only in
the antiparallel configuration but also in the parallel one. In
the antiparallel configuration the shifts of the Fermi level for a
given spin orientation are equal on both islands, whereas in the
parallel configuration they are opposite. The effects due to
discrete charging lead to an oscillatory behavior of the Fermi
level shift, in a similar way as in the case of single-island FM
SETs with nonmagnetic islands discussed above. The currents
flowing through the system in both magnetic configurations are
shown in Fig.~\ref{Sec3:Fig3}b. Due to nonequilibrium spin
accumulation induced on the islands, these currents are different,
which leads to nonzero tunnel magnetoresistance, as displayed in
Fig.~\ref{Sec3:Fig3}c. It is interesting to note, that now TMR
changes sign in certain transport voltage regions. These effects
are clearly due to magnetic moments accumulated on the islands. If
the spin relaxation time becomes shorter than the time between
successive tunnelling events, spin accumulation disappears and,
consequently, TMR also vanishes.

\begin{figure}[t]
\begin{center}
\includegraphics[width=0.85\columnwidth]{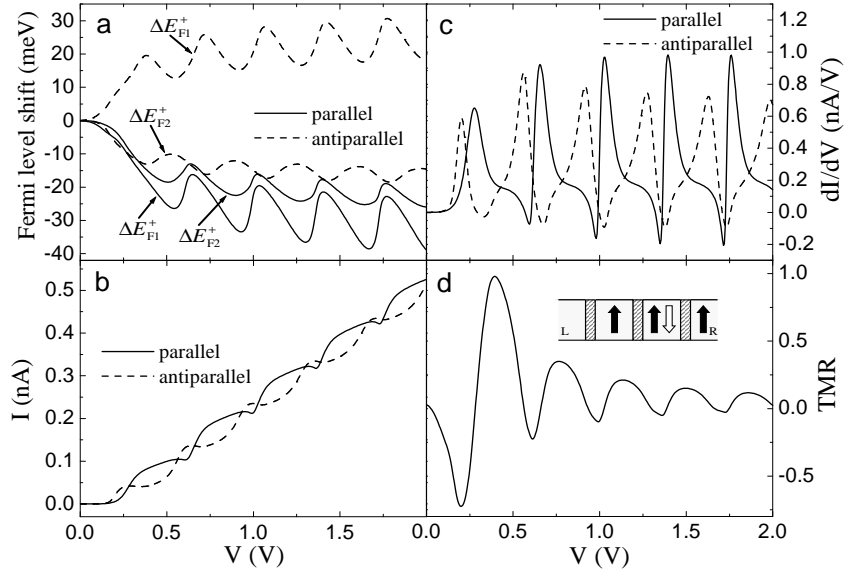}
  \caption{\label{Sec3:Fig4} The
  shifts of the Fermi levels for majority-spin electrons (a), currents
  (b) and differential conductance (c)
  in the parallel and antiparallel configurations,
  and the resulting TMR (d) as a function of the
  bias voltage.
  The parameters are: $T=140$ K, $C_{L}=0.45$ aF,
  $C_{M}=0.2$ aF, $C_{R}=0.35$ aF,
  $C_{g1}=C_{g2}=0$,
  $\alpha_{L}^2=\alpha_{M}=\alpha_{R}=25$,
  and $\beta_1=\beta_2=0.2$, whereas $\tau_{sf1}=\tau_{sf2}=\infty$.
  The total junction resistances are $R_{L}=3500$ M$\Omega$,
  $R_{M}=R_{R}=1$ M$\Omega$.
  In the antiparallel configuration, $R_{L \sigma}^{\rm AP}=R_{L \sigma}^{\rm
  P}$, and $R_{r \uparrow}^{\rm AP}=R_{r \downarrow}^{\rm AP}
  = (R_{r \uparrow}^{\rm P}R_{r \downarrow}^{\rm P})^{1/2}$, for $r={M,R}$.
  (After Ref.~\cite{weymannPRB06_DSET})}
\end{center}
\end{figure}

Recently, several experiments on spin-polarized transport through
granular systems were reported
\cite{yakushijiAPL01,yakushijiJAP02,yakushijiN05,imamura00,takanashi00}.
For example, in Ref.~\cite{imamura00}, the tunnelling current was
driven from a tip of scanning tunnelling microscope through
ferromagnetic grains to ferromagnetic electrode. Transport
measurements of such devices showed pronounced Coulomb steps in
the current-voltage characteristics. Moreover, negative
differential conductance was observed. It was further proposed
that the negative differential conductance could be a consequence
of a nonequilibrium spin accumulation. Such systems can be
modelled theoretically by double-island devices whose two islands
and the right electrode are ferromagnetic, whereas the left
electrode is nonmagnetic, corresponding to the nonmagnetic tip of
scanning tunnelling microscope, see the inset of
Fig.~\ref{Sec3:Fig4}c. Transport characteristics of such a device
are displayed in Fig.~\ref{Sec3:Fig4} for the parameters taken
from Ref.~\cite{imamura00}. The shifts of the Fermi level for
spin-up electrons are shown in Fig.~\ref{Sec3:Fig4}a. These shifts
are different in both magnetic configurations. Generally, spin
accumulation on the first island is larger than the accumulation
on the second island. The reason for this is the fact that the
rate for electron tunnelling from the first island to the left
lead is smaller than the rate for tunnelling of electrons to or
from the second island, which is due to asymmetry of barriers. The
currents flowing through the system in the parallel and
antiparallel configurations are illustrated in
Fig.~\ref{Sec3:Fig4}b. Moreover, negative differential conductance
occurs in both magnetic configurations, however, it is more
pronounced in the antiparallel configuration, as shown in
Fig.~\ref{Sec3:Fig4}c. It can be seen that negative differential
conductance increases with increasing the bias voltage. The
resulting TMR effect is displayed in Fig.~\ref{Sec3:Fig4}d. It is
interesting to note that TMR oscillates between negative and
positive values.

\begin{figure}[t]
\begin{center}
\includegraphics[width=0.45\columnwidth]{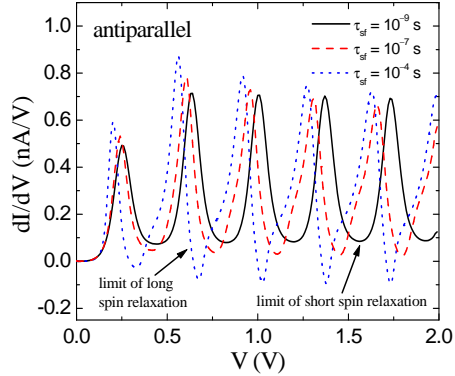}
  \caption{\label{Sec3Fig:G_AP} (Color online) Differential conductance in
  the antiparallel configuration as a function
  of the bias voltage calculated for different values of the spin relaxation time
  $\tau_{sf1}={\tau_{sf2}}=\tau_{sf}$
  and for $\rho_{I1}^+ \Omega_{I1}=\rho_{I2}^+ \Omega_{I2}=1000$/eV.
  The other parameters as well as magnetic configuration of the system
  are the same as in Fig.~\ref{Sec3:Fig4}.
  (After Ref.~\cite{weymannPRB06_DSET})}
\end{center}
\end{figure}
\begin{figure}[t]
\begin{center}
  \includegraphics[width=0.45\columnwidth]{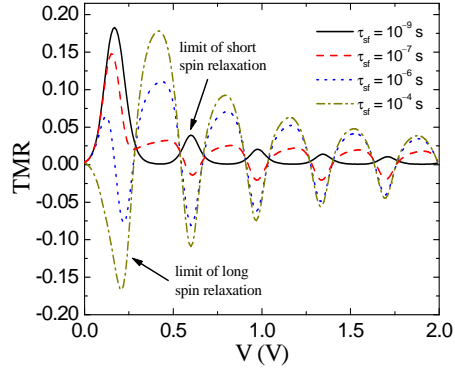}
  \caption{\label{Sec3Fig:TMR} (Color online)  TMR  as a function of
  the bias voltage calculated for different values of spin relaxation time
  $\tau_{sf1}={\tau_{sf2}}=\tau_{sf}$
  and for $\rho_{I1}^+ \Omega_{I1}=\rho_{I2}^+ \Omega_{I2}=1000$/eV.
  The other parameters as well as magnetic configuration of the system
  are the same as in Fig.~\ref{Sec3:Fig4}.
  (After Ref.~\cite{weymannPRB06_DSET})}
\end{center}
\end{figure}

Oscillations of the sign of TMR and differential conductance
result from spin accumulations in both islands and are absent in
the limit of fast spin relaxation (no spin accumulation) in the
islands. The results presented in Fig.~\ref{Sec3:Fig4} correspond
to the long spin relaxation limit. In such a limit, some spin
accumulation may occur even for a very small current flowing
through the system, giving rise to NDC and oscillations of the TMR
sign for small bias voltages. However, both effects disappear when
the spin relaxation time is shorter than the time between
successive tunnelling events. Thus, for a finite relaxation time,
one may expect absence of  NDC and  TMR oscillations for small
voltages and the onset of these effects at larger voltages. This
is because at some voltage there is a crossover from the fast to
slow spin relaxation limits. In fact such a behavior is consistent
with experimental data of Ref.~\cite{imamura00}.

The disappearance of NDC with decreasing spin relaxation time
$\tau_{\rm sf}$ is shown explicitly in Fig.~\ref{Sec3Fig:G_AP},
where the bias dependence of differential conductance is presented
for different values of the spin relaxation time. This figure
clearly shows that NDC disappears when spin relaxation time
decreases, in agreement with the  above discussion. In the limit
of fast spin relaxation time, the differential conductance is
positive, although its periodic modulation still remains.

Similarly, periodic oscillations of the sign of TMR also disappear
with decreasing spin relaxation time $\tau_{sf}$. This behavior is
shown in Fig.~\ref{Sec3Fig:TMR}, where the bias dependence of TMR
is shown for several values of $\tau_{sf}$. First, the transitions
to negative TMR disappear with decreasing $\tau_{sf}$. The TMR
becomes then positive, although some periodic modulations survive.
Second, the phase of the modulations shifts by about $\pi$ when
the spin relaxation varies from fast to slow limits.

In the case of analyzed systems, the negative differential
conductance occurs due to nonequilibrium spin accumulation on the
islands. It is however worth noting that the negative differential
conductance may also exist in single-electron devices built of
nonmagnetic materials \cite{nakashima97,nguyen04}.


\section{Spin polarized transport through single-level quantum
dots connected to ferromagnetic leads}


In the previous two sections we considered spin-dependent
transport through single-electron devices, whose central
electrodes (islands) were described by continuous (or discrete
with small level separation) energy spectrum, and the most
relevant and dominant energy scale was the electrostatic charging
energy. In ultra-small metallic islands (nanoparticles) or in
semiconducting quantum dots, the level spacing is comparable with
the charging energy or even larger. This also happens in the case
of a molecule attached to metallic leads. In this limit one
arrives at slightly more sophisticated single-electron devices
\cite{book:kouwenhoven97,kastner93,kouwenhoven95}.

In this section we consider a FM SET based on a semiconductor
quantum dot coupled to ferromagnetic leads. This model also
applies to molecules attached to ferromagnetic electrodes.
Single-electron transistors based on quantum dots are of current
interest not only because of new and interesting physics emerging
in those systems, but, more importantly, due to possible future
applications and due to the possibility of manipulation of a
single electron charge and a single electron spin
\cite{book:kouwenhoven97,deshmukh02,kouwenhoven98,hanson03}.
Furthermore, quantum dots are also interesting for future
applications in quantum computing \cite{lossPRA98,lossPRL00}.

Transport properties of quantum dots coupled to nonmagnetic leads
have already been extensively studied both theoretically and
experimentally \cite{book:kouwenhoven97,geerlings90,beenakker91,
kouvenhoven91,kouwenhoven98,hanson03,meir93,meir94}. However,
further interesting effects occur in the case of quantum dots
coupled to ferromagnetic leads, e.g. spin accumulation, parity
effect on tunnel magnetoresistance, zero-bias anomaly in the
Coulomb blockade regime, exchange field, splitting of the Kondo
anomaly, and others. Most of the works concerned theoretical
description of spin-polarized transport in the weak coupling
regime, as well as in the strong coupling regime, where the Kondo
physics emerges \cite{surgueevPRB02,lopezPRL03,martinekPRL03,
franssonPRB05,franssonEPL05,utsumiPRB05,swirkowiczPRB06,eto06,simon06}.
Sequential transport through a single-level quantum dot coupled to
ferromagnetic leads was studied for both collinear
\cite{bulka00,rudzinski01} and non-collinear
\cite{koenigPRL03,braun04,rudzinski05,braig04, wetzels05,braun06}
configurations of the electrodes' magnetic moments. Spin-polarized
transport in the cotunneling regime has also been addressed for
collinear systems
\cite{hartmann03,weymannPRBBR05,weymannPRB05,weymannPRB06}, as
well as for systems magnetized non-collinearly
\cite{pedersen05,braig05,weymannEPJ05,weymann07noncol}.
Furthermore, the resonant tunnelling  was also considered
\cite{utsumiPRB05,swirkowicz02}.

Quantum dots coupled to ferromagnetic leads may be realized
experimentally in various ways, including ultrasmall metallic
(e.g. aluminum) nanoparticles \cite{deshmukh02}, single molecules
\cite{pasupathy04}, granular structures \cite{zhang05},
self-assembled dots in ferromagnetic semiconductors
\cite{chyePRB02}, carbon nanotubes
\cite{tsukagoshi99,zhaoAPL02,zhaoJAP02,jensen03,
sahooNP05,jensenPRB05,manPRB06,liuPRB06,
nagabhiravaAPL06,cottetSST06,schonenbergerSST06}, and magnetic
tunnel junctions \cite{fertAPL06}. Quite recently, semiconductor
quantum dots based on two-dimensional electron gas were
successfully attached to ferromagnetic leads
\cite{hamayaAPL07a,hamayaAPL07b,hamayaAPL07c,hamayaPRB08}.


\subsection{Quantum dots weakly coupled to ferromagnetic leads: collinear
magnetizations}


Let us begin our discussion with the case when the dot is weakly
coupled to the leads. One can then use a perturbative approach in
which the coupling is considered as a small perturbation to the
unperturbed quantum dot and the leads. Transport is then dominated
either by the first order or by the second order (cotunneling)
contributions. Later on we consider a more general situation when
the coupling may be strong and a new Kondo physics emerges at low
temperatures.

\subsubsection{Model and method}

\begin{figure}[h]
\begin{center}
  \includegraphics[width=0.5\columnwidth]{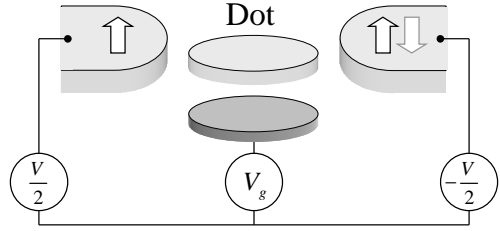}
  \caption{\label{Sec4:Fig1} Schematic of a quantum dot coupled
  to ferromagnetic leads. The magnetic moments of external
  electrodes can be aligned either in parallel or antiparallel. The system
  is symmetrically biased and there is a gate voltage attached to
  the dot.}
\end{center}
\end{figure}

A schematic of a quantum dot coupled to ferromagnetic leads is
presented in Fig. \ref{Sec4:Fig1}. The magnetizations of the leads
can be either parallel or antiparallel. There is also a gate
voltage attached to the dot. The system is modelled by an
Anderson-like Hamiltonian of the general form \cite{anderson61}
\begin{equation}\label{Sec4Eq:QDHamiltonian}
  H=H_{\rm L}+H_{\rm R}+ H_{\rm D}+H_{\rm T},
\end{equation}
where the first and second terms describe the left and right
reservoirs of noninteracting electrons, $H_{r}=\sum_{{\mathbf
q}\sigma} \varepsilon_{r {\mathbf q}\sigma}c^{\dagger}_{r {\mathbf
q}\sigma} c_{r {\mathbf q}\sigma}$ for $r ={\rm L,R}$ [see
Eq.~(\ref{Eq:hamiltonian2})]. The third term of the Hamiltonian,
$H_{\rm D}$, represents the dot and includes two components: one
describes noninteracting electrons in the dot level and the other
represents the Coulomb interaction of two electrons residing in
this level,
\begin{equation}\label{Sec4Eq:H_D}
  H_{\rm D}=\sum_{\sigma=\uparrow,\downarrow}
  \varepsilon_{\sigma}d^{\dagger}_{\sigma} d_{\sigma}+U
  d^{\dagger}_{\uparrow} d_{\uparrow}
  d^{\dagger}_{\downarrow}d_{\downarrow}\,,
\end{equation}
where $U$ is the correlation energy,
$\varepsilon_{\sigma}=\varepsilon\mp \Delta/2$ is the energy of an
electron in the dot with spin $\sigma$, and $d^{\dagger}_{\sigma}$
($d_{\sigma}$) is the corresponding creation (annihilation)
operator. The position of the dot level can be tuned by the gate
voltage, but is independent of the symmetrically applied transport
voltage. Generally, the dot level may be spin-split, for example
due to a stray field of the electrodes or due to an external
magnetic field. The corresponding level splitting is denoted by
$\Delta$, whereas $\varepsilon$ is the energy of the
spin-degenerate dot level. In general, four different states of
the dot are possible: empty dot ($\chi =0$), singly occupied dot
with a spin-up ($\chi =\uparrow$) or spin-down ($\chi
=\downarrow$) electron, and doubly occupied dot ($\chi={\rm d}$),
where $\ket{\chi}$ are the corresponding eigenfunctions.

Interaction between the leads and quantum dot is incorporated in
the tunnelling Hamiltonian, $H_{\rm T}$, given by
\begin{equation}\label{Sec4Eq:H_T}
  H_{\rm T}=\sum_{r={\rm L,R}}\sum_{{\mathbf q}\sigma}\left( t_{r
  {\mathbf
q}\sigma}c^{\dagger}_{r {\mathbf q}\sigma} d_{\sigma}+t^{*}_{r
  {\mathbf
q}\sigma}d_{\sigma}^{\dagger}c_{r {\mathbf q}\sigma}\right) \, ,
\end{equation}
where $t_{r {\mathbf q}\sigma}$ are the tunnel matrix elements.
Tunnelling gives rise to an intrinsic broadening $\Gamma^{\sigma}$
of the dot levels, $\Gamma^{\sigma}=\sum_{r={\rm
L,R}}\Gamma_r^{\sigma}$. The parameters $\Gamma^{\uparrow}_r$ and
$\Gamma^{\downarrow}_r$ describe contributions to the level widths
due to coupling of the dot to the lead $r$. The respective
contribution $\Gamma_r^{\sigma}$ can be expressed in terms of the
Fermi golden rule as $\Gamma^{\sigma}_r= 2\pi \sum_{{\bf q}} |t_{r
{\mathbf q}\sigma}|^2 \delta(\omega-\varepsilon_{r {\mathbf
q}\sigma})$. Assuming the tunnel matrix elements $t_{r {\mathbf
q}\sigma}$ to be independent of the wave vector $\mathbf q$, one
can write
\begin{equation}\label{Sec4Eq:Gamma}
\Gamma_{r}^{\sigma}= 2\pi |t_{r\sigma}|^2 \rho_{r \sigma} \,.
\end{equation}
The coupling parameters are usually expressed in terms of the spin
polarization $p_{r}$ of the lead $r$, defined by
Eq.~(\ref{Eq:polarization}), as
$\Gamma_{r}^{\uparrow(\downarrow)}=\Gamma_{r}(1\pm p_{r})$, where
$\Gamma_{r}= (\Gamma_{r}^{\uparrow} +\Gamma_{r}^{\downarrow})/2$.
As reported in \cite{kogan04}, typical values of the dot-lead
coupling strength $\Gamma$ in the weak coupling regime are of the
order of tens of $\mu$eV.

In order to investigate transport properties of the system in the
whole range of parameters, one has to use a method which is more
sophisticated than that based on the Fermi golden rule. Two such
techniques are commonly used: the method based on the equation of
motion for the electron Green functions, and the real-time
diagrammatic technique. Discussion in this section is based mainly
on the latter one
\cite{schoeller94,koenigdiss,koenig96,koenigPRL96,thielmann03,
thielmannPRL05}. This technique is based on a systematic
perturbation expansion of expectation value of the current
operator and the density matrix order by order in the dot-lead
coupling strength $\Gamma$.  Even, if we limit further
considerations based on this technique to the second order
processes, this allows us to describe transport in the resonance
regime, where the usual second order perturbation term diverges.
One  also could go beyond the second order theory and describe
more subtle effects.

Within the real-time diagrammatic technique, the density matrix
elements of the quantum dot, $P_{\chi_2}^{\chi_1}$, are given by a
kinetic equation in the Liouville space
\begin{equation}\label{Sec4Eq:master}
  0=(\varepsilon_{\chi_1}-\varepsilon_{\chi_2})P_{\chi_2}^{\chi_1}+
  \sum_{\chi_1^\prime \chi_2^\prime}\Sigma_{\chi_2
  \chi_2^\prime}^{\chi_1
  \chi_1^\prime}P_{\chi_2^\prime}^{\chi_1^\prime},
\end{equation}
where $\Sigma_{\chi_2^\prime \chi_2}^{\chi_1^\prime \chi_1}$ is
the irreducible self-energy corresponding to transition forward in
time from state $\ket{\chi_1^\prime}$ to $\ket{\chi_1}$ and then
backward in time from state $\ket{\chi_2}$ to
$\ket{\chi_2^\prime}$.

When the tunnelling processes are spin-conserving and magnetic
moments of the leads are collinear, the density matrix is
diagonal. After performing the perturbation expansion, one gets
the following first-order (sequential tunnelling) and second-order
(cotunneling) master equations:
\begin{eqnarray}
  \label{Sec4Eq:master_1}
  0&=&\sum_{\chi} \Sigma_{\chi^\prime\chi}^{(1)} P_{\chi}^{(0)}\,,\\
  \label{Sec4Eq:master_2}
  0&=&\sum_{\chi} \Sigma_{\chi^\prime\chi}^{(2)} P_{\chi}^{(0)}+
  \Sigma_{\chi^\prime\chi}^{(1)} P_{\chi}^{(1)}\,,
\end{eqnarray}
respectively, where the probabilities obey the normalization
condition  $\sum_{\chi} P^{(m)}_{\chi}=\delta_{m,0}$.

Systematic perturbation expansion of electric current with respect
to the coupling strength $\Gamma$ can be performed in a similar
way, and the first-order and second-order contributions to current
are given by  the expressions \cite{koenigdiss,thielmann03}
\begin{eqnarray}
  I^{(1)}&=&-\frac{i e}{2\hbar}\sum_{\chi\chi^\prime}
  \Sigma_{\chi^\prime\chi}^{\rm I(1)} P_{\chi}^{(0)}\, ,
  \label{Sec4Eq:current1}\\
  I^{(2)}&=&-\frac{i
  e}{2\hbar}\sum_{\chi\chi^\prime} \Sigma_{\chi^\prime\chi}^{\rm I(2)}
  P_{\chi}^{(0)}+
  \Sigma_{\chi^\prime\chi}^{\rm I(1)} P_{\chi}^{(1)} \, ,
  \label{Sec4Eq:current2}
\end{eqnarray}
where the coefficients $\Sigma_{\chi^\prime\chi}^{\rm I(1)}$ and
$\Sigma_{\chi^\prime\chi}^{\rm I(2)}$ are the first- and
second-order self-energies, modified as compared to
$\Sigma_{\chi^\prime\chi}^{\rm (m)}$, to account for the number of
electrons transferred through the barriers \cite{weymannPRB05}.

\begin{figure}[t]
\begin{center}
\includegraphics[width=0.35\columnwidth]{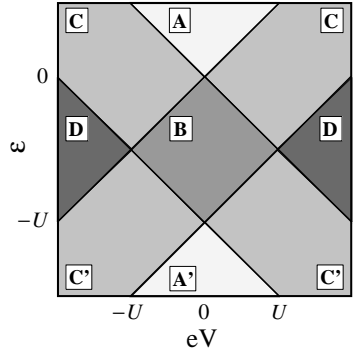}
\caption{\label{Sec4Fig:regimes}
  A sketch illustrating different transport regimes.
  The respective regimes are separated by solid lines and labelled
   correspondingly.}
\end{center}
\end{figure}

Transport characteristics of quantum dots are generally studied in
both the linear and nonlinear response regimes
\cite{weymannPRB05}. It is thus important to distinguish between
different transport regimes of quantum dots, which are sketched in
Fig.~\ref{Sec4Fig:regimes} and labelled by the corresponding
capital letters.

First of all, by changing position of the dot level (by the gate
voltage for instance) or applying the bias voltage, one can cross
over from one regime to another. The three regions (A, B and A')
around zero bias correspond to the regime where sequential
tunnelling is exponentially suppressed and current flows mainly
due to cotunneling. The charge state of the dot is then fixed
(strictly at zero temperature) to zero electrons in the regime A,
one electron in the regime B, and two electrons in the regime A'.
The first-order tunnelling processes are possible once the bias
voltage is increased above the threshold voltage, allowing for
finite occupation of two adjacent charge states (zero and one for
regime C, and one and two for regime C'). In the regime D all four
dot states $\chi=0,\uparrow,\downarrow, {\rm d}$ are possible. By
performing a particle-hole transformation, the behavior in regime
A' and C' can be mapped to that in regime A and C, respectively.

\subsubsection{Case of nonmagnetic leads}

To discuss the role of second-order processes we first briefly
describe the case of a quantum dot coupled to nonmagnetic leads
($p_{\rm L}=p_{\rm R}=0$).
\begin{figure}[h]
\begin{center}
  \includegraphics[width=0.45\columnwidth]{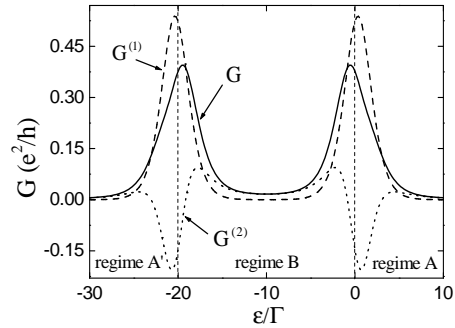}
  \caption{\label{Sec4Fig:nonmagncond} Linear conductance as a function of
  the level position. The dashed line corresponds to the first-order
  contribution $G^{(1)}$, the dotted line presents the
  second-order conductance $G^{(2)}$ and the solid line shows the
  sum $G^{(1)}+G^{(2)}$. The different transport regimes are also
  specified. The parameters are: $k_{\rm B}T=\Gamma$, $U=20\Gamma$,
  and $p=0$. (After Ref.~\cite{weymannPRB05})}
\end{center}
\end{figure}
Figure~\ref{Sec4Fig:nonmagncond} shows the first-order (dashed
line) and second-order (dotted line) contributions to the linear
conductance as well as the total conductance (solid line). The
conductance is shown there as a function of the dot's level
energy.  When the dot level crosses the Fermi level of electrodes,
there is a resonance peak in the linear conductance. Another
resonance appears when $\varepsilon+U$ crosses the Fermi level.
The resonance peaks acquire a certain width as a result of the
level broadening due to coupling to the leads (thermal
fluctuations also contribute). It is interesting to note, that the
second-order contribution becomes negative at resonances, which
indicates that the second-order processes renormalize the
first-order (sequential) contribution. Except for resonances, the
dot is either in the empty (regime A) or doubly occupied (regime
A') state, or in the Coulomb blockade regime (regime B). In all
these three cases the cotunneling contribution to electric current
becomes dominant. It is also worth noting, that the second-order
processes lead to renormalization of the dot level energy
\cite{koenigdiss}.

\subsubsection{Case of ferromagnetic leads}

When the external electrodes are ferromagnetic, qualitatively new
features appear in the transport characteristics. These are
particularly pronounced in the tunnel magnetoresistance, defined
by Eq.~(\ref{Eq:TMR}). Below we describe some of them.

\begin{figure}[h]
\begin{center}
  \includegraphics[width=0.47\columnwidth]{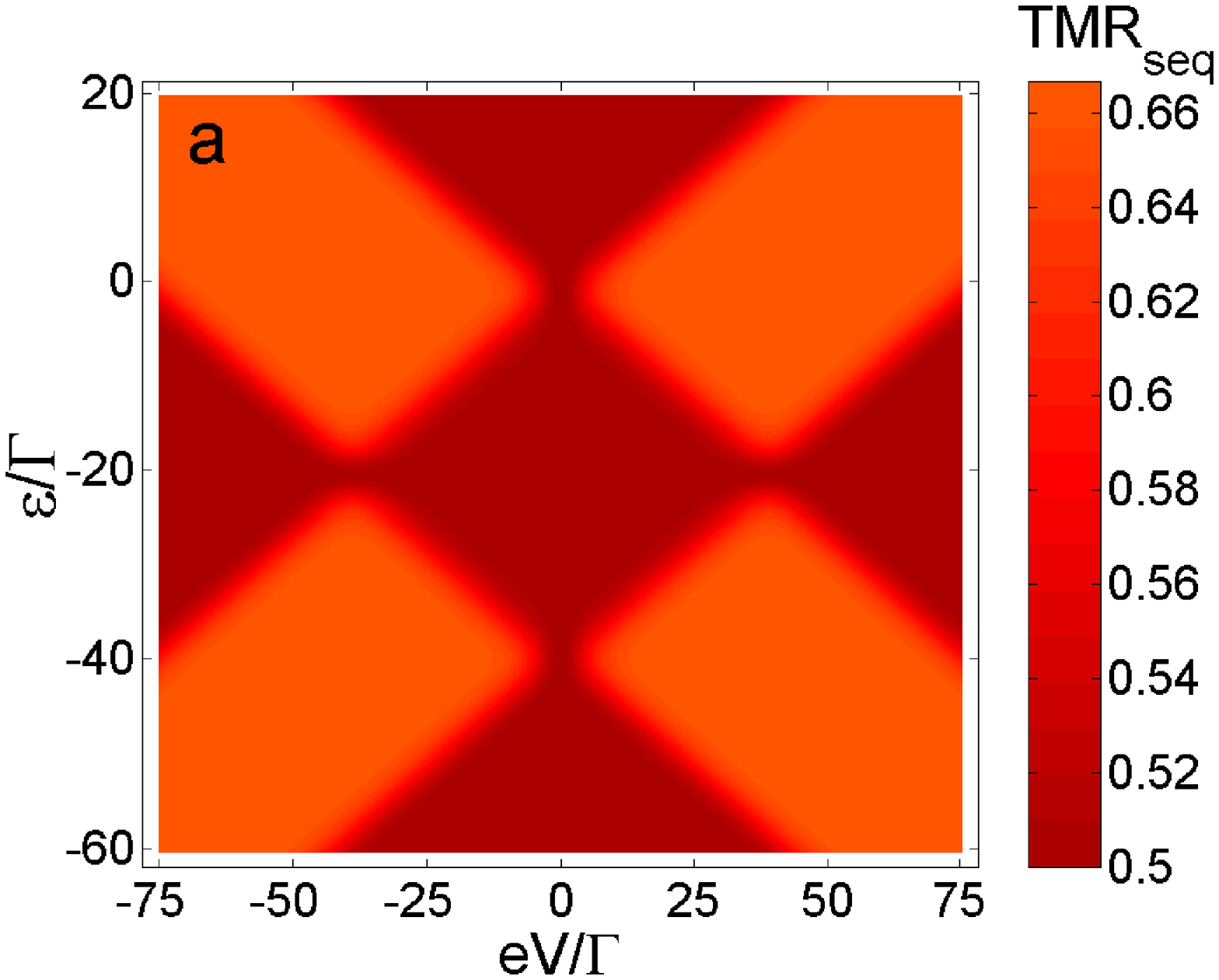}
  \hspace{0.5cm}
  \includegraphics[width=0.47\columnwidth]{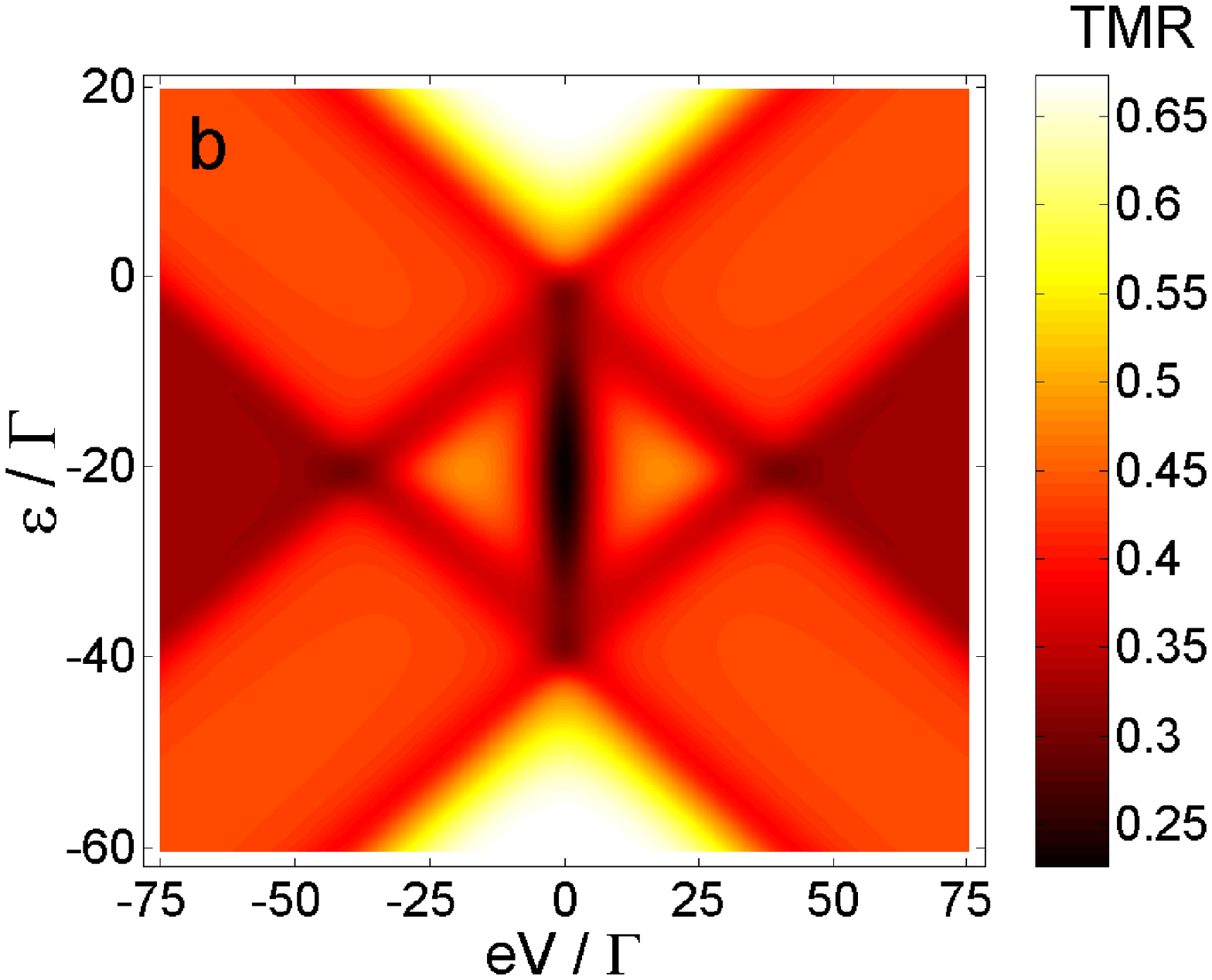}
  \caption{\label{Sec4Fig:tmr3d}
  (Color online) The first-order (a) and
  first- plus second-order (b) tunnel magnetoresistance
  as a function of the bias and gate voltages for the parameters:
  $k_{\rm B}T = 1.5\Gamma$, $U=40\Gamma$, and $p=0.5$. (After Ref.~ \cite{weymannPRB05})}
\end{center}
\end{figure}

Tunnel magnetoresistance calculated in the first order of the
dot-lead coupling strength $\Gamma$  (sequential transport limit)
is shown in Fig.~\ref{Sec4Fig:tmr3d}a as a function of the bias
and gate voltages. It is evident that the TMR generally acquires
then two different values, depending on the transport region. For
the regions A (and A'), B and D, the TMR value is
\begin{equation}\label{Sec4Eq:tmrseqABD}
  {\rm TMR}_{\rm seq}^{\rm A,B,D} = \frac{p^2}{1-p^2} = \frac{1}{2}
  {\rm TMR}^{\rm Jull} \, ,
\end{equation}
while for the region C (and C') it is
\begin{equation}\label{Sec4Eq:tmrseqC}
  {\rm TMR}_{\rm seq}^{\rm C} = \frac{4p^2}{3(1-p^2)} = \frac{2}{3}
  {\rm TMR}^{\rm Jull} \, .
\end{equation}
This behavior can be accounted for as follows. For the regions A
(A') and B the first-order linear conductance in the parallel and
antiparallel configurations is
\begin{equation}
G_{\rm P}^{(1)}\sim \Gamma/2 \hspace{0.6cm} {\rm and}
\hspace{0.6cm} G_{\rm AP}^{(1)}\sim \Gamma(1-p^2)/2 \;,
\end{equation}
which leads to the magnetoresistance equal to ${\rm TMR}={\rm
TMR}_{\rm seq}^{\rm A,B} = p^2/(1-p^2)$. To account for the
behavior of TMR in the regions C and D let us consider the zero
temperature limit. In the region C there are then three dot states
taking part in transport: $\chi=0,\uparrow, \downarrow$ (because
of the particle-hole symmetry, the results are also applicable to
regime C'). On finds then the first-order currents in the parallel
and antiparallel configurations to be $I_{\rm P}^{(1)}\sim
\Gamma/3$ and $I_{\rm AP}^{(1)}\sim \Gamma(1-p^2)/(3+p^2)$, which
leads to  ${\rm TMR}_{\rm seq}^{\rm C}=4p^2/3(1-p^2)$. Similar
analysis for the region D gives ${\rm TMR}_{\rm seq}^{\rm D} =
p^2/(1-p^2)$.

As we already know from previous sections, first-order
(sequential) transport does not describe properly the blockade
regions (the regions A, A' and B). Transport in these regions is
expected to be strongly modified by the second-order (cotunneling)
processes. Accordingly, the above picture is also strongly
modified. The corresponding TMR is displayed in Fig.
\ref{Sec4Fig:tmr3d}b. Since the sequential tunnelling dominates
transport above the threshold voltage (outside the blockade
regions, i.e. in the regions C and D), the total TMR is only
slightly modified there as compared to that in the first-order. In
the regions A, A' and B, however, the first-order tunnelling
processes are exponentially suppressed and the current is
dominated by the second-order (cotunneling) processes. As follows
from Fig.~\ref{Sec4Fig:tmr3d}b, cotunneling has then a significant
influence on TMR. Apart from this, TMR in the regions A and A'
behaves differently from that in the region B.

\begin{figure}[h]
\begin{center}
  \includegraphics[width=0.45\columnwidth]{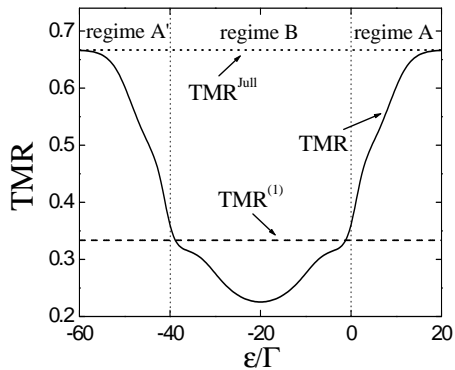}
  \caption{\label{Sec4Fig:regimeABA}
  The total linear tunnel magnetoresistance (solid
  line) as a function of the level position. The dashed line
  represents the first-order tunnel magnetoresistance,
  whereas the dotted line corresponds to the Julli\`{e}re's value.
  The parameters are $k_{\rm B}T=1.5\Gamma$,
  $U=40\Gamma$, and $p=0.5$. (After Ref.~\cite{weymannPRB05})}
\end{center}
\end{figure}

Let us look now in more details at the  vertical cross-section of
Fig.~\ref{Sec4Fig:tmr3d}b along the zero bias (linear response).
The corresponding linear TMR as a function of the level position
(gate voltage) is shown in Fig.~\ref{Sec4Fig:regimeABA}. By
changing position of the dot level, one crosses over from the
regime A through the regime B to the regime A'. For comparison,
the Julli\`{e}re's value of TMR, as well as the first-order
contribution to TMR (denoted as ${\rm TMR}^{(1)}$) are shown
there. Note, that the first-order TMR is independent of the gate
voltage and is equal to a half of the Julli\`{e}re's value. First
of all, one can note that the second-order processes modify the
TMR substantially. Unlike the first-order term, the total TMR
(first-order plus second-order contributions) does depend on the
level position. The interesting feature of TMR shown in
Fig.~\ref{Sec4Fig:regimeABA}b is a strong parity effect. The TMR
reaches maximum when there is an even number of electrons on the
dot (zero for the regime A or two for the regime A'), and has
minimum for an odd (one in the regime B) number of electrons. A
universal feature is that for even electron number, TMR exactly
coincides with the Julli\`{e}re's value. The system behaves then
like a single ferromagnetic tunnel junction. Such a situation can
take place when tunnelling processes are coherent. The only
second-order processes that contribute to conductance in these
regions are the non-spin-flip cotunneling processes, in which the
electron spin is conserved. Such processes indeed are fully
coherent. The corresponding cotunneling rates are proportional to
the product of the density of states of the left and right leads,
thus, one can express the second-order linear conductance as
\begin{equation}
  G_{\rm P}^{(2)}\sim\frac{\Gamma^2}{2}(1+p^2) \hspace{0.6cm} {\rm and}
  \hspace{0.6cm}
  G_{\rm AP}^{(2)}\sim \frac{\Gamma^2}{2}(1-p^2) \;,
\end{equation}
for the parallel and antiparallel configuration, respectively. As
a consequence, the TMR is then equal to that of a single planar
ferromagnetic tunnel junction
\begin{equation}
  {\rm TMR}^{\rm A} = \frac{2p^2}{1-p^2} = {\rm TMR}^{\rm Jull} \,
  .
\end{equation}
Because of the particle-hole symmetry, the same result can be
obtained for regime A'.

The situation becomes, however, more complex for an odd number of
electrons on the dot. Apart from the non-spin-flip processes,
there are also spin-flip cotunneling processes that change spin of
the dot. Such second-order processes give rise to spin relaxation
on the dot and lead to reduction of TMR in the regime B. The
dependence of TMR on the position of the energy level in the
regime B reflects the relative relation of the spin-flip to
non-spin-flip cotunneling processes. The minimum value of TMR
appears for $\varepsilon=-U/2$, see Fig.~\ref{Sec4Fig:regimeABA}b,
where one finds
\begin{equation}\label{Sec4Eq:TMRB_min}
  {\rm TMR}^{\rm B}_{\rm min}=\frac{2p^2}{3(1-p^2)} =
  \frac{1}{3}{\rm TMR}^{\rm Jull}\,.
\end{equation}

\begin{figure}[h]
\begin{center}
\includegraphics[width=0.45\columnwidth]{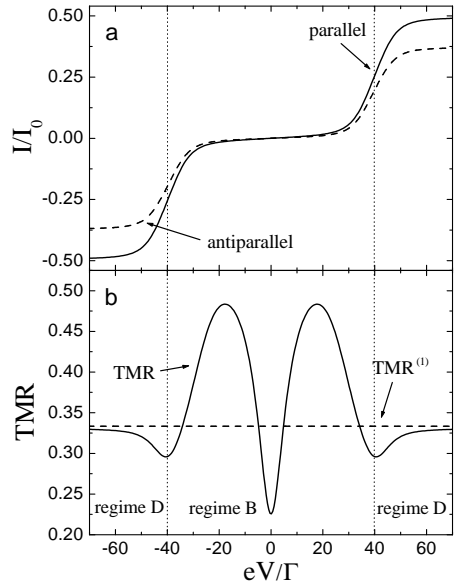}
  \caption{\label{Sec4Fig:regimeDBD}The total current (a) in the parallel
  (solid line) and antiparallel (dashed line) magnetic
  configurations as a function of the bias voltage
  in units of $I_0 = e\Gamma/\hbar$. Part (b) shows
  the first-order contribution to the TMR (dashed line) and the total
  TMR (solid line).
  The parameters are: $k_{\rm B}T=1.5\Gamma$,
  $\varepsilon=-20\Gamma$,
  $U=40\Gamma$, and $p=0.5$. (After Ref.~\cite{weymannPRB05}) }
\end{center}
\end{figure}

Electric current flowing through the system in the nonlinear
response regime is shown in Fig.~\ref{Sec4Fig:regimeDBD}a for both
magnetic configurations and for a symmetric Anderson model
($\varepsilon=-U/2$). The corresponding TMR is also shown there,
see Fig.~\ref{Sec4Fig:regimeDBD}b. The contribution due to the
second-order tunnelling processes is significant (or even
dominant) in the Coulomb blockade regime. The sequential TMR is
shown by a dashed line in Fig.~\ref{Sec4Fig:regimeDBD}b. One can
see that sequential TMR is constant as a function of the bias
voltage. However, this is not a universal behavior and occurs only
for symmetric Anderson model. Generally, first-order TMR depends
on the applied bias voltage, as illustrated in
Fig~\ref{Sec4Fig:tmr3d}a. The second-order processes lead to a
strong and nontrivial dependence of the total TMR on the bias
voltage. For large values of the transport voltage (regime D), the
first-order tunnelling processes dominate, and therefore the total
TMR is only slightly modified as compared to that in the
sequential transport regime. However, for voltages below the
threshold voltage (regime B), the cotunneling processes lead to a
strong bias dependence of TMR in the Coulomb blockade regime, and
to a deep minimum of TMR in the zero-bias limit. In the case of
metallic islands, cotunneling processes usually lead to an
enhancement of the TMR effect in the Coulomb blockade regime
\cite{takahashi98}. Here, we have the opposite situation, i.e.,
suppression of the effect. At low bias voltage, $|eV| \ll k_{\rm
B}T$, the single-barrier spin-flip processes reduce the TMR. This
is however no longer the case for nonlinear response regime, $|eV|
\gg k_{\rm B}T$, where the spin accumulation diminishes the amount
of spin-flip processes and the TMR increases. Consequently, the
TMR effect in regime B increases with rising the bias voltage
within the limits ${\rm TMR}^{\rm Jull}/3 \le {\rm TMR}^{\rm B}
\le {\rm TMR}^{\rm Jull}$. The minimal value is reached at $V=0$
and $\varepsilon=-U/2$, whereas the maximal value is approached
for bias voltages large as compared to thermal energy but still
far away from the onset of sequential tunnelling.

\begin{figure}[t]
\begin{center}
  \includegraphics[width=0.46\columnwidth]{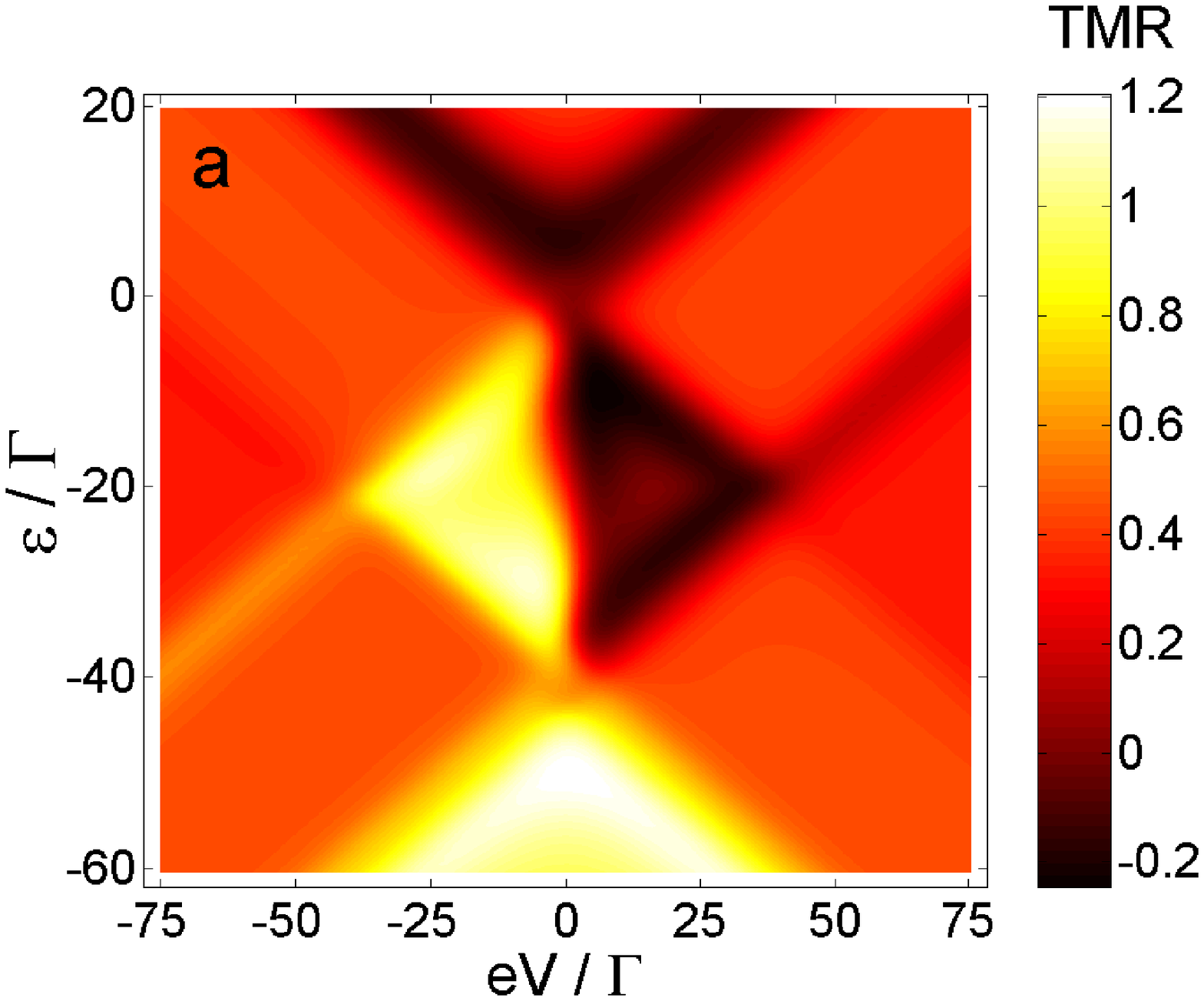}
  \hspace{0.5cm}
  \includegraphics[width=0.46\columnwidth]{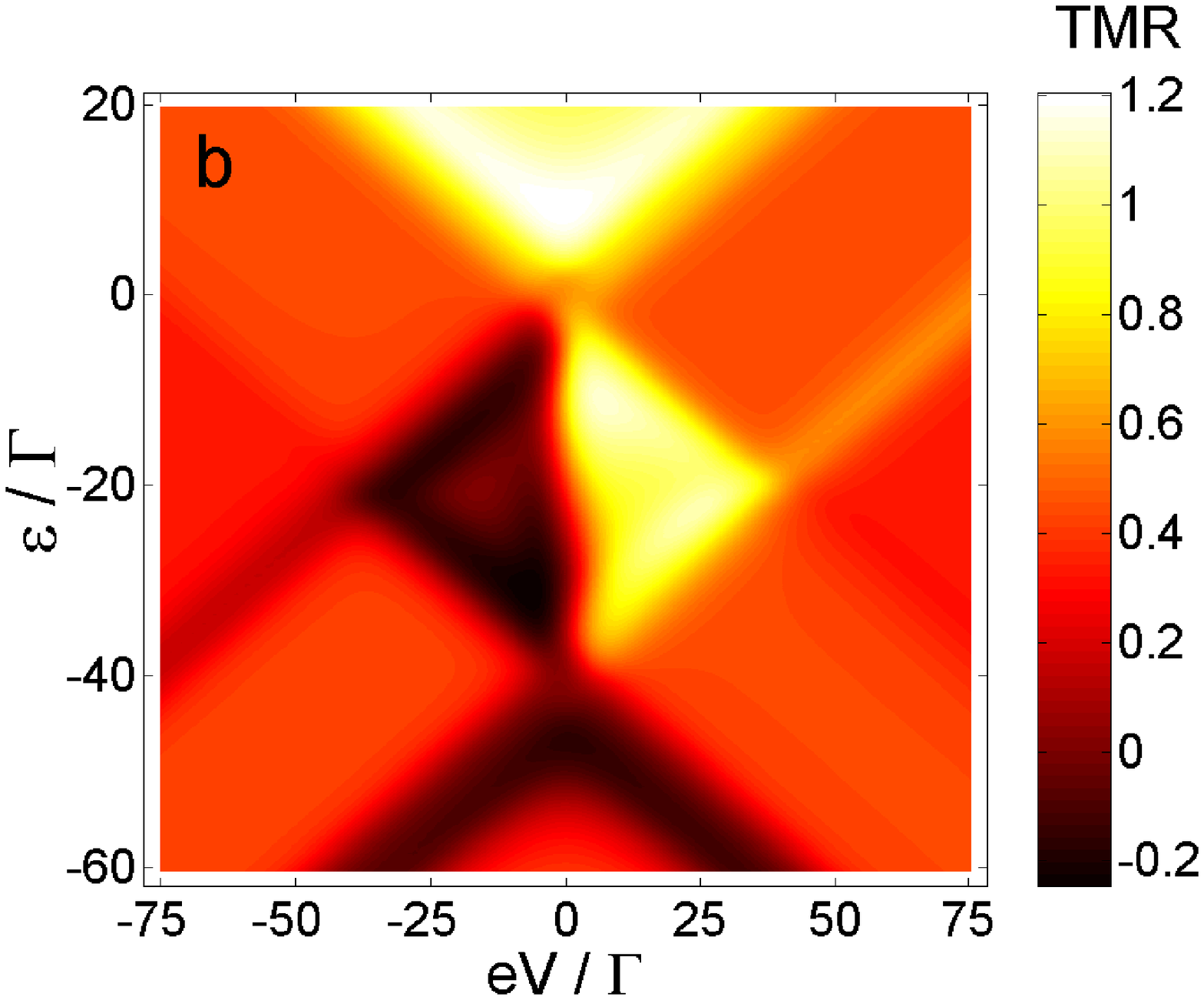}
  \caption{\label{Sec4Fig:TMR2Dfield}
  (Color online) The total tunnel magnetoresistance
  as a function of the bias and gate voltages in the presence
  of external magnetic field for (a) $\Delta=4\Gamma$
  and (b) $\Delta=-4\Gamma$. The parameters are:
  $k_{\rm B}T=1.5\Gamma$,
  $U=40\Gamma$, and $p=0.5$.}
\end{center}
\end{figure}

New features of transport characteristics appear when the spin
degeneration of the dot level is lifted, $\varepsilon_\uparrow
\neq \varepsilon _\downarrow$, e.g. due to an external magnetic
field applied to the system. It turns out that  a finite Zeeman
splitting, $\Delta=\varepsilon_\downarrow - \varepsilon_\uparrow$,
changes the transport characteristics substantially.
Figure~\ref{Sec4Fig:TMR2Dfield}a illustrates the gate and bias
voltage dependence of tunnel magnetoresistance for $\Delta =
4\Gamma$. First of all, a finite Zeeman splitting affects the TMR
mainly in the regimes A (A') and B, where the second-order
processes dominate. Furthermore, TMR exhibits a distinctively
different behavior in the regions A and A'. It is enhanced in the
case of empty dot, and reduced in the case of doubly occupied dot
(it may even become negative in the latter case). This effect
depends on the orientation of applied magnetic field; when the
field is applied in the opposite direction, $\Delta = -4\Gamma$,
there is an enhancement of TMR in regime A' and suppression in
regime A, as illustrated in Fig.~\ref{Sec4Fig:TMR2Dfield}b. Thus,
by changing the sign of the Zeeman splitting (magnetic field
orientation) or applying a gate voltage, which changes the dot
occupation, it is possible to reduce or enhance the TMR effect
considerably. Another interesting feature of TMR displayed in
Fig.~\ref{Sec4Fig:TMR2Dfield}a appears in the region B. There is a
strong asymmetry of TMR with respect to the bias reversal. The TMR
is decreased for positive and increased for negative bias voltage.
This can also be seen in Fig.~\ref{Sec4Fig:TMR2Dfield}b, but for
positive transport voltage TMR is then larger than for negative
bias voltage. The crossover between those two values of TMR takes
place roughly at the zero bias. As a consequence, by changing the
bias voltage in the small range one can substantially tune
magnitude of the TMR effect.


\subsection{Transport through quantum dots in the Coulomb blockade
regime: collinear magnetizations}


As we know from previous subsection, the most interesting features
due to ferromagnetism of the electrodes, e.g. the zero-bias
anomaly, occur in the blockade regime, where sequential tunnelling
processes play a minor role. This anomaly appears only in the case
when the dot is singly occupied at equilibrium
($\varepsilon_\sigma<0$, $\varepsilon_\sigma+U>0$) and the system
is in a deep Coulomb blockade regime; $\Gamma,k_{\rm B}T \ll
|\varepsilon_\sigma|,\varepsilon_\sigma+U$. In such a case the
sequential tunnelling is exponentially suppressed, and cotunneling
gives the dominant contribution to electric current
\cite{averin90,weymannPRBBR05,weymannPRB06,kangPRB97,
franceschi01,zumbuhl04}. Therefore, the sequential tunnelling can
be completely ignored. In this subsection we look more carefully
at these features taking into account only second-order
contribution.

In order to calculate the cotunneling current in the deep blockade
regime one can employ a simplified second-order perturbation
theory and master equation for the occupation probabilities. The
rate of a cotunneling processes from lead $r$ to lead $r^\prime$,
which change the dot state from $\ket{\chi}$ to
$\ket{\chi^\prime}$, can be written as
\begin{equation}\label{Eq:cotunnelingrate}
  \gamma_{\rm rr^\prime}^{\chi\rightarrow\chi^\prime} = \frac{2\pi}{\hbar}
  \left|\sum_{v}\frac{\bra{\Phi_{r^\prime}^{\chi^\prime}}H_{\rm T}\ket{\Phi_v}\bra{\Phi_v}
  H_{\rm T} \ket{\Phi_{r}^\chi}} {\varepsilon_i-\varepsilon_{v}}\right|^2\delta
  (\varepsilon_i-\varepsilon_f),
\end{equation}
with $\varepsilon_i$ and $\varepsilon_f$ denoting the energies of
initial and final states, $\ket{\Phi_r^\chi}$ being the state of
the system with an electron in the lead $r$ and the dot in state
$\ket{\chi}$, whereas $\ket{\Phi_v}$ is a virtual state with
$\varepsilon_{v}$ denoting the corresponding energy. Among
different cotunneling processes one can distinguish the
single-barrier ($r=r^\prime$) and double-barrier ($r\neq
r^\prime$) cotunneling as well as spin-flip ($\chi\neq
\chi^\prime$) and non-spin-flip ($\chi= \chi^\prime$) cotunneling.
The spin-flip processes change the spin state of the dot, whereas
the non-spin-flip processes  do not change the dot state. The
current flows through the system due to double-barrier cotunneling
processes. On the other hand, the single-barrier processes do not
contribute directly to electric current, however, they can change
the dot occupations, and this way also influence the current.

The cotunneling current flowing through the system from the left
to right lead is given by
\begin{equation}\label{Eq:cotunnelingcurrent}
  I = e \sum_{\chi\chi^\prime} P_\chi \left[
  \gamma^{\chi \rightarrow \chi^\prime}_{\rm LR} -
  \gamma^{\chi \rightarrow \chi^\prime}_{\rm RL} \right],
\end{equation}
where $P_\chi$ denotes the corresponding occupation probability.
The probabilities $P_\chi$ can be found from the master equation,
\begin{equation}\label{Eq:cotunnelingmaster}
   0 = \sum_{rr^\prime}\sum_{\chi^\prime} \left[
   -\gamma_{rr^\prime}
   ^{\chi\rightarrow\chi^\prime} P_{\chi} + \gamma_{rr^\prime}
   ^{\chi^\prime\rightarrow\chi}
   P_{\chi^\prime}
   \right] \,,
\end{equation}
together with the normalization condition $\sum_{\chi}P_\chi = 1$.

\subsubsection{Zero-bias anomaly and its physical mechanism}

\begin{figure}[h]
\begin{center}
  \includegraphics[width=0.45\columnwidth]{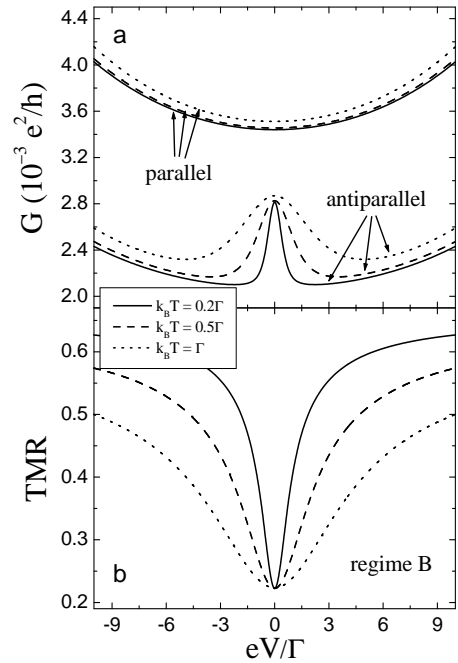}
  \caption{\label{Sec4Fig:regimeB}
  The differential conductance (a)
  for the parallel and antiparallel
  configurations and the tunnel magnetoresistance (b) as a function
  of the bias voltage for different values of temperature. The maximum in
  differential conductance for antiparallel configuration at zero bias is clearly
  demonstrated. The other parameters are the same as in Fig.~\ref{Sec4Fig:regimeDBD}.
  Figure was generated using the scheme for
  the perturbation expansion in the Coulomb blockade regime. (After Ref.~\cite{weymannPRB05})}
\end{center}
\end{figure}

In order to elucidate and understand the anomalous behavior of TMR
in the Coulomb blockade regime, we show in
Fig.~\ref{Sec4Fig:regimeB}a the differential conductance in the
small bias regime for both parallel and antiparallel
configurations and  for several temperatures.
Figure~\ref{Sec4Fig:regimeB}b displays the corresponding TMR
effect. First of all, the TMR effect in the regime B for $|eV| \gg
k_{\rm B}T$ increases with lowering temperature and approaches the
Julli\`{e}re's value, whereas the minimum at zero bias does not
depend on temperature. The differential conductance in the
parallel alignment has characteristics typical of the cotunneling
regime, with a smooth parabolic dependence on the bias voltage.
For antiparallel configuration, on the other hand, differential
conductance has a local maximum at zero bias, followed by local
minimum with increasing bias, as illustrated in
Fig.~\ref{Sec4Fig:regimeB}a. This zero-bias anomaly stems from the
interplay of the spin-flip and non-spin-flip single-barrier and
double-barrier cotunneling processes \cite{weymannPRBBR05}. The
minimum in the TMR effect is a direct consequence of this
anomalous behavior of differential conductance in the antiparallel
configuration.

The zero-bias anomaly in the cotunneling regime is qualitatively
similar to the anomaly due to the Kondo effect, which occurs in
the strong coupling limit \cite{cronenwett98,gores00,sasaki00}.
There are, however, some distinct differences. First of all,
processes responsible for the zero-bias anomaly in the cotunneling
regime are of the second order in tunnelling processes, while
these leading to the Kondo effect are of higher order. The
conductance in the cotunneling regime is much smaller than in the
Kondo regime, where almost perfect transmission ($G=e^2/h$)
through the dot is possible owing to the Kondo peak in the density
of states at the Fermi level. Furthermore, the Kondo peak occurs
at temperatures lower than the so-called Kondo temperature, $T
\lesssim T_{\rm K}$, and exists also in the parallel configuration
\cite{pasupathy04,martinekPRL03}.

\begin{figure}[h]
\begin{center}
\includegraphics[width=0.65\columnwidth]{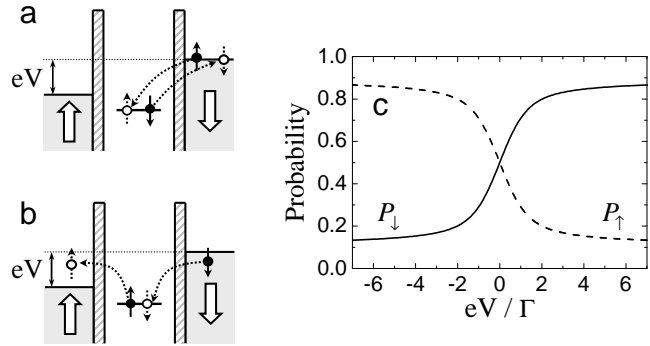}
  \caption{\label{Sec5Fig:ZBA2}
  Single-barrier (a) and double-barrier (b) cotunneling processes,
  and the occupation probabilities for spin-up and spin-down
  electrons in the antiparallel configuration (c).
  The parameters are $k_{\rm B}T= 0.5 \Gamma$, $\varepsilon =-U/2$,
  $U=30 \Gamma$, and $p=0.5$. (After Ref.~\cite{weymannPRBBR05})}
\end{center}
\end{figure}

To understand the mechanism of the zero-bias anomaly it is crucial
to distinguish between different types of cotunneling processes:
the single-barrier cotunneling processes shown in
Fig.~\ref{Sec5Fig:ZBA2}a and double-barrier cotunneling processes
illustrated in Fig.~\ref{Sec5Fig:ZBA2}b. Both single-barrier and
double-barrier processes can be either spin-flip or non-spin-flip
ones. The current flows due to double-barrier cotunneling, whereas
the single-barrier cotunneling can influence the current in an
indirect way, by changing the spin state of the dot. In the
antiparallel configuration, there is a finite spin accumulation on
the dot, as presented in Fig.~\ref{Sec5Fig:ZBA2}c.
The different occupation probabilities for spin-up and spin-down
electrons appear due to spin asymmetry in tunnelling processes. In
equilibrium, both rates are equal and there is no spin
accumulation, $P_\uparrow =P_\downarrow$. When a bias voltage is
applied, and the system is in the antiparallel configuration, the
relative amount of single-barrier cotunneling is diminished as
compared to the double-barrier cotunneling. This is because the
rate of single-barrier cotunneling is proportional to thermal
energy, whereas that of double-barrier cotunneling is proportional
to the bias voltage. In the nonlinear response regime, the
magnetic state of the dot is mainly determined by the spin-flip
processes that transfer an electron from the left to the right
leads. The one shown in Fig.~\ref{Sec5Fig:ZBA2}b changes the dot
spin from $\ket{\uparrow}$ to $\ket{\downarrow}$. Because the rate
of this process is proportional to a product of densities of
states for majority electrons, the corresponding rate is larger
than that of the other process that changes the dot spin from
$\ket{\downarrow}$ to $\ket{\uparrow}$, where only the minority
spins are involved. This results in a nonequilibrium spin
accumulation, $P_\downarrow > P_\uparrow$, that increases with
increasing voltage, as shown in Fig.~\ref{Sec5Fig:ZBA2}c. The
initial state for the dominant spin-flip cotunneling process that
contributes to current is $\ket{\uparrow}$, as sketched in
Fig.~\ref{Sec5Fig:ZBA2}b. Thus, the conductance is diminished by
spin accumulation. This is the mechanism by which spin
accumulation gives rise to nonzero tunnel magnetoresistance
effect, $G^{\rm P} > G^{\rm AP}$, see Fig.~\ref{Sec4Fig:regimeB}b.

Since the spin accumulation reduces electronic transport, any
spin-flip process that reduces the spin accumulation should
enhance the conductance. In particular, single-barrier spin-flip
cotunneling is an example of such a process. As pointed above, the
rate of single-barrier processes scales with $k_{\rm B} T$ while
that of double-barrier cotunneling is proportional to $\max\{|eV|,
k_{\rm B}T \}$, which explains the mechanism of the zero-bias
anomaly. At low bias voltage, $|eV| \lesssim k_{\rm B}T$,
single-barrier spin-flip processes play a significant role -- they
decrease the spin accumulation opening this way the system for the
fastest double-barrier cotunneling. As a consequence, the current
increases relatively fast with applied bias, leading to a maximum
in differential conductance. For $|eV| \gg k_{\rm B}T$, on the
other hand, the relative role of single-barrier processes is
negligible as compared to double-barrier cotunneling, and the
conductance is reduced. Thus, the interplay between the rates of
double-barrier and single-barrier cotunneling processes  leads to
the maximum in the differential conductance at the zero bias.


\subsubsection{Effects of spin relaxation in the dot
on cotunneling transport}


Intrinsic spin relaxation in the dot can result, for instance,
from spin-orbit interaction, coupling of the electron spin to
nuclear spins, etc. However, we will not consider a particular
microscopic mechanism of the intrinsic spin-flip processes, but
simply assume that the spin-flip relaxation is described by a
spin-relaxation time $\tau_{\rm sf}$, and is  taken into account
{\it via} a relaxation term in the appropriate master equation for
the occupation probabilities \cite{weymannPRB06},
\begin{equation}
  0=\sum_{\nu,\nu^\prime = {\rm L,R}}\left(-\gamma_{\nu\nu^\prime}
  ^{\sigma\rightarrow\bar{\sigma}} P_{\sigma} +
  \gamma_{\nu\nu^\prime} ^{\bar{\sigma}\rightarrow\sigma}
  P_{\bar{\sigma}}\right)
  -\frac{2}{\tau_{\rm sf}}\frac{P_\sigma e^{\beta\varepsilon_\sigma} -
   P_{\bar{\sigma}} e^{\beta\varepsilon_{\bar{\sigma}}} }
   {e^{\beta\varepsilon_\sigma}+e^{\beta\varepsilon_{\bar{\sigma}}}},
\end{equation}
where $\beta=1/(k_{\rm B}T)$, $P_\sigma$ denotes the probability
that the dot is occupied by a spin-$\sigma$ electron, and
$\gamma_{\nu\nu^\prime} ^{\sigma\rightarrow\bar{\sigma}}$ is the
cotunneling rate from lead $\nu$ to lead $\nu^\prime$ with a
change of the dot spin from $\sigma$ to $\bar{\sigma}$
($\bar{\sigma}=-\sigma$). The last term describes the spin
relaxation processes, which in the case of spin-degenerate dot
level reduces to $-(P_\sigma - P_{\bar{\sigma}})/\tau_{\rm sf}$.

As before, (see section 2) one can distinguish between the fast
and slow spin relaxation limits. The former (latter) limit
corresponds to the situation when the time between successive
cotunneling events, $\tau_{\rm cot}$, is significantly longer
(shorter) than the intrinsic spin relaxation time $\tau_{\rm sf}$.
A typical spin relaxation time for quantum dots can be relatively
long, up to $\mu$s \cite{fujisawa02,elzerman04}. On the other
hand, the time between successive cotunneling events can be
estimated taking into account the fact that the rate of spin-flip
cotunneling is generally larger than that of non-spin-flip
cotunneling (for a finite parameter $U$). Assuming
$\varepsilon_\uparrow=\varepsilon_\downarrow=\varepsilon$, one
then finds
\begin{equation}\label{Sec5Eq:tcot}
  \tau_{\rm cot}\approx
  \frac{h\varepsilon^2(\varepsilon+U)^2}{AU^2\Gamma^2}\,,
\end{equation}
with $A={\rm max}\{|eV|,k_{\rm B}T \}$ and $h=2\pi\hbar$. Assuming
typical parameters \cite{kogan04}, one can roughly estimate
$\tau_{\rm cot}$ to range from $10^{-3}$ ns to $1$ ns.

\begin{figure}[t]
  \centering
  \includegraphics[width=0.8\columnwidth]{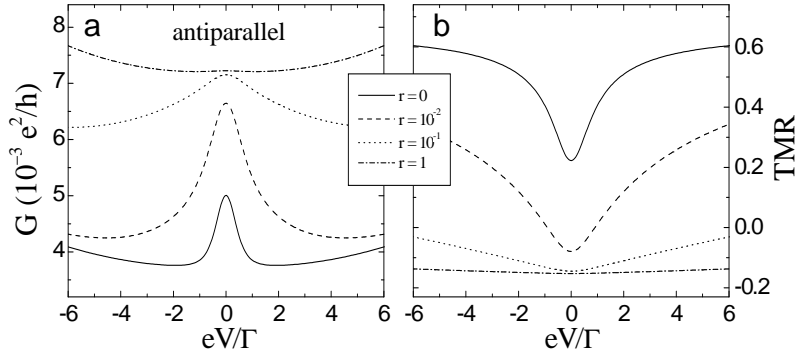}
  \caption{\label{Sec5Fig1} The differential conductance in the antiparallel
  configurations (a) and tunnel magnetoresistance
  (b) as a function of the bias voltage for different spin
  relaxation $r=h/(\tau_{\rm sf}\Gamma)$. The parameters are:
  $k_{\rm B}T=0.2\Gamma$,
  $\varepsilon =-15\Gamma$, $U=30\Gamma$,
  and $p_{\rm L}=p_{\rm R}=0.5$. (After Ref.~\cite{weymannPRB06})}
\end{figure}

The anomalous behavior of differential conductance in the
antiparallel configuration results from spin asymmetry of the
tunnelling processes, which gives rise to spin accumulation,
$P_\uparrow \neq P_\downarrow$. The  intrinsic spin-flip processes
in the dot reduce the spin accumulation and this way suppress the
anomaly. This behavior is displayed in Fig.~\ref{Sec5Fig1}(a),
where different curves correspond to different values of the
parameter $r$ defined as $r=h/(\tau_{\rm sf}\Gamma)$. Thus, $r=0$
describes the case with no intrinsic spin relaxation, whereas the
curves corresponding to nonzero $r$ describe the influence of
intrinsic relaxation processes. First of all, one can note that
small amount of intrinsic spin-flip processes enhances the
zero-bias anomaly [see the curve for $r=10^{-2}$ in
Fig.~\ref{Sec5Fig1}(a)]. This is because such processes play then
a role similar to that of single-barrier spin-flip cotunneling. In
the case of fast spin relaxation, on the other hand, the spin
accumulation is suppressed and the anomaly disappears, as can be
seen in Fig.~\ref{Sec5Fig1}(a) for $r=1$. It is also worth noting
that spin-flip processes in the dot enhance the overall
conductance in the antiparallel configuration. In the parallel
configuration, however, the differential conductance does not
depend on intrinsic relaxation.

Modifying the conductance in the antiparallel configuration, spin
relaxation processes suppress the tunnel magnetoresistance, as
shown in Fig. \ref{Sec5Fig1}(b). Since the zero-bias maximum in
conductance is suppressed in the fast spin relaxation limit, the
corresponding dip in TMR at small voltages is also suppressed by
the intrinsic relaxation processes. More specifically, the dip in
TMR broadens with increasing $r$ and disappears in the limit of
fast relaxation (see the curve for $r=1$). An interesting feature
of TMR in the presence of spin-flip scattering in the dot is the
crossover from positive to negative values when $r$ increases, as
illustrated in Fig.~\ref{Sec5Fig1}(b). Thus, the difference
between conductances in the parallel and antiparallel magnetic
configurations persists even for fast spin relaxation in the dot,
contrary to the sequential tunnelling regime, where such a
difference disappears \cite{rudzinski01}. This seemingly
counterintuitive behavior can be understood by taking into account
the following two facts: (i) absence of spin accumulation in the
dot for fast spin relaxation ($P_\uparrow =P_\downarrow $), and
(ii) difference in the fastest cotunneling processes contributing
to the current in the two magnetic configurations. The fastest
double-barrier cotunneling processes involve only the
majority-spin electrons of the two leads --  thus, in the parallel
configuration the fastest cotunneling processes are the
non-spin-flip ones. They take place either {\it via} the empty-dot
virtual state (for one orientation of the dot spin) or {\it via}
the doubly occupied dot virtual state (for the second orientation
of the dot spin). The dominant contribution to the current is then
proportional to $1/\varepsilon^2 + 1/(\varepsilon +U)^2$. On the
other hand, in the antiparallel magnetic configuration the fastest
cotunneling processes are the spin-flip ones, which can occur only
for one particular orientation of the dot spin. However, for this
spin orientation cotunneling can take place {\it via} both empty
and doubly occupied dot virtual states. The corresponding dominant
contribution to electric current is then proportional to
$[1/\varepsilon- 1/(\varepsilon +U)]^2=1/\varepsilon^2 +
1/(\varepsilon +U)^2 -2/[\varepsilon (\varepsilon +U)]$. It is
thus clear that the difference in currents flowing through the
system in the antiparallel and parallel configurations is equal to
$-2/[\varepsilon (\varepsilon +U)]$, which results from the
interference term. Since $\varepsilon <0$ and $\varepsilon +U>0$,
this interference contribution is positive. As a result, the
current in the antiparallel configuration is larger than the
current in the parallel configuration.

The minimum in TMR at zero bias in the case of a symmetric
Anderson model can be expressed as
\begin{equation}
  {\rm TMR_{min} } = \frac{2p^2 \left(4k_{\rm B}T\Gamma^2 -\varepsilon^2
  h/\tau_{\rm sf} \right)}{12(1-p^2)k_{\rm B}T\Gamma^2+(3+p^2)\varepsilon^2
  h/\tau_{\rm sf}},
\end{equation}
whereas for $|eV|\gg k_{\rm B}T$ and $r\ll 1$ one finds
\begin{equation}
  {\rm TMR_{max} } = \frac{2p^2 \left[ 2(3-p^2)k_{\rm B}T\Gamma^2
  -\varepsilon^2 h/\tau_{\rm sf} \right]}
  {2(1-p^2)(3-p^2)k_{\rm B}T\Gamma^2+(3+p^2)\varepsilon^2
  h/\tau_{\rm sf}}.
\end{equation}
The latter formula approximates the value of TMR corresponding to
the bias voltage at which the differential conductance has a local
minimum. In the slow spin relaxation limit one finds ${\rm
TMR_{min} }=2p^2/(3-3p^2)$, and ${\rm TMR_{max} }=2p^2/(1-p^2)$.
However, in the limit of fast spin relaxation TMR becomes negative
and is given by
\begin{equation}
{\rm TMR_{min} }={\rm TMR_{max} }=-2p^2/(3+p^2).
\end{equation}

\subsubsection{Effects of  external
magnetic field on cotunneling transport}

\begin{figure}[t]
  \centering
  \includegraphics[width=0.4\columnwidth]{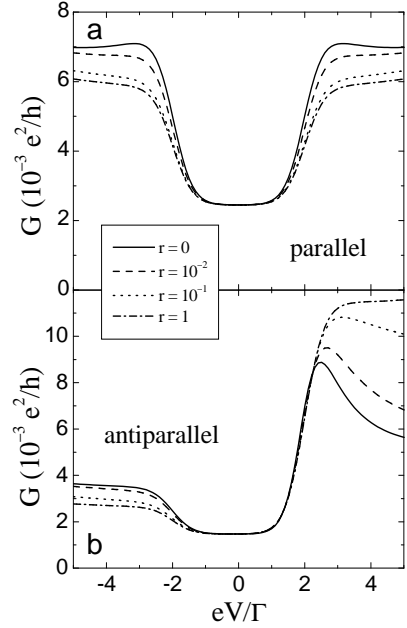}
  \caption{\label{Sec5Fig4} The differential conductance in the
  nonlinear response regime for different spin relaxation in the
  parallel (a) and antiparallel (b) configurations. The parameters
  are: $k_{\rm B}T=0.2\Gamma$, $\varepsilon_{\uparrow}=-16\Gamma$,
  $\varepsilon_{\downarrow}=-14\Gamma$, $U=30\Gamma$ and $p=0.5$.
  (After Ref.~\cite{weymannPRB06})}
\end{figure}

The discussion up to now was limited to the case of degenerate dot
level. The situation changes when
$\varepsilon_\uparrow\ne\varepsilon_\downarrow$, e.g., due to an
external magnetic field. The level splitting is described by the
parameter $\Delta=\varepsilon_\downarrow- \varepsilon_\uparrow$,
where the magnetic field is assumed to be along the magnetic
moment of the left electrode. In Fig.~\ref{Sec5Fig4} we show the
bias voltage dependence of the differential conductance in the
parallel and antiparallel configurations for different values of
parameter $r$.
In the limit of no intrinsic spin relaxation in the dot (solid
line in Fig.~\ref{Sec5Fig4}) and at low bias voltage, the dot is
occupied by a spin-up electron and the current flows mainly due to
non-spin-flip cotunneling. The spin-flip cotunneling processes are
suppressed for $\vert \Delta\vert \gtrsim \vert eV\vert, k_{\rm
B}T$, which results in the steps in differential conductance at
$\vert\Delta\vert \simeq \vert eV\vert$. The suppression of
spin-flip inelastic cotunneling was recently used as a tool to
determine the spectroscopic $g$-factor \cite{kogan04}. When $\vert
eV\vert$ becomes larger than $\vert\Delta\vert$, spin-flip
cotunneling is allowed, consequently the conductance increases.
However, there is a large asymmetry of differential conductance in
the antiparallel configuration with respect to the bias reversal.
To understand this asymmetry, it is crucial to realize that when
the splitting $\Delta =\varepsilon_\downarrow -
\varepsilon_\uparrow$ is larger than $k_{\rm B}T$, the
single-barrier spin-flip cotunneling processes can occur only when
the dot is occupied by a spin-down electron. Thus, the
single-barrier processes can assist the fastest double-barrier
cotunneling processes, but only for positive bias. This is because
the fastest processes can occur when the dot is occupied by a
spin-down electron for negative bias and by a spin-up electron for
positive bias, leading to larger conductance for positive than for
negative bias voltage. No such asymmetry occurs in the parallel
configuration [see Fig.~\ref{Sec5Fig4}(a)], as now the system is
fully symmetric with respect to bias reversal.

The situation changes when intrinsic spin-flip relaxation
processes occur in the dot. For the parameters assumed in
Fig.~\ref{Sec5Fig4}, the spin relaxation in the dot affects the
conductance only for $|eV|\gtrsim|\Delta|$, while for
$|eV|\lesssim|\Delta|$ the conductance is basically independent of
$r$ (see Fig.~\ref{Sec5Fig4}). This is because for
$|eV|\lesssim|\Delta|$ and $|\Delta|\gg k_{\rm B}T$, the dot is
predominantly occupied by a spin-up electron and the transitions
to the spin-down state due to relaxation processes are
energetically forbidden. As a consequence, the current flows
mainly due to non-spin-flip cotunneling, irrespective of spin
relaxation time. This scenario holds for both magnetic
configurations of the system.

When $|eV|\gtrsim|\Delta|$, the spin-flip cotunneling processes
can take place and the dot can be either in the spin-up or
spin-down state. In the parallel configuration,
Fig.~\ref{Sec5Fig4}(a), the conductance is slightly reduced by the
spin-flip relaxation processes. This can be understood by
realizing the fact that the fastest non-spin-flip cotunneling
processes in the parallel configuration are more probable when the
dot is occupied by a spin-down electron than by a spin-up one [due
to smaller energy denominator, see
Eq.~(\ref{Eq:cotunnelingrate})]. Since the spin-down state (as
that of larger energy) relaxes relatively fast to the spin-up
state (which has definitely smaller energy), this leads to a
reduction in the conductance. On the other hand, the differential
conductance in the antiparallel configuration is enhanced by the
relaxation processes for positive bias and diminished for negative
bias voltages. Consider first the situation for positive bias. As
already discussed above for $r=0$, an important role in that
transport regime is played by the single-barrier spin-flip
cotunneling processes, which open the system for the fast
double-barrier cotunneling by reversing spin of the dot from the
spin-down to the spin-up state. The relaxation processes play a
role similar to that of the single-barrier cotunneling, and lead
to a certain increase in the conductance. For negative bias
voltage, in turn, the fast double-barrier cotunneling processes
occur when the dot is occupied by a spin-down electron. The
probability of such events is decreased by spin relaxation,
leading to a reduced conductance. An interesting consequence of
the enhancement (reduction) of the differential conductance for
positive (negative) bias voltage is an increase of the asymmetry
with respect to the bias reversal -- see the curve for $r=1$ in
Fig.~\ref{Sec5Fig4}(b).

\subsubsection{Asymmetric situations}

\begin{figure}[t]
  \centering
  \includegraphics[width=0.4\columnwidth]{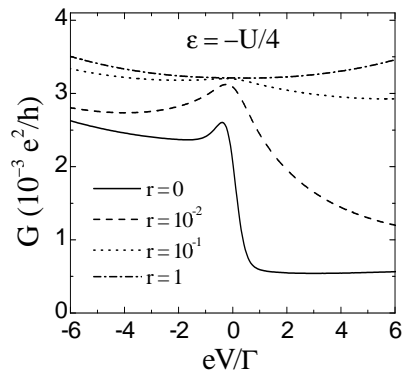}
  \caption{\label{Sec5Fig5} The differential conductance in the
  nonlinear response regime for the asymmetric
  Anderson model, $\varepsilon=-U/4$, for different spin relaxation.
  The parameters are: $k_{\rm B}T=0.2\Gamma$,
  $U=60\Gamma$, $p_{\rm L}=0.95$ and $p_{\rm R}=0$. (After Ref.~\cite{weymannPRB06})}
\end{figure}

An interesting situation occurs when the quantum dot is coupled
asymmetrically to the left and right leads $(p_{\rm L}\neq p_{\rm
R})$. The differential conductance for a system with one electrode
nonmagnetic and the other one made of a ferromagnet with large
spin polarization (in the following referred to as strong
ferromagnet) is shown in Fig.~\ref{Sec5Fig5} for the case when the
dot is described by an asymmetric Anderson model
($\vert\varepsilon\vert\ne\varepsilon +U$).

Consider first the situation in the absence of intrinsic spin
relaxation in the dot (solid curves in Fig.~\ref{Sec5Fig5}). When
$\vert eV\vert \gg k_{\rm B}T$, the influence of single-barrier
cotunneling can be neglected \cite{weymannPRBBR05}. The
cotunneling processes which transfer charge from one lead to
another take place {\it via} two possible virtual states -- empty
dot (an electron residing in the dot tunnels to one of the leads
and another electron from the second lead enters the dot) and
doubly occupied dot (an electron of spin opposite to that in the
dot enters the dot and then one of the two electrons leaves the
dot). Consider first positive bias ($eV>0$, electrons flow from
right to left, i.e., from normal metal to strong ferromagnet), and
assume for clarity of discussion that the strong ferromagnet is a
half-metallic one with full spin polarization (only spin-up
electrons can then tunnel to the left lead). When a spin-down
electron enters the dot, it has no possibility to leave the dot
for a long time. The allowed cotunneling processes occur then {\it
via } doubly occupied dot virtual states. In the absence of
intrinsic spin relaxation in the dot, the only processes which can
reverse the dot spin are the single-barrier cotunneling ones,
which however play a minor role when $k_{\rm B}T\ll \vert
eV\vert$. Thus, the current flows due to non-spin-flip cotunneling
{\it via} doubly-occupied dot virtual states, whereas cotunneling
through empty-dot virtual states is suppressed. The situation is
changed for negative bias (electrons flow from strong ferromagnet
to normal metal). Now, the dot is mostly occupied by a spin-up
electron, which suppresses cotunneling {\it via} doubly-occupied
dot virtual state and the only contribution comes from cotunneling
{\it via} empty-dot virtual state. The ratio of cotunneling rates
through the empty dot and doubly occupied dot virtual states is
approximately equal to $\xi=[\varepsilon/ (\varepsilon+U) ]^{-2}$.
In the situation presented in Fig.~\ref{Sec5Fig5} one finds
$\xi\gg 1$. Accordingly, the conductance for negative bias is much
larger than for positive bias voltage.

When $\vert eV \vert$ becomes of the order of $k_{\rm B}T$ or
smaller, the rate of single-barrier cotunneling is of the order of
the rate of double-barrier cotunneling. Therefore, the
single-barrier processes can play an important role in transport.
More precisely, single-barrier cotunneling processes can reverse
spin of an electron in the dot and thus can open the system for
the fast cotunneling processes.

Intrinsic spin-flip processes in the dot have similar influence on
electronic transport as in the case discussed in the previous
section. As before, relaxation processes remove the asymmetry with
respect to the bias reversal and suppress the zero-bias anomaly.
Thus, the diode-like behavior can appear only in the limit of slow
spin relaxation, and is suppressed in the limit of fast spin
relaxation, as shown in Fig.~\ref{Sec5Fig5} by the curves
corresponding to $r=1$.


\subsection{Systems with noncollinear magnetizations}


New features of transport characteristics occur when the leads'
magnetizations form an arbitrary noncollinear magnetic
configuration \cite{surgueevPRB02,franssonPRB05,franssonEPL05,
koenigPRL03,braun04,rudzinski05,weymann07noncol,koenig_springer04,
rudzinskiJMMM04,rudzinskiJMMM05,flensberg04,rudzinskiPSS05,mu06}.
Such a configuration can be controlled by a weak external magnetic
field, weak enough to neglect the corresponding Zeeman splitting
of the dot level and weaker than the effective exchange field
exerted on the dot by ferromagnetic leads. In particular, the
differential conductance is significantly modified by the exchange
field in noncollinear configurations. This also applies to the TMR
effect, which for arbitrarily aligned leads' magnetizations  can
be defined as, ${\rm TMR} = [I_{\rm P} - I(\varphi)]/I(\varphi)$,
where $\varphi$ is an angle between the leads' magnetic moments.

\begin{figure}[h]
  \centering
  \includegraphics[width=0.4\columnwidth]{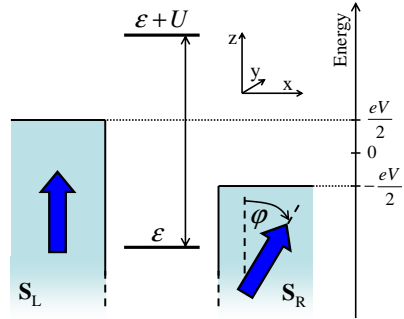}
  \caption{\label{Sec6Fig:1}
  (color online) Schematic of a quantum dot coupled to
  ferromagnetic leads with non-collinearly aligned
  magnetizations. The net spin moments of the left
  $\mathbf{S}_{\rm L}$ and right $\mathbf{S}_{\rm R}$ lead
  form an angle $\varphi$. There is a symmetric bias voltage
  applied to the system.}
\end{figure}

The system, whose spin moments $\mathbf{S}_{\rm L}$ and
$\mathbf{S}_{\rm R}$ of the left and right lead, respectively,
form an arbitrary configuration is shown in Fig.~\ref{Sec6Fig:1}.
The corresponding Hamiltonian, written for the quantization axis
of the dot equivalent to that of the left lead, is given by Eq.
(\ref{Sec4Eq:QDHamiltonian}), where the tunnel Hamiltonian can be
decomposed into two terms $H_{\rm T} = H_{\rm TL} + H_{\rm TR}$.
The term $H_{\rm TL}$ describes tunnelling processes between the
left electrode and the dot and takes the same form as for
collinear configurations, while $H_{\rm TR}$ describes  tunnelling
processes between the dot and right lead, and acquires the form
\begin{eqnarray}
\lefteqn{  H_{\rm TR} = \sum_{k} \left[\left(
    t_{{\rm R} +}c_{{\rm R} k+}^{\dagger} \cos\frac{\varphi}{2}
    - t_{{\rm R} -}c_{{\rm R} k-}^{\dagger} \sin\frac{\varphi}{2} \right)
    d_{\uparrow}  \right. } \nonumber\\
    &&\left. + \left(
    t_{{\rm R} +}c_{{\rm R} k+}^{\dagger} \sin\frac{\varphi}{2}
    + t_{{\rm R} -}a_{{\rm R} k-}^{\dagger} \cos\frac{\varphi}{2} \right)
    d_{\downarrow} + {\rm h.c.} \right] \,.
\end{eqnarray}

The real-time diagrammatic method described at the beginning of
this section for collinear magnetic configurations can be extended
to include the effects due to noncollinearity of the leads'
magnetic moments. The density matrix of the quantum dot for an
arbitrary magnetic configuration is given by
\begin{equation}
  \hat{\rho}_{\rm D} = \left(\begin{array}{cccc}
    P_0^0 & 0 & 0 & 0\\
    0 & P_{\uparrow}^{\uparrow} & P_\downarrow^\uparrow  & 0\\
    0 & P^\downarrow_\uparrow & P_\downarrow^\downarrow  & 0\\
    0 & 0 & 0 & P_{\rm d}^{\rm d}\\
  \end{array} \right) .
\end{equation}
The diagonal elements of the density matrix correspond to the
respective occupation probabilities, while the off-diagonal
elements $P_{\uparrow}^{\downarrow}$ and
$P_{\downarrow}^{\uparrow}$ describe the dot spin $\vec{S}$, with
$S_x={\rm Re}P_\downarrow^\uparrow$, $S_y={\rm
Im}P_\downarrow^\uparrow$, and
$S_z=\left(P_\uparrow^\uparrow-P_\downarrow^\downarrow \right)/2$.
The density matrix elements can be determined from the
corresponding kinetic equation, which in the steady state and for
spin-degenerate dot level can be written as
\cite{schoeller94,weymann07noncol,koenigdiss,koenig96}
\begin{equation} \label{Sec6Eq:master}
  0 = \sum_{\chi_1^\prime,\chi_2^\prime} P_{\chi^\prime_2}^{\chi^\prime_1}
    \Sigma_{\chi_2^\prime \chi_2}^{\chi_1^\prime \chi_1} \,.
\end{equation}
By expanding the self-energies and density matrix elements, one
can calculate the sequential and cotunneling current.

In the following we focus on  transport in the Coulomb blockade
regime, $|\varepsilon|,|\varepsilon+U|\gg \Gamma,k_{\rm B}T$, and
when the dot is singly occupied, $\varepsilon<0<\varepsilon+U$.
>From the discussion above we know that there is an anomalous
behavior of the differential conductance in the small bias regime
for antiparallel  magnetic configuration. Now, we will consider
how this anomaly changes when magnetic configuration varies
continuously from antiparallel to parallel alignment.

\subsubsection{Symmetric Anderson model}

We consider first the symmetric Anderson model,
$\varepsilon=-U/2$. The exchange field vanishes then for an
arbitrary magnetic configuration, and both differential
conductance and TMR change monotonically when going from
antiparallel to parallel magnetic configurations. However, for
$\varepsilon \neq -U/2$, the effects of exchange field become
important and lead to nontrivial behavior of transport
characteristics, as will be discussed later.

\begin{figure}[t]
  \centering
  \includegraphics[width=0.46\columnwidth]{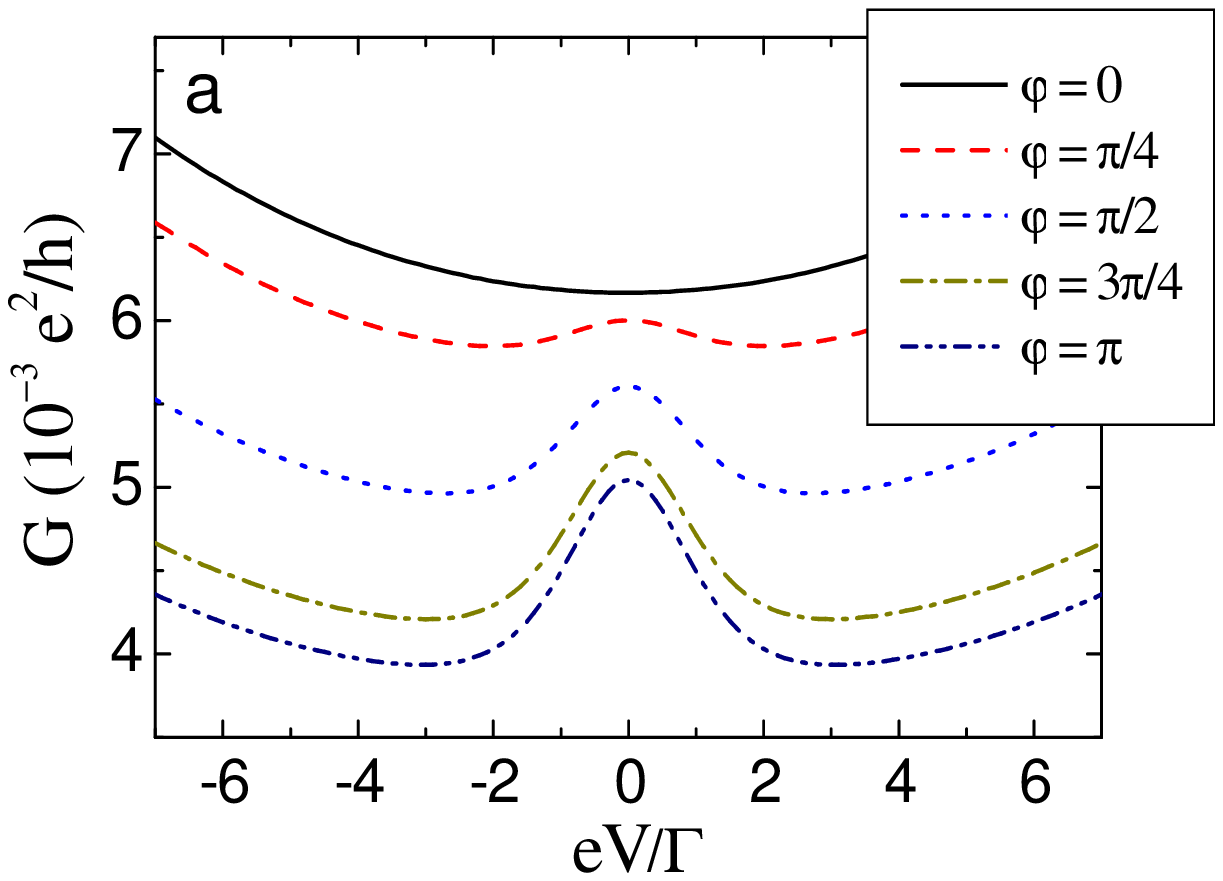}
  \vspace{0.7cm}
  \includegraphics[width=0.46\columnwidth]{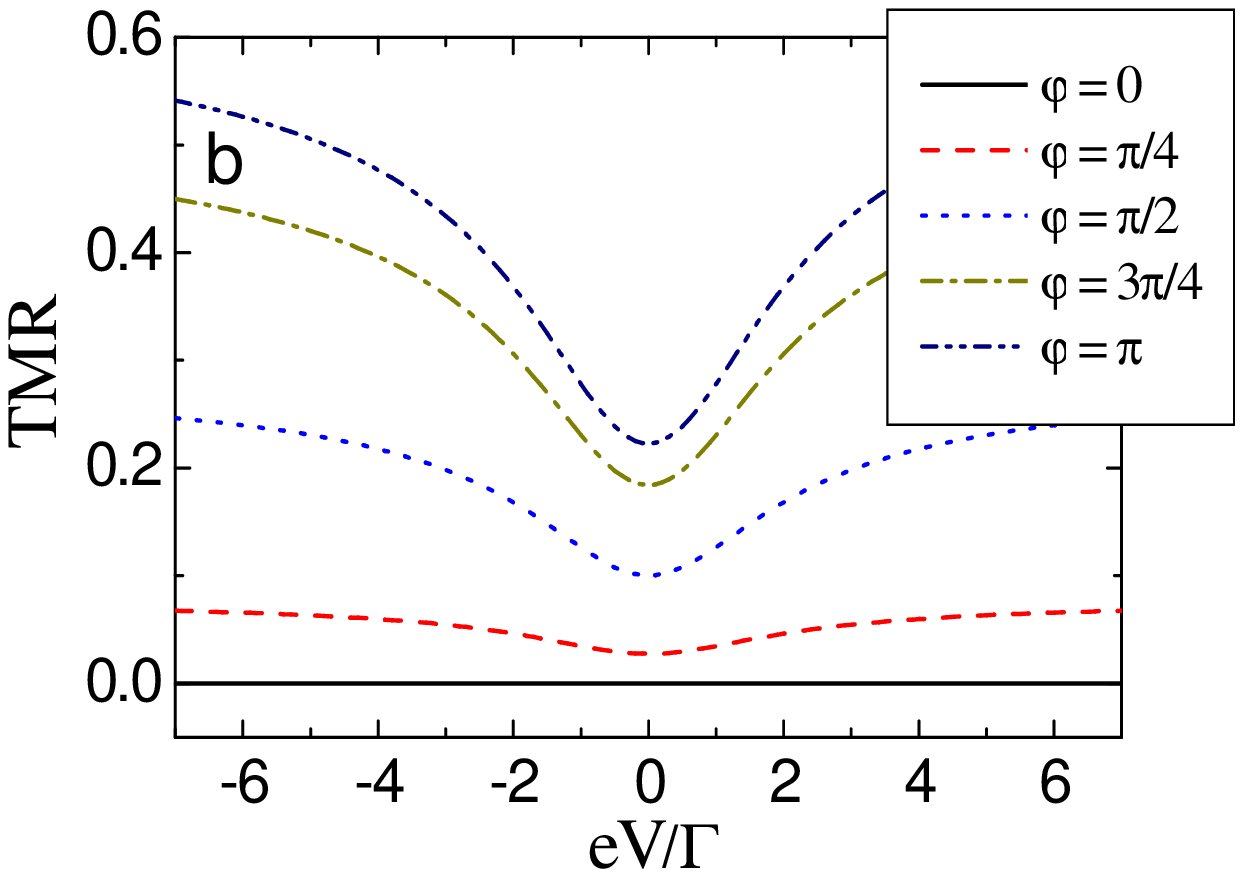}
  \caption{\label{Sec6Fig:2} (color online)
  Differential conductance $G=dI/dV$ (a)
  and tunnel magnetoresistance TMR (b) as a function
  of the bias voltage $V$ for indicated values of the angle
  $\varphi$ between magnetic moments, calculated for
  symmetric Anderson model.
  The parameters are:
  $k_{\rm B} T=0.5 \Gamma$,
  $\varepsilon=-15\Gamma$, $U=30\Gamma$, and
  $p_{\rm L} = p_{\rm R}\equiv p=0.5$. (After Ref.~\cite{weymann07noncol})
  }
\end{figure}

The differential conductance $G$ as a function of the bias voltage
is shown in Fig.~\ref{Sec6Fig:2}(a) for several values of the
angle $\varphi$. The conductance varies monotonically with the
angle between leads' magnetic moments. When the angle increases
from zero to $\pi$, the anomaly emerges at small values of
$\varphi$ and its relative height increases with increasing angle,
reaching a maximum value at $\varphi =\pi$.

The exchange field for a symmetric Anderson model is negligible,
and the average dot spin tends to zero in the linear response
regime. The angular dependence of the linear conductance can be
then expressed as
\begin{equation}\label{Sec6Eq:Glin}
  G = \frac{e^2\Gamma^2}{2h\varepsilon^2}\left[
  3-p^2\left( 1+2\sin^2 \frac{\varphi}{2}\right)
  \right] \,.
\end{equation}
The variation of $G$ with $\varphi$ at low bias is thus
characterized by the factor $1+2\sin^2 (\varphi/2)$, which leads
to maximum (minimum) conductance in the parallel (antiparallel)
magnetic configuration. Such behavior is typical of a normal
spin-valve effect.

The bias dependence of the associated TMR is shown in
Fig.~\ref{Sec6Fig:2}(b) for several values of the angle $\varphi$.
The zero-bias anomaly in the differential conductance [see
Fig.~\ref{Sec6Fig:2}(a)] leads to the corresponding anomaly (dip)
in TMR at small bias voltages. The dip in TMR decreases when
magnetic configuration departs from the antiparallel alignment,
and eventually disappears in the parallel configuration. The
variation of TMR with the angle is monotonic, similarly as the
angular variation of the differential conductance. The dependence
of TMR on the angle $\varphi$ at zero bias is given approximately
by the formula
\begin{equation}
  {\rm TMR} = \frac{2p^2 \sin^2 \frac{\varphi}{2}}
  {3-p^2\left( 1+2\sin^2 \frac{\varphi}{2}\right)} \,.
\end{equation}
Now, the angular dependence of TMR is governed by $\sin^2
(\varphi/2)$, which gives maximum TMR in the antiparallel
configuration and zero TMR in the parallel one.

\subsubsection{Asymmetric Anderson model}

\begin{figure}[t]
  \centering
  \includegraphics[width=0.46\columnwidth]{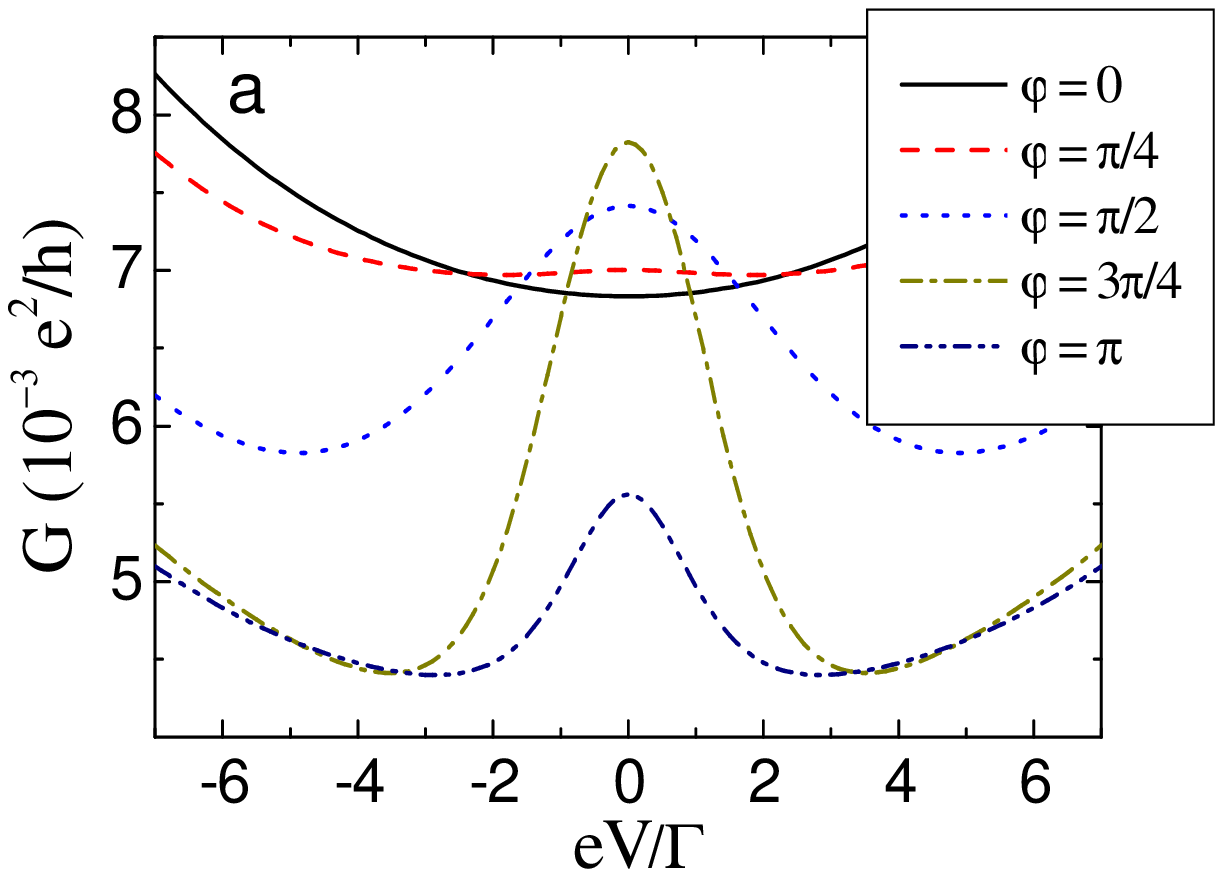}
  \vspace{0.7cm}
  \includegraphics[width=0.46\columnwidth]{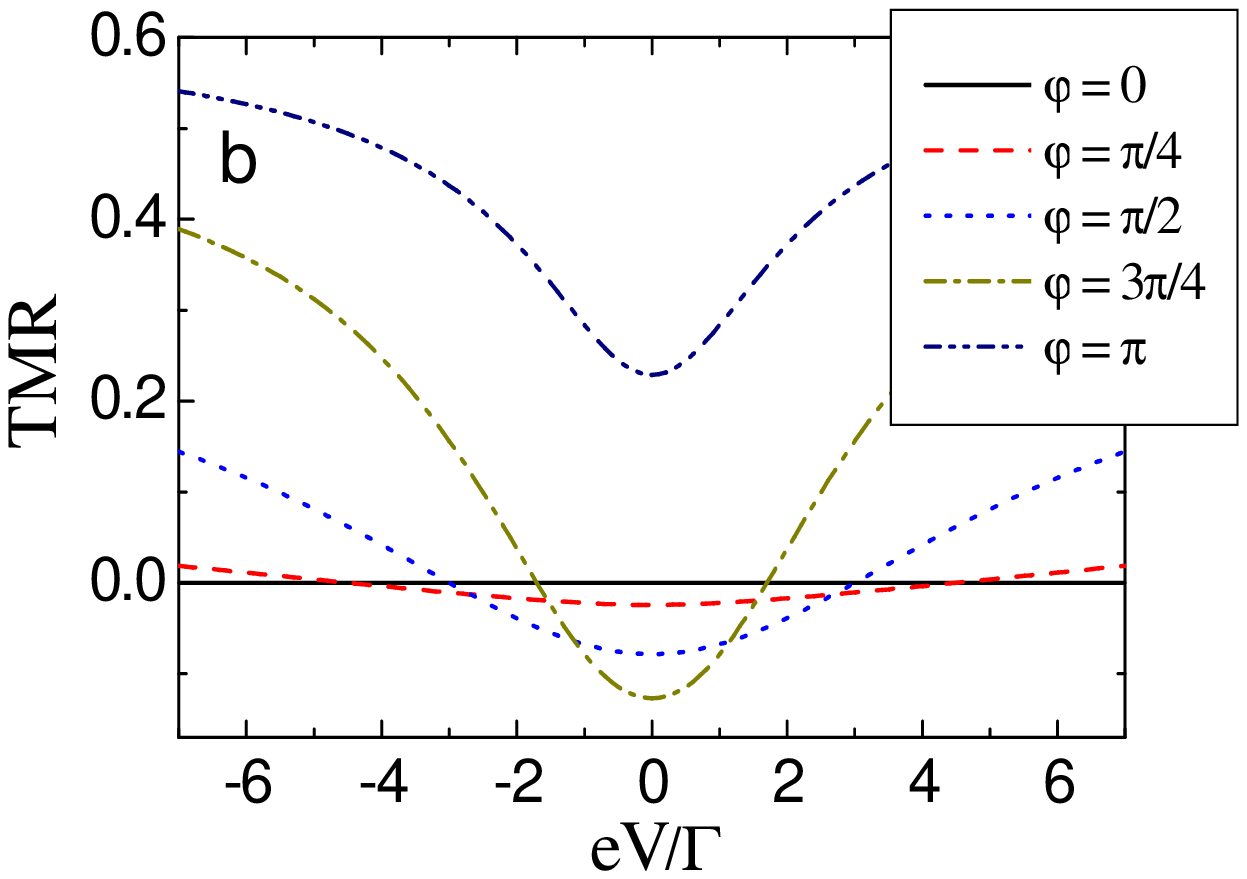}
  \caption{\label{Sec6Fig:3} (color online)
  The differential conductance $G=dI/dV$ (a) and
  tunnel magnetoresistance TMR (b) as a function
  of the bias voltage $V$ for indicated  values of
  the angle $\varphi$   for asymmetric Anderson model
  $\varepsilon=-12\Gamma$ and $U=30\Gamma$.
  The other parameters are the same as in Fig.~\ref{Sec6Fig:2}.
  (After Ref.~\cite{weymann07noncol})  }
\end{figure}

The transport characteristics, described above for the symmetric
Anderson model, are significantly modified when the model becomes
asymmetric, i.e. when $\varepsilon \neq -U/2$. This can be
realized by shifting the dot level position by a gate voltage
applied to the dot. Different contributions to the exchange field
do not cancel then, and the resulting effective exchange field
becomes nonzero. And this exchange field has a significant
influence on transport properties. The strength of effective
exchange field grows with deviation of the Anderson model from the
symmetric one, described quantitatively by $2\varepsilon+U$ (with
$2\varepsilon+U=0$ for the symmetric model).

The leading contribution to the exchange field comes from the
first-order diagrams of the perturbation expansion in terms of the
real-time diagrammatic technique, whereas the current flows due to
the second-order tunnelling processes. This leads to two different
time scales which determine transport characteristics. Transport
properties are thus a result of the interplay of the first- and
second-order processes. Although the first-order processes do not
contribute directly to current, they influence transport {\it via}
modification of the dot spin.

The differential conductance is shown in Fig.~\ref{Sec6Fig:3}(a)
as a function of bias voltage for an asymmetric Anderson model,
$\varepsilon\neq -U/2$, and for different values of $\varphi$. The
low-bias differential conductance becomes enhanced in a certain
range of the angle $\varphi$, leading to a nonmonotonic dependence
of the conductance $G$ on the angle between the leads'
magnetizations. This nonmonotonic behavior is also visible in TMR,
see Fig.~\ref{Sec6Fig:3}(b). There is also a range of the angle
$\varphi$, where TMR changes sign and becomes negative in the
small bias region, i.e. the corresponding conductance is larger
than that in the parallel configuration.

The most characteristic features of transport characteristics in
the presence of exchange field are the enhanced differential
conductance at a non-collinear alignment and its rapid drop when
the system approaches the antiparallel configuration. The key role
in this behavior is played by the first-order processes giving
rise to the exchange field. These processes lead to the precession
of spin in the dot, which facilitates tunnelling processes and
leads to an increase in the conductance as compared to that in the
parallel configuration. When the configuration becomes close to
the antiparallel one, the first-order processes become suppressed
and the conductance drops to that for antiparallel alignment. The
nonmonotonic behavior of the differential conductance with
$\varphi$ leads to a nonmonotonic dependence of TMR. The effects
due to exchange field give rise to a local minimum in TMR at a
non-collinear magnetic configuration. Moreover, in this transport
regime TMR changes sign and becomes negative. When the magnetic
configuration is close to the antiparallel one, TMR starts to
increase rapidly reaching maximum for $\varphi=\pi$. The negative
TMR and its sudden increase when the configuration tends to the
antiparallel one are a consequence of the processes leading to
nonmonotonic behavior of the differential conductance, as
described above.

To understand more intuitively the above presented behavior of the
differential conductance and TMR at low bias voltage and close to
the antiparallel configuration, one should consider two different
time scales. One time scale, $\tau_{\rm prec}$, is established by
the virtual first-order processes responsible for the spin
precession due to exchange field,
\begin{equation} \label{Sec6Eq:prec_rate}
  \left|\tau^{-1}_{\rm prec}\right| \approx \frac{\Gamma}{2h} p
  \sin\varphi \ln \left|\frac{\varepsilon}{\varepsilon+U}
  \right|\,.
\end{equation}
The second time scale, $\tau_{\rm cot}$, is associated with
second-order processes which drive the current through the system.
At low temperature and low bias voltage, the cotunneling rate can
be expressed as
\begin{equation} \label{Sec6Eq:cot_rate}
  \tau^{-1}_{\rm cot} \approx \frac{\Gamma^2}{4h}
  (1+p)(1-p\cos\varphi)
  \frac{|eV|U^2}{\varepsilon^2(\varepsilon+U)^2}\,.
\end{equation}
The rate $\tau_{\rm cot}$ depends linearly on the applied voltage,
whereas $\tau_{\rm prec}$ is rather independent of $V$. As a
consequence, at low bias and for non-collinear configuration, the
exchange field plays an important role leading to a nonmonotonic
dependence of differential conductance on the angle between the
leads' magnetizations. When magnetic configuration is close to the
antiparallel one, the spin precession rate is deceased
($\left|\tau^{-1}_{\rm prec}\right|\sim\sin\varphi$) and, at
certain angle, the rate of spin precession becomes comparable to
the cotunneling rate. This gives rise to a sudden drop (increase)
in differential conductance (TMR). We note that the nonmonotonic
dependence of differential conductance and magnetoresistance has
also been observed in quantum dots in the strong coupling limit
\cite{franssonPRB05,franssonEPL05}.


\section{Transport through multi-level quantum dots connected to ferromagnetic leads}


Up to now we discussed theoretical aspects of electronic transport
through the simplest quantum dots, i.e. the dots with only one
orbital level. Real dots are usually more complex and their
electronic spectrum includes many orbital levels taking part in
electronic transport. This may significantly change transport
characteristics and also may lead to qualitatively new features
\cite{thielmann05,belzigPRB05,elste06,weymannJPCM07,weymannEPL06,
cottetPRB06}. In this section we will consider some of the new
effects in transport through multi-dot systems.


\subsection{Sequential transport in two-level dots}

\begin{figure}[t]
\centering
  \includegraphics[width=0.4\columnwidth]{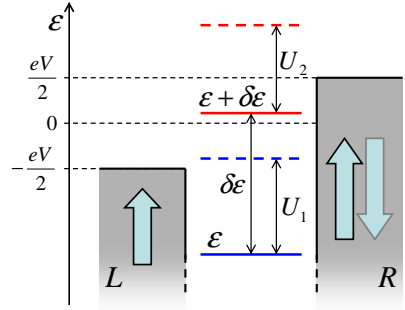}
  \caption{\label{Sec7Fig:1} (color online) Energy diagram of a two-level
  quantum dot coupled to ferromagnetic leads.
  For clarity reasons the energy diagram is shown here for $\Delta =U^\prime
  =0$. The leads' magnetizations can form either
  parallel or antiparallel configurations.
  The arrows indicate the net spin of the leads.}
\end{figure}

The schematic of a two-level quantum dot coupled to ferromagnetic
leads with collinear magnetizations is shown in
Fig.~\ref{Sec7Fig:1}.
The quantum dot is described by the following Hamiltonian
\begin{eqnarray}\label{Sec7Eq:Hamiltonian}\fl
  \hat{H}_{\rm D} =\sum_{j\sigma} \varepsilon_{j} n_{j\sigma}
  + U \sum_j n_{j\uparrow} n_{j\downarrow}
  + U^\prime \sum_{\sigma\sigma^\prime}
  n_{1\sigma}n_{2\sigma^\prime}
  -\frac{\Delta}{2} \sum_j
  \left(n_{j\uparrow}-n_{j\downarrow}\right) \,,
\end{eqnarray}
where $n_{j\sigma}$ is the particle number operator, $n_{j\sigma}
= d^{\dagger}_{j\sigma}d_{j\sigma}$, $d^{\dagger}_{j\sigma}$
($d_{j\sigma}$) is the creation (annihilation) operator of an
electron with spin $\sigma$ on the $j$th level ($j=1,2$), and
$\varepsilon_{j}$ is the corresponding single-particle energy. The
on-level Coulomb repulsion between two electrons of opposite spins
is described by $U$, whereas the inter-level repulsion energy is
denoted by $U^\prime$. The forth term in
Eq.~(\ref{Sec7Eq:Hamiltonian}) describes the Zeeman energy, with
$\Delta=g\mu_B B$ being the Zeeman splitting of the energy levels
($B$ is an external magnetic field along the magnetic moment of
the left electrode). To present the main features of transport
characteristics, it is convenient to introduce the level spacing
$\delta\varepsilon= \varepsilon_2 -\varepsilon_1$ and also define
$\varepsilon_1\equiv \varepsilon$.

In the following we analyze the current $I$, differential
conductance $G$, and the Fano factor $F$ in the parallel and
antiparallel magnetic configurations, as well as the corresponding
TMR for two-level quantum dots. The Fano factor, $F=S/S_p$,
describes the deviation of the zero-frequency shot noise $S$ from
the Poissonian shot noise $S_p=2e|I|$. The presented results have
been obtained within the real-time diagrammatic technique
\cite{thielmann03,thielmann05,weymannJPCM07}.

\subsubsection{Quantum dots symmetrically coupled to ferromagnetic
leads}

\begin{figure}[h]
\centering
  \includegraphics[width=0.9\columnwidth]{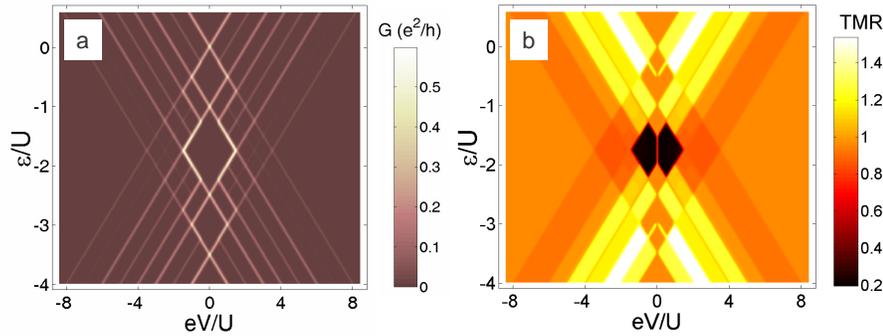}
  \caption{\label{Sec7Fig:2} (color online)
  The differential conductance $G=dI/dV$ as a function
  of the bias voltage and level position in
  the parallel magnetic configuration (a) and TMR (b)
  for the parameters:
  $k_{\rm B} T=\Gamma$, $\delta\varepsilon=25\Gamma$,
  $U=50\Gamma$, $\Delta=0$, $p_{\rm L} = p_{\rm R}\equiv p=0.7$, and
  $\Gamma_{rj}\equiv\Gamma/2$ ($r={\rm L,R}$, $j=1,2$).
  (After Ref.~ \cite{weymannJPCM07})}
\end{figure}

Typical variation of the differential conductance with the bias
voltage $V$ and level position $\varepsilon$ (gate voltage) is
shown in Fig.~\ref{Sec7Fig:2}(a) for the parallel magnetic
configuration. Conductance in the antiparallel configuration is
qualitatively similar to that in the parallel configuration, but
generally smaller, which leads to nonzero TMR effect. The diamonds
in Fig.~\ref{Sec7Fig:2}(a) around $V=0$ correspond to the Coulomb
blockade regions. When lowering position of the dot levels, the
charge of the dot changes successively. The dot is empty for
$\varepsilon\gtrsim 0$, occupied by one electron for
$0\gtrsim\varepsilon\gtrsim -U$, doubly occupied for
$-U\gtrsim\varepsilon\gtrsim -(2U+\delta\varepsilon)$, occupied by
three electrons for $-(2U+\delta\varepsilon)
\gtrsim\varepsilon\gtrsim -(3U+\delta\varepsilon)$, and the two
orbital levels of the dot are fully occupied for
$-(3U+\delta\varepsilon) \gtrsim\varepsilon$. In all these
transport regions the dot is in a well-defined charge state, and
the sequential tunnelling is exponentially suppressed. If the bias
voltage is increased above a certain threshold voltage, the
current flows due to first-order (sequential) tunnelling
processes. When the thermal energy is low enough, one observes
then a well-resolved step in the current as a function of the bias
voltage. In the density plots shown in Fig.~\ref{Sec7Fig:2}a, this
can be seen in the form of lines that clearly separate the Coulomb
blockade regions from transport regions associated with
consecutive charge states taking part in transport. When the bias
voltage increases further, additional steps [and consequently
lines in Fig.~\ref{Sec7Fig:2}(a)] arise at voltages where new
states becomes active in transport.

The corresponding TMR as a function of the bias and gate voltages
is shown in Fig.~\ref{Sec7Fig:2}(b). It is worth noting that TMR
takes now several well-defined values. Such behavior of TMR is
significantly different from that for a single-level quantum dot,
where  TMR in the sequential tunnelling regime acquires only two
values \cite{weymannPRB05}. As in single-level dots, TMR in the
linear response regime is independent of the gate voltage and is
given by $p^2/(1-p^2)$, i.e. half of the TMR in the Julliere model
\cite{julliere}. Figure~\ref{Sec7Fig:2} also shows that, when
increasing the bias voltage $V$ and keeping constant position of
the dot levels, the current and TMR acquire some specific and
well-defined values in different transport regions. As shown in
Ref. \cite{weymannJPCM07},  the current and TMR at these plateaus
can be approximated by simple analytical formulas.

\begin{figure}[t]
\centering
  \includegraphics[height=8.2cm]{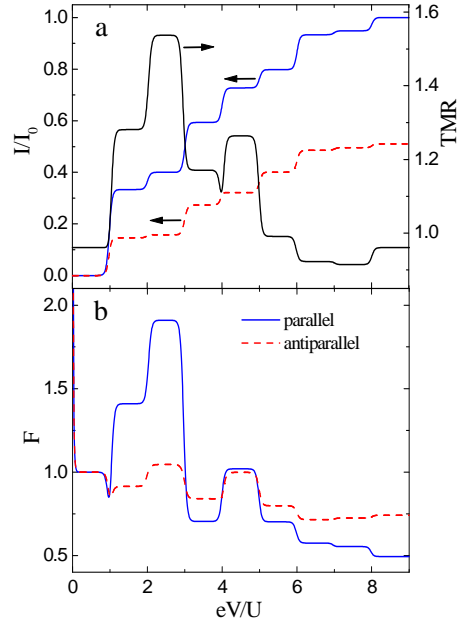}
  \caption{\label{Sec7Fig:4} (color online)
 Current (a) in the units of $I_0=e\Gamma/\hbar$
  and Fano factor (b) in the
  parallel (solid line) and antiparallel
  (dashed line) configurations
  as well as tunnel magnetoresistance (a) as a function
  of the bias voltage for $\varepsilon=U/2$. The
  other parameters are the same as in Fig.~\ref{Sec7Fig:2}.
  (After Ref.~\cite{weymannJPCM07})}
\end{figure}

This behavior is presented in more details in
Fig.~\ref{Sec7Fig:4}(a), where the bias voltage dependence of the
current and TMR is shown explicitly for the case when the dot
level is above ($\varepsilon = U/2$) the Fermi level of the leads
at equilibrium. The current in both magnetic configurations and
the associated TMR exhibit characteristic plateaus which
correspond to different transport regions. The corresponding Fano
factors $F_{\rm P}$ and $F_{\rm AP}$ in the parallel and
antiparallel magnetic configurations are shown in
Fig.~\ref{Sec7Fig:4}b.  Similarly as current and TMR, the Fano
factor acquires roughly constant values, different in different
transport regions. Due to the spin asymmetry in the coupling of
the dot to external leads, the bias dependence of the Fano factor
is significantly different from that in the corresponding
nonmagnetic situations \cite{thielmann05}. For both magnetic
configurations of the system, the Fano factor depends on the
polarization factor $p$ (differently in the two configurations, in
general). If $|eV|\ll k_{\rm B}T$, the Fano factor becomes
divergent due to the thermal noise, which dominates in this
transport regime; in the case of $V=0$, the noise is given by $S =
4k_{\rm B}T G^{\rm lin}$, with $G^{\rm lin}$ being the linear
conductance, leading to a divergency of the Fano factor
\cite{blanterPR00,sukhorukovPRB01}. Apart from this, in some
transport regions we find $F_{\rm P}>F_{\rm AP}$, while in the
other ones $F_{\rm P} < F_{\rm AP}$. The ratio $F_{\rm P}/F_{\rm
AP}$ depends generally on the spin polarization of the leads $p$
\cite{weymannJPCM07}. Furthermore, we note that if the leads are
half-metallic, the Fano factor in the parallel configuration
diverges as $p\rightarrow 1$. This increase of the Fano factor is
due to the enhanced spin asymmetry in transport processes through
the dot \cite{bulka00,cottetPRL04,cottetPRB04}. On the other hand,
in the antiparallel configuration the Fano factor tends to unity
for $p\rightarrow 1$, except for the Coulomb blockade regime with
two electrons trapped in the dot, where the shot noise is
super-Poissonian.

\subsubsection{Quantum dots asymmetrically coupled to the leads}

\begin{figure}[t]
\centering
  \includegraphics[width=0.45\columnwidth]{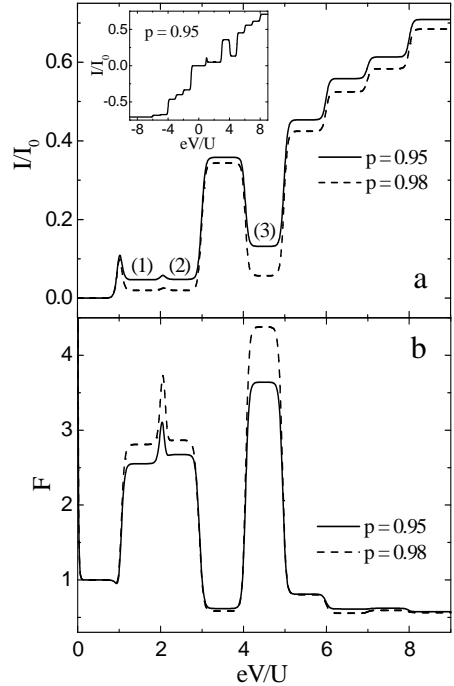}
  \caption{\label{Sec7Fig:5}
  The current (a) in units of $I_0=e\Gamma/\hbar$
  and Fano factor (b)
  as a function of the bias voltage for
  $p_{\rm L} \equiv p = 0.95, 0.98$, $p_{\rm R}=0$, while the
  other parameters are
  the same as in Fig.~\ref{Sec7Fig:2}. The inset
  in part (a) shows the current in the whole range of
  the bias voltage. (After Ref.~\cite{weymannJPCM07})}
\end{figure}

When one of the leads is half-metallic ($p=1$) and the other one
is nonmagnetic ($p=0$), transport characteristics become
asymmetric with respect to the bias reversal. Furthermore, the
current can be suppressed in certain  bias regions, and this
suppression is accompanied by the occurrence of NDC. This
basically happens when the electrons residing in the dot have spin
opposite to that of electrons in the half-metallic drain
electrode. In Fig.~\ref{Sec7Fig:5} we show the current and Fano
factor for a quantum dot coupled to half-metallic (left) and
nonmagnetic (right) lead as a function of the bias voltage. For
the situation shown in Fig.~\ref{Sec7Fig:5}, current is suppressed
in certain regions of positive bias voltage. As follows from
Fig.~\ref{Sec7Fig:5}(a), there are three such blockade regions,
labelled with the consecutive numbers. On the other hand, for
negative bias voltage, the current changes monotonically with the
transport voltage, as shown in the inset of
Fig.~\ref{Sec7Fig:5}(a).

The mechanism of the blockade can be described as follows. In the
blockade region (1), $1\lesssim eV/U\lesssim 2$, the dot is in the
state $\ket{\downarrow}\ket{0}$ and the spin-down electron
residing in the dot has no possibility to tunnel further to the
left lead, which leads to suppression of the current. Here, the
first (second) {\it ket} corresponds to the first (second) orbital
level of the dot. The blockade region (2), $2\lesssim eV/U\lesssim
3$, is associated with the occupation of the states
$\ket{\downarrow}\ket{0}$ and $\ket{0}\ket{\downarrow}$. The
current is then prohibited due to the full occupation of the
single-particle spin-down states. When increasing the bias voltage
further, $3\lesssim eV/U\lesssim 4$, the blockade of the current
becomes suppressed [see the plateau between regions (2) and (3) in
Fig.~\ref{Sec7Fig:5}(a)], which is due to a finite occupation of
state $\ket{\uparrow\downarrow}\ket{0}$. Although tunnelling of
spin-down electrons is then blocked, the current is still carried
by spin-up electrons. In turn, the blockade region (3) occurs for
$4\lesssim eV/U\lesssim 5$, where the dot is in the triplet state
$\ket{\downarrow}\ket{\downarrow}$ and tunnelling is also
suppressed. Thus, the current is blocked when the total dot spin
$S_z$ is either $S_z=-1/2$ or $S_z=-1$, i.e., the spin of
electrons on the dot is opposite to that of electrons in the
half-metallic lead. There is no suppression of current for
negative voltage. This is because the electrons residing in the
dot can always tunnel to the nonmagnetic drain electrode. It is
also interesting to note that a pure triplet state is formed in
the region (3) \cite{weymannJPCM07,franssonNano06}.

The current blockade in the regions (1) to (3) in
Fig.~\ref{Sec7Fig:5}(a) for positive bias is not due to the
charging effects as in the Coulomb blockade regime, but due to a
particular occupation of the dot spin state. Such blockade is
frequently referred to as the Pauli spin blockade, and has already
been found in single-dot and double-dot systems
\cite{weinmann95,franssonPRB06,franssonNJP06,ono02,inarreaPRB07}.
It is worth noting that the current blockade in
Fig.~\ref{Sec7Fig:5}(a) is not complete -- there is a small
leakage current in each blockade region, which results from the
fact that the assumed spin polarization of the half-metallic lead
is not exactly equal to unity. When spin polarization is
increased, the current in the blockade regions decreases (compare
the curves for $p=0.95$ and $p=0.98$).

The spin blockade of charge current leads to the super-Poissonian
shot noise, i.e., the corresponding Fano factor is larger than
unity. On the other hand, the Fano factor outside the spin
blockade regions is sub-Poissonian (smaller than unity), as shown
in Fig.~\ref{Sec7Fig:5}(b). The enhancement of the shot noise in
the Pauli blockade regions is a consequence of large spin
asymmetry in the tunnelling processes. The occurrence of a
spin-down electron on the dot prevents further tunnelling
processes for a longer time, while spin-up electrons on the dot
can escape much faster, allowing further tunnelling processes.
This bunching of fast tunnelling processes gives rise to large
current fluctuations, and consequently also to Fano factors much
larger than unity.

Recently an interacting three-terminal quantum dot with
ferromagnetic leads was considered in Ref.
\cite{cottetPRL04,cottetPRB04}. The dot operated as a beam
splitter - one contact was a source and the other two acted as
drains. The authors found a dynamical spin blockade
(spin-dependent bunching of tunnelling events) and positive
zero-frequency cross-correlations of the current in the drain
electrodes.

\subsection{Cotunneling in two-level quantum dots}

In all transport characteristics discussed above only sequential
tunnelling processes were taken into account, while higher order
processes, in particular the cotunneling ones, have been
neglected. From our discussion on single-level dots we already
know that higher order contributions are particularly pronounced
in the blockade regions, where they have a significant influence
on TMR and Fano factor. The role of cotunneling processes in
two-level dots was studied in a recent paper \cite{weymannPRB08}.
It has been shown there that the cotunneling processes lead to a
significant enhancement (or reduction) of TMR in the blockade
regime. Outside the blockade regions, however, TMR is determined
mainly by sequential transport, so the contribution from
cotunneling processes is rather minor. Similarly, the cotunneling
processes significantly modify the Fano factor in the blockade
regions, while outside these regions the Fano factor is only
weakly sensitive to the cotunneling. More specifically, the Fano
factor in the blockade regions is reduced in comparison to that
obtained in the sequential transport limit. However, it is still
larger than one, indicating super-Poissonian shot noise in the
blockade regime due to bunching of the fastest cotunneling
processes.

There is also another interesting phenomenon in the blockade
regime, which resembles the zero bias anomaly observed in
single-level dots in the antiparallel magnetic configuration. This
takes place when each orbital level of the dot at equilibrium is
occupied by a single electron, and the system is in the Coulomb
blockade regime \cite{weymannEPL06}. The Hamiltonian $H$ of the
system is given by Eq. (\ref{Sec4Eq:QDHamiltonian}), where $H_{\rm
D}$ additionally includes the exchange term and can be expressed
by the formula \cite{izumidaPRL01,dongPRB02}
\begin{equation} \fl
  H_{\rm D} =\sum_{j\sigma} \varepsilon_{j} n_{j\sigma}
  + U \sum_j n_{j\uparrow} n_{j\downarrow}
  + U^\prime \sum_{\sigma\sigma^\prime} n_{1\sigma}n_{2\sigma^\prime}
  - J \sum_{\alpha\beta\gamma\delta} d^\dagger_{1\alpha}
    d_{1\beta} d^\dagger_{2\gamma} d_{2\delta}
    \vec{\sigma}_{\alpha\beta} \vec{\sigma}_{\gamma\delta} \,.
\end{equation}
The last term in $H_D$ corresponds to the exchange energy
according to the Hund's rule, with $J$ being the respective
exchange coupling and $\vec{\sigma}$ denoting a vector of Pauli
spin matrices.

Six different two-particle states of the dot are possible, these
are: three singlets $\ket{S=0,M=0}_1 =
(\ket{\uparrow}\ket{\downarrow} -
\ket{\downarrow}\ket{\uparrow})/\sqrt{2}$, $\ket{0,0}_2 =
\ket{\uparrow\downarrow} \ket{0}$, $\ket{0,0}_3 = \ket{0}
\ket{\uparrow\downarrow}$, and three triplets
$\ket{1,0}=(\ket{\uparrow}\ket{\downarrow} +
\ket{\downarrow}\ket{\uparrow})/\sqrt{2}$,
$\ket{1,1}=\ket{\uparrow}\ket{\uparrow}$ and
$\ket{1,-1}=\ket{\downarrow}\ket{\downarrow}$. In the case of
finite level spacing, $\delta\varepsilon>k_{\rm B}T,\Gamma$, and
$J<\delta\varepsilon$, the lowest singlet state is $\ket{0,0}_2$.
Transport characteristics strongly depend on the ground state of
the system. It is therefore useful to introduce the difference
between the energy of the lowest lying singlet $(\varepsilon_{\rm
S})$ and triplet $(\varepsilon_{\rm T})$ states, $\Delta_{\rm ST}
= \varepsilon_{\rm S} - \varepsilon_{\rm T} = J -
\delta\varepsilon$.

For $\Delta_{\rm ST}<0$, the ground state of the dot is a singlet,
$\ket{0,0}_2$, whereas for $\Delta_{\rm ST}>0$, the ground state
is a triplet, which is three-fold degenerate, $\ket{1,0}$,
$\ket{1,1}$, $\ket{1,-1}$. On the other hand, for $\Delta_{\rm
ST}=0$, the dot is in a mixed state and the occupation of singlet
and each triplet is equal at equilibrium and given by $1/4$.

\begin{figure}[h]
  \center
  \includegraphics[width=0.7\columnwidth]{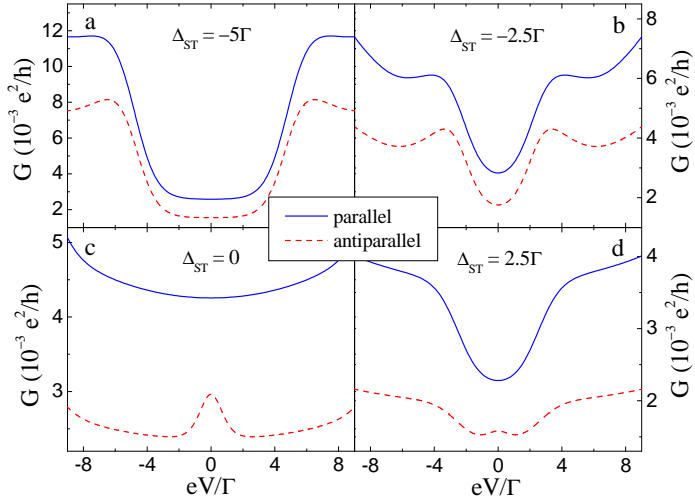}
  \caption{\label{Sec7Fig:8} (color online) The differential conductance
  in the parallel (solid line) and antiparallel (dashed line)
  magnetic configuration of the system
  as a function of the bias voltage for several values of
  $\Delta_{\rm ST}= J-\delta\varepsilon$
  as indicated: (a) $\Delta_{\rm ST}=-5\Gamma$,
  (b) $\Delta_{\rm ST}=-2.5\Gamma$, (c) $\Delta_{\rm ST}=0$,
  (d) $\Delta_{\rm ST}=2.5\Gamma$.
  The other parameters are: $k_{\rm B}T=0.5\Gamma$,
  $\varepsilon=-60\Gamma$, $\delta\varepsilon = 5\Gamma$,
  $U = U^\prime = 40\Gamma$, and $p=0.5$. (After Ref.~\cite{weymannEPL06})}
\end{figure}

Figure~\ref{Sec7Fig:8} presents the bias voltage dependence of
differential conductance $G$ in the parallel and antiparallel
magnetic configuration and for several values of $\Delta_{\rm ST}
= J- \delta\varepsilon$. Let us consider the situation in
Fig~\ref{Sec7Fig:8}(c), which corresponds to degenerate singlet
and triplet states, $\Delta_{\rm ST}=0$. The differential
conductance in the parallel configuration exhibits a smooth
parabolic dependence on the bias voltage, whereas in the
antiparallel configuration there is a maximum at the zero bias
voltage. This effect bears a resemblance to that found in the case
of singly occupied one-level quantum dots discussed in previous
sections \cite{weymannPRBBR05}. Here, however, the mechanism
leading to the maximum is different -- the zero-bias peak appears
due to cotunneling through the singlet and triplet states of the
dot. In the case of $\Delta_{\rm ST}=0$ and at low bias voltages,
all the four dot states, i.e. $\ket{0,0}_2$, $\ket{1,0}$,
$\ket{1,-1}$, $\ket{1,1}$, participate in transport on an equal
footing. Consequently, the current flows due to both spin-flip and
non-spin-flip cotunneling processes. To understand the mechanism
leading to the zero-bias peak, one should bear in mind that in the
antiparallel configuration the spin-majority electrons of one lead
tunnel to the spin-minority electron band of the other lead. For
example, for positive bias voltage (electrons tunnel then from the
right to left lead), the spin-$\uparrow$ electrons can easily
tunnel to the left lead (the spin-$\uparrow$ electrons are the
majority ones), while this is more difficult for the
spin-$\downarrow$ electrons (they tunnel to the minority electron
band). Thus, the occupation of state $\ket{1,-1}$
($\ket{\downarrow}\ket{\downarrow}$) is increased with increasing
bias voltage, while the occupation of state $\ket{1,1}$
($\ket{\uparrow}\ket{\uparrow}$) decreases. The unequal
occupations of these triplet states lead to a nonequilibrium spin
accumulation in the dot, $P_{\ket{1,1}}-P_{\ket{1,-1}} < 0$. It is
further interesting to note that in the antiparallel configuration
the possible non-spin-flip cotunneling processes are proportional
to $\Gamma_{\rm L}^{+}\Gamma_{\rm R}^{-}$ and $\Gamma_{\rm
L}^{-}\Gamma_{\rm R}^{+}$, whereas the spin-flip cotunneling is
proportional to $\Gamma_{\rm L}^{+}\Gamma_{\rm R}^{+}$ and
$\Gamma_{\rm L}^{-}\Gamma_{\rm R}^{-}$. It is clear that the
fastest cotunneling processes are the ones involving only the
majority spins, i.e. $\Gamma_{\rm L}^{+}\Gamma_{\rm R}^{+}$.
However, because of nonequilibrium spin accumulation, with
increasing the bias ($V>0$), the dot becomes dominantly occupied
by majority electrons of the right lead,
$P_{\ket{1,-1}}>P_{\ket{1,1}}$, and the processes proportional to
$\Gamma_{\rm L}^{+}\Gamma_{\rm R}^{+}$ are suppressed. As a
consequence, the differential conductance drops with the bias
voltage, leading to the zero-bias peak. This is thus the
nonequilibrium accumulation of spin $S=1$ which is responsible for
the maximum in $G$ at low bias voltage. Because the dot is coupled
symmetrically to the leads, there is no spin accumulation in the
parallel configuration and the differential conductance exhibits a
smooth parabolic dependence.

The maximum in the conductance at zero bias disappears for
$\Delta_{\rm ST}<0$ as well as for $\Delta_{\rm ST} >0$. Instead
of the narrow maximum, a relatively broad minimum develops then in
the conductance for both magnetic configurations, as shown in
Fig~\ref{Sec7Fig:8}(a,b) or for one magnetic configuration as in
Fig~\ref{Sec7Fig:8}(d). Consider for instance the case
$\Delta_{\rm ST}<0$,  shown in Fig.~\ref{Sec7Fig:8}(a,b). At low
bias voltage the dot is occupied by two electrons on the lowest
energy level and the ground state is singlet, $S=0$. Current is
mediated then only by non-spin-flip cotunneling processes. Once
$|eV|\gtrsim |\Delta_{\rm ST}|$, the triplet states start
participating in transport leading to an increase in the
differential conductance at $|eV|\approx|\Delta_{\rm ST}|$. Thus,
the suppression of cotunneling through $S=1$ states gives rise to
a broad minimum in the differential conductance. The width of this
minimum is determined by the splitting between the singlet and
triplet states, $2|\Delta_{\rm ST}|$.

\begin{figure}[h]
  \center
  \includegraphics[width=0.4\columnwidth]{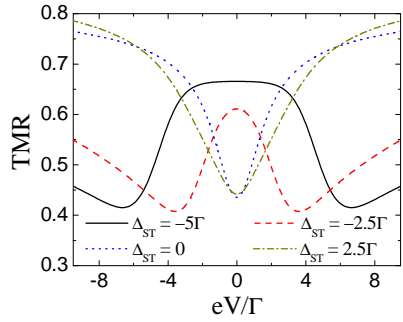}
  \caption{\label{Sec7Fig:10} (color online)
  Tunnel magnetoresistance
  as a function of the bias voltage for
  $\Delta_{\rm ST} = -5,-2.5,0,2.5 \Gamma$,
  as indicated in the figure.
  The other parameters are the same as in Fig.~\ref{Sec7Fig:8}.
  (After Ref.~\cite{weymannEPL06})}
\end{figure}

The bias voltage dependence of TMR is shown in
Fig.~\ref{Sec7Fig:10}. Generally, TMR displays a nontrivial
dependence on the ground state of the dot. When the dot is
occupied by a singlet, $\Delta_{\rm ST}<0$, TMR  displays a
maximum plateau at low bias, which can be approximated by, ${\rm
TMR}^{S=0} = 2p^2/(1-p^2)$ \cite{weymannEPL06}. This is the
Julliere value of TMR. Here, it results from the fact that in this
transport regime current flows due to non-spin-flip cotunneling.
On the other hand, for $\Delta_{\rm ST} \geq 0$, the TMR exhibits
a minimum at zero bias, which can be seen in Fig.~\ref{Sec7Fig:10}
for $\Delta_{\rm ST}=0$ and $\Delta_{\rm ST} = 2.5\Gamma$.


\subsection{Multi-level quantum dots based on single-wall
carbon nanotubes}


When the number of discrete levels in the dot (relevant for
electronic transport) becomes larger than two, transport
characteristics of the dots attached to ferromagnetic leads become
more complex and reveal further interesting features. Such
features were observed experimentally mainly in molecule-based
quantum dots. When a large natural molecule is weakly attached to
metallic leads, it can be treated simply as a multi-level QD
\cite{liangPRL02,sapmazPRB05,onacPRL06}. As a specific case, we
discuss transport properties of a single-wall metallic carbon
nanotube (CNT) weakly coupled to ferromagnetic leads
\cite{weymann2007CNT,mayrhoferEPJB07,koller07,
weymannAPL08,weymannNT08}.

Sequential transport in CNTs weakly coupled to nonmagnetic and
ferromagnetic leads was considered in a recent paper
\cite{weymann2007CNT}, where TMR and the Fano factor have been
calculated within the real time diagrammatic technique. The system
was modelled by the Hamiltonian in the form introduced by Oreg
{\it et al.} \cite{oregPRL00},
\begin{eqnarray}\label{Eq:HNT}
   H_{\rm QD} = \sum_{\mu j\sigma}
   \varepsilon_{\mu j} n_{\mu j\sigma} + \frac{U}{2}
   \left[ \sum_{\mu j\sigma} n_{\mu j\sigma} - N_0 \right]^2 \nonumber \\
   + \delta U \sum_{\mu j} n_{\mu j\uparrow} n_{\mu j\downarrow}
   + J \sum_{\mu j, \mu^\prime j^\prime}
   n_{\mu j\uparrow} n_{\mu^\prime j^\prime\downarrow}
   \,,
\end{eqnarray}
where $n_{\mu j\sigma} = d^{\dagger}_{\mu j\sigma}d_{\mu
j\sigma}$, and $d^{\dagger}_{\mu j\sigma}$ ($d_{\mu j\sigma}$) is
the creation (annihilation) operator of an electron with spin
$\sigma$ on the $j$th level in the subband $\mu$ ($\mu=1,2$). The
corresponding energy $\varepsilon_{\mu j}$ of the $j$th discrete
level in the subband $\mu$ is given by $\varepsilon_{\mu j} =
j\Delta + (\mu-1)\delta$, where $\Delta$ is the mean level spacing
and $\delta$ describes the energy mismatch between the discrete
levels corresponding to the two subbands. The second term in Eq.
(\ref{Eq:HNT}) stands for the electrostatic energy of a charged
CNT, with $U$ denoting the charging energy and $N_0$ being the
charge on the nanotube induced by gate voltages. The next term
corresponds to the on-level Coulomb interaction with $\delta U$
being the relevant on-site Coulomb parameter. Finally, the last
term in Eq. (\ref{Eq:HNT}) describes the exchange energy, with $J$
being the relevant exchange parameter.

\begin{figure}[t]
\center
  \includegraphics[width=0.6\columnwidth]{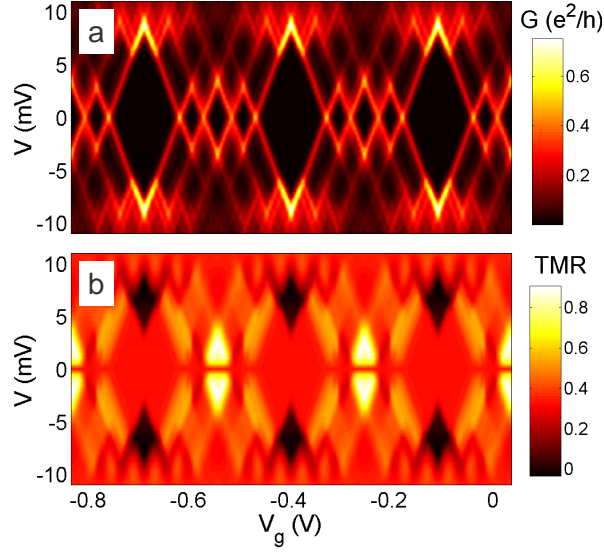}
  \caption{\label{Sec7Fig:11} (color online)
  The differential conductance in the parallel configuration (a)
  and tunnel magnetoresistance (b)
  as a function of bias and gate voltages.
  The parameters are: $\Delta = 8.4$ meV, $U/\Delta = 0.26$,
  $J/\Delta = 0.12$, $\delta U/\Delta=0.04$,
  $\delta/\Delta = 0.1$, $k_{\rm B}T/\Delta = 0.025$,
  $p_{L} = p_{R} = 0.5$, $x=0.14$, and $\Gamma = 0.2$ meV.}
\end{figure}

In Fig.~\ref{Sec7Fig:11} we show the differential conductance in
the parallel configuration and associated TMR, calculated in the
first order approximation. The results are for a specific choice
of parameters, namely $\delta U+J>\delta$. The latter condition
indicates that the following sequence of the ground states is
realized when filling the CNT with electrons while sweeping the
gate voltage (for $V=0$): $S=0,1/2,1,1/2$, where $S$ is the total
spin of the nanotube. This means that in a certain transport
regions, the systems is in the triplet state at equilibrium.
Fig.~\ref{Sec7Fig:11}(a) clearly reveals the fourfold pattern of
the conductance spectra, associated with filling of the
consecutive levels in the two electron subbands of the CNT. The
corresponding TMR is shown in Fig.~\ref{Sec7Fig:11}(b), and
reveals a complex variation with both gate and bias voltages.

As shown in Ref.~\cite{weymann2007CNT}, shot noise calculated in
the first order approximation is super-Poissonian in the blockade
regions, and sub-Poissonian in the transport regions were
sequential tunnelling dominates. The corresponding Fano factor in
the latter case is close to 1/2, while in the former one is above
1. Detailed calculations taking into account second order
(cotunneling) processes show that the Fano factor in the blockade
regions is generally larger than 1, although it is reduced in
comparison to that derived in the first order approximation,
similarly as in the case of two-level dot.


\section{Kondo effect in quantum dots coupled to ferromagnetic leads}


When a quantum dot or a large molecule is strongly coupled to
external leads, low-temperature transport characteristics reveal
features that are typical of the Kondo phenomenon
\cite{pasupathy04,martinekPRL03,cronenwett98,gores00,sasaki00,lopez03}.
To analyze the Kondo effect \cite{kondo}, one may use various
theoretical techniques. For instance, one can use the perturbative
methods, like  the real-time diagrammatic technique discussed in
previous sections. However, one has to go beyond the
approximations used in the discussion of the sequential and
cotunneling transport. Another approximate method which is
frequently used to study electronic transport through quantum
dots, including the Kondo regime, is the equation of motion method
for the relevant Green function. Alternatively, one may employ
nonperturbative methods, such as for example the numerical
renormalization group method \cite{wilson75,krishna80a,
krishna80b,bulla07}.

The Kondo effect \cite{book:hewson} in electronic transport
through a quantum dot has been first predicted theoretically in
Refs.~\cite{glazman88,ng88}. Since then, it has become well
documented experimentally
\cite{cronenwett98,goldhaber,simmel99,schmid00,wiel00}. The dot
with an odd number of electrons has a local spin which at low
temperatures, $k_{\rm B}T \le k_{\rm B}T_{\rm K} \ll \Gamma$, and
in the presence of strong coupling to the electrodes behaves
effectively like a magnetic Kondo impurity.
Exchange coupling with the leads' electrons gives rise to spin
fluctuations in the dot and screening of the dot's spin. This, in
turn, leads to the formation of a singlet state and Kondo
resonance peak at the Fermi level in the dot's density of states.
This Kondo-correlated state leads to an increased transmission
through the dot, and also gives rise to a sharp zero-bias anomaly
in the current-voltage characteristics.
The successful observation of the Kondo effect in
semiconductor-based quantum dots, as well as in molecular quantum
dots based for instance on carbon nanotubes \cite{cobden,bachtold}
or other single molecules \cite{parkNat02,liangNat02} attached to
metallic electrodes opened new possibility to study the influence
of ferromagnetism  on the Kondo phenomenon
\cite{hamayaAPL07c,pasupathy04,nygard}.

In this section we briefly discuss how the Kondo effect is
modified by ferromagnetism of the electrodes. First, we note that
in the extreme case of half-metallic leads, i.e. in the absence of
minority-spin electrons, the screening of the dot spin is not
possible in the parallel magnetic configuration. Consequently, no
Kondo-correlated state can be then formed in this particular
magnetic configuration.

The possibility of the Kondo effect in a quantum dot attached to
ferromagnetic electrodes was discussed in a number of papers
\cite{surgueevPRB02,martinekPRL03,swirkowiczPRB06,lopez03,zhang,bulka,
martinekPRL03a,choi,swirkowicz04,swirkowicz05,sindelPRB07}. It was
shown, that the Kondo resonance in the parallel configuration is
split and suppressed in the presence of ferromagnetic leads
\cite{martinekPRL03,martinekPRL03a,choi,sindelPRB07,swirkowicz05}.
However, it was also demonstrated that this splitting can be
compensated by an appropriate external magnetic field which
restores the Kondo effect \cite{martinekPRL03,martinekPRL03a}.

Let us begin with the collinear (parallel and antiparallel)
magnetic configurations. For the antiparallel configuration and
vanishing magnetic field and bias voltage, the model is equivalent
to a quantum dot coupled to a single lead with density of states
$\rho_{L\uparrow}+\rho_{R\uparrow} =
\rho_{L\downarrow}+\rho_{R\downarrow}$ \cite{glazman88,ng88}.
Thus, the usual Kondo resonance forms in such a case, which is the
same as for nonmagnetic electrodes \cite{book:hewson}.
The situation changes in the parallel configuration, where there
is an overall asymmetry for up and down spins, say
$\rho_{L\uparrow}+\rho_{R\uparrow} >
\rho_{L\downarrow}+\rho_{R\downarrow}$. This significantly reduces
(or even suppresses) the Kondo effect.


\subsection{Poor man's scaling approach}
\label{sec:pertur}


Some basic relations concerning the Kondo effect, and particularly
those concerning the Kondo temperature, can be derived from a
simple poor man's scaling \cite{anderson} performed in two stages
\cite{haldane}.
First one reduces the energy scale of the effective electron
bandwidth $D$ starting from $D_0$. This leads to a renormalization
of the level position $\epsilon_{\rm d \sigma}$ according to the
scaling equations \cite{martinekPRL03}
\begin{eqnarray}
  \frac{d \epsilon_{\rm d
\sigma} }{d \ln( D_0/D)}
  = \frac{ \Gamma_{ \bar{\sigma} }}{2 \pi} \, .
\label{eq:Hal_scaling}
\end{eqnarray}
This leads to a spin splitting of the level, which in the presence
of a magnetic field simply adds to the initial Zeeman splitting
$\Delta \epsilon_{\rm d }$. As a result, one finds
\begin{eqnarray}
 \Delta \widetilde{\epsilon}_{\rm d } = \widetilde
\epsilon_{\rm d\uparrow} - \widetilde \epsilon_{\rm d \downarrow}
= - (1/\pi) P \Gamma \ln(D_0/D) + \Delta \epsilon_{\rm d} \; .
 \label{eq:scalingresult}
\end{eqnarray}
The scaling of Eq.~(\ref{eq:Hal_scaling}) is terminated at
$\widetilde{D} \sim -\widetilde{\epsilon}_{\rm d}$ \cite{haldane}.

The strong-coupling limit can be reached by tuning the external
magnetic field $B$ in such a way that the total effective Zeeman
splitting vanishes, $\Delta \widetilde{\epsilon}_{\rm d} = 0$. In
the second stage \cite{anderson}, spin fluctuations are integrated
out. The Schrieffer-Wolff transformation \cite{book:hewson} allows
to map the Anderson model, with renormalized parameters
$\widetilde{D}$ and $\widetilde{\epsilon}_{\rm d}$, to the
effective Kondo Hamiltonian \cite{martinekPRL03}
\begin{eqnarray}
  H_{\rm Kondo} &=& J_+ S^+ \sum_{rr'kq}a^\dagger_{rk\downarrow}a_{r'q\uparrow}
  + J_- S^- \sum_{rr'kq} a^\dagger_{rq\uparrow}a_{r'k\downarrow}
\nonumber \\
  &&
  + S^z \left( J_{z\uparrow} \sum_{rr'qq'}
  a^\dagger_{rq\uparrow} a_{r'q'\uparrow}
  - J_{z\downarrow} \sum_{rr'kk'}
  a^\dagger_{rk\downarrow}a_{r'k'\downarrow}
  \right)
\, , \label{Kondo}
\end{eqnarray}
with $J_+ = J_- = J_{z\uparrow} = J_{z\downarrow} = {|T|^2
/|\widetilde \epsilon_{\rm d}|} \equiv J_0$ in the large-$U$
limit. Although initially identical, the three coupling constants
$J_+ = J_- \equiv J_\pm$, $J_{z\uparrow}$, and $J_{z\downarrow}$
are renormalized differently during the second stage of scaling.
The relevant scaling equations are

\begin{eqnarray}
   \frac{ d (\rho_\pm J_\pm) }{d \ln(\widetilde D/D) }
   &=& \rho_\pm J_\pm
   (\rho_\uparrow J_{z \uparrow} + \rho_\downarrow J_{z \downarrow} )
\label{eq:scaling1}
\\
   \frac{ d (\rho_\sigma J_{z\sigma}) }{d \ln(\widetilde D/D)} &=&
   2 (\rho_\pm J_\pm)^2
\label{eq:scaling2}
\end{eqnarray}
All the coupling constants reach the stable strong-coupling fixed
point $J_\pm = J_{z\uparrow} = J_{z\downarrow} = \infty$ at the
Kondo energy scale, $D \sim k_B T_K$. For the parallel
configuration, the Kondo temperature in leading order depends on
the polarization $p$ in the leads following the formula
\begin{equation}
  T_{\rm K} (p) \approx \widetilde D \exp \left\{ - {1\over
    (\rho_\uparrow + \rho_\downarrow) J_0} \, {{\rm arctanh} (p) \over p}
  \right\} \; .
\label{eq:Kondo_temperature}
\end{equation}
The Kondo temperature is maximal for nonmagnetic leads, $p = 0$,
and vanishes for $p \rightarrow 1$. The unitary limit for the
parallel configuration can be reached by tuning the magnetic
field, as discussed above. In this case, the maximum conductance
through the quantum dot is the same as for nonmagnetic leads, i.e.
$G_{{\rm max},\sigma}^P = e^2/h$ per spin.

The scaling procedure was extended in Ref.
\cite{eto06,simon06,wawrzyniak08} to noncollinear configurations,
where instead of Eq.~(\ref{eq:Kondo_temperature}) one finds
\begin{equation}
  T_{\rm K} (p) \approx \widetilde D \exp \left\{ - {1\over
    (\rho_\uparrow + \rho_\downarrow) J_0} \, {{\rm arctanh}
    (p\cos (\theta /2)) \over p\cos (\theta /2)}
  \right\} \; .
\end{equation}
This formula shows explicitly how the Kondo temperature varies
with the angle $\theta $ between magnetic moments of the leads.


\subsection{Numerical renormalization group}


Numerical renormalization-group (NRG) technique \cite{book:hewson,
wilson75,bulla07} is one of the most powerful and accurate methods
available currently to study strongly-correlated systems in the
Kondo regime. This technique has been adapted recently to the case
of a quantum dot coupled to ferromagnetic leads
\cite{simon06,martinekPRL03a,choi}. A simple way to model the
ferromagnetic leads in the standard NRG procedure is to take the
density of states in the leads to be constant and independent of
electron spin, $\rho_{r \sigma} (\omega) \equiv \rho$, the
bandwidths to be equal $D_{\uparrow} = D_{\downarrow}$, and lump
all the spin-dependence into the spin-dependent coupling
parameter, $\Gamma_{r \sigma} (\omega)$, which can be taken as
independent of energy.

The NRG method, with recent improvements related to high-energy
features and finite magnetic field
\cite{costi1,hofstetter2,costiPRL00,costiPRB01}, is a
well-established method to study the Kondo impurity (quantum dot)
physics. Using this technique one can calculate the level
occupation $ n_{\rm d \sigma}  \equiv \langle d_{\sigma}^{\dagger}
d_{\sigma}\rangle $, and the spin-dependent single-particle
spectral density $ A_{ \sigma}(\omega )$ for arbitrary temperature
$T$, magnetic field $B$ and polarization $p$. Using the spectral
function one can find the spin-resolved linear conductance
\begin{equation}
G_{\sigma} = {e^2 \over \hbar } \frac{ \Gamma_{\rm L\sigma}
\Gamma_{\rm R\sigma} }{( \Gamma_{\rm L\sigma} + \Gamma_{\rm
R\sigma} )}
 \int_{- \infty } ^{ \infty}
d \omega A_{\sigma}(\omega) (-{ \partial f( \omega ) \over
\partial \omega } )
\label{eq:conductance}
\end{equation}
with $f( \omega )$ denoting the Fermi function.


\subsection{Nonequilibrium transport: zero-bias anomaly}


The Kondo effect is usually observed as a zero bias anomaly in the
differential conductance. This requires theoretical methods
applicable to nonequilibrium situations. One of such techniques is
based on the nonequilibrium Green functions. The latter are
usually determined from the relevant equation of motion and the
appropriate decoupling scheme. To describe the main features of
the Kondo effect in transport through quantum dots connected to
ferromagnetic leads (at least qualitatively), one may use the
decoupling scheme introduced in Ref.~\cite{meir93,meir94}, but
generalized by a self-consistent determination of the level energy
to account for the exchange field
\cite{martinekPRL03,swirkowicz04}. Alternatively, one may
calculate the exchange field first and then include it in the
formalism by hand \cite{swirkowicz05}.

\begin{figure}[t]
\center
  \includegraphics[width=0.6\columnwidth]{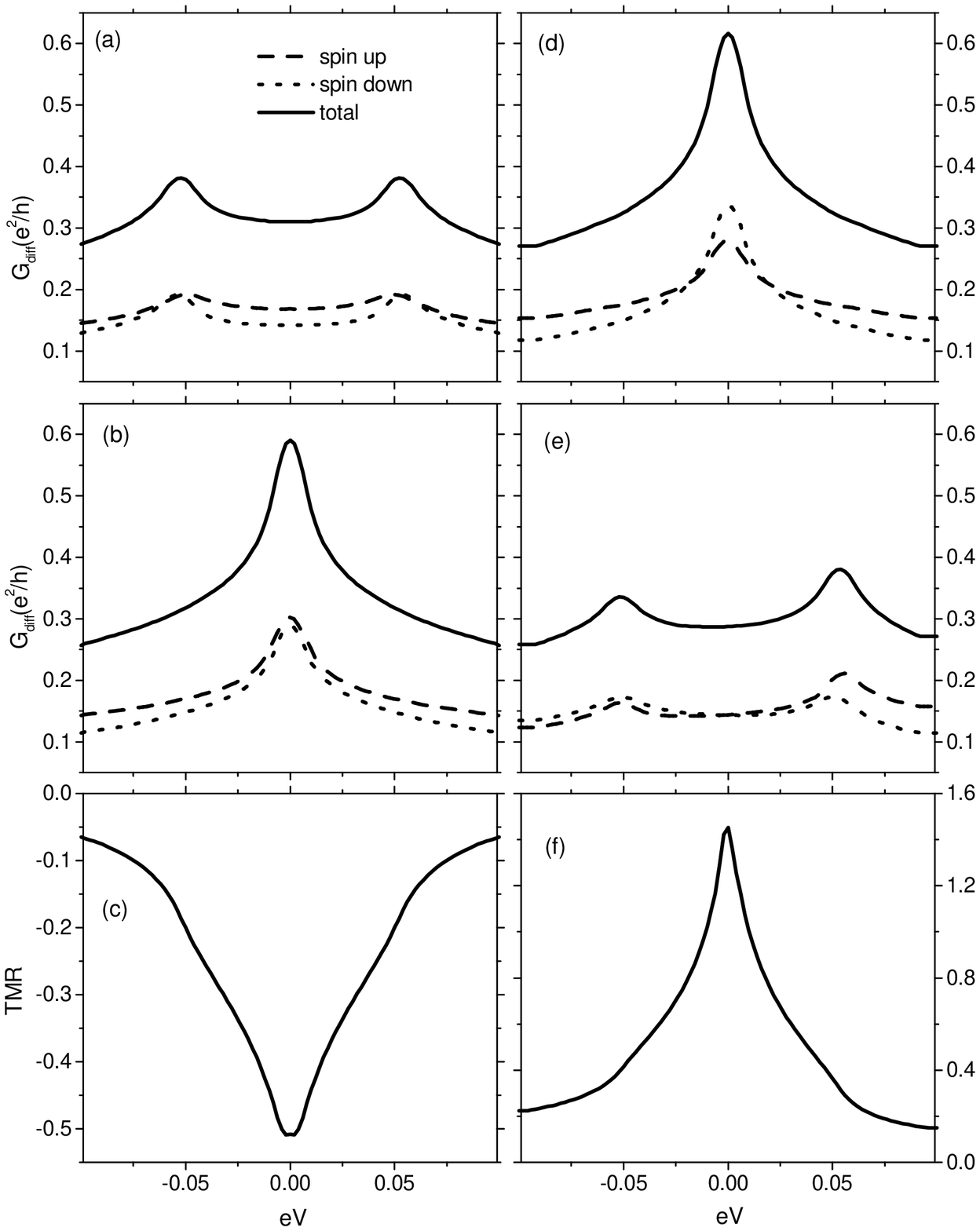}
  \caption{\label{Sec8Fig:3}
Differential conductance in the P (a,d) and AP (b,e)
configurations, and the corresponding TMR (c,f). Contributions to
the conductance from spin-up (dashed lines) and spin-down (dotted
lines) channels are also indicated. Left column (a,b,c)
corresponds to zero magnetic field, whereas the right one (d,e,f)
to the case when the compensating magnetic field $B=14.53$ T is
applied (for the Lande factor $g=0.152$). The numerical results
are for large $U$ limit, while the other parameters are:
$k_BT/\Gamma =0.01$, $p=0.2$, $\Gamma /D=0.001$. (After
Ref.~\cite{swirkowicz04})}
\end{figure}

Figure~\ref{Sec8Fig:3} shows the differential conductance as a
function of the transport voltage. For zero magnetic field there
is a pronounced splitting of the peak in the parallel
configuration [Fig.~\ref{Sec8Fig:3}(a)], which can be tuned away
by an appropriate external magnetic field
[Fig.~\ref{Sec8Fig:3}(b)]. In the antiparallel configuration, in
turn, the opposite happens, i.e. there is no splitting at $B=0$
[Fig.~\ref{Sec8Fig:3}(d)] and a  finite splitting at $B>0$
[Fig.~\ref{Sec8Fig:3}(e)] with an additional asymmetry in the peak
amplitudes as a function of the bias voltage.
Figure~\ref{Sec8Fig:3}(c,f) shows the corresponding TMR, which
becomes negative without magnetic field and changes sign to
positive when a compensating magnetic field is applied.

Within the EOM approach the effect of spin-dependent quantum
charge fluctuations is accounted in the self-consistent but
intuitive manner. The real-time diagrammatic technique
\cite{schoeller94,koenig96} enables one to construct a systematic
approach, where the effect of ferromagnetic electrodes can be
analyzed without any additional assumptions. Recently the resonant
tunnelling approximation (RTA) was extended by Utsumi {\it et al.}
\cite{utsumiPRB05} to account for influence of the electrodes'
ferromagnetism on the Kondo phenomenon. This technique gives more
reasonable results and allows for further systematic insight into
the physics of the transport through quantum dots in the Kondo
regime.

The Kondo phenomenon in noncollinear magnetic configuration was
considered in Refs \cite{swirkowiczPRB06,eto06,simon06}. The
results show that in symmetric systems, the Kondo anomaly
gradually disappears when the magnetic configuration varies from
antiparallel to parallel one. However, the main drop of the Kondo
peak appears already at small deviations from the antiparallel
alignment.

Recently Pasupathy et al.~\cite{pasupathy04} studied electronic
transport through a single ${\rm C}_{60}$ molecule attached to
ferromagnetic Nickel electrodes in the Kondo regime. It was shown
that the Kondo correlations appear even in the presence of
ferromagnetic leads. The zero-bias anomaly in the nonequilibrium
conductance, however, was split for the parallel alignment of the
leads magnetization in agreement with theoretical predictions. For
the antiparallel alignment, on the other hand,  no splitting of
the zero-bias anomaly was observed. Some residual splitting
observed in this geometry for some samples can be interpreted as
an effect of asymmetric coupling $\Gamma_L \neq \Gamma_R $.
Similar behavior has also been observed in electronic transport
through carbon nanotubes coupled to ferromagnetic leads
\cite{lindelof}.


\section{Concluding remarks}

In this review we addressed certain aspects of electronic
transport in mesoscopic tunnel junctions consisting of magnetic
metallic nanoparticles, semiconducting quantum dots, or molecules
attached to ferromagnetic electrodes. We are aware that the review
does not cover all the problems considered in such systems and
many references have not been included. There are some  other
aspects of electronic and spin transport which have been omitted.
One of such aspects is spin torque acting on the island and/or
electrodes in ferromagnetic single electron transistors based on
magnetic metallic nanoparticles \cite{kowalik}. Another topic
omitted in this review is the role of electron coupling to a
phonon bath. Such a coupling leads to new interesting phenomena in
transport characteristics \cite{galperin}. For instance, the
electron-phonon coupling leads to phonon satellite peaks in the
differential conductance. However, the most important spin effects
generated by spin-dependent tunnelling through the barriers
separating the central part from the ferromagnetic electrodes have
been addressed.

We note that spin polarized transport through nanostructures,
where the Coulomb effects become important, is rapidly developing
in recent years. This applies not only to theoretical part of the
subject, but also to the experimental side, where recent progress
in nanotechnology allows to attach external leads to individual
nanoparticles, quantum dots, and single molecules. An interesting
and new field within this topic is the spintronics based on
molecular magnets. Such molecules are good candidates for magnetic
memory cells and also are considered as potential candidates for
information processing devices. There is no doubt, that the rapid
progress is the physics and technology of spin polarized
electronic transport through nanostructures is stimulated by
application possibilities in magnetoelectronics, spintronics, and
information technology.


\ack

We acknowledge discussions with J. K\"onig, J. Martinek, W.
Rudzi\'nski, R. \'Swirkowicz, and M. Wilczy\'nski. This work was
supported by funds of the Ministry of Science and Higher Education
as a research project in years 2006-2009. I. W. also acknowledges
support from the Foundation for Polish Science.


\section*{References}


\end{document}